\documentclass[12pt]{article}
\pdfoutput=1

\usepackage{putex}
\usepackage{amsmath,amssymb,amsfonts}
\usepackage{psfrag}
\usepackage{footmisc}
\usepackage{url}
\usepackage{mathtools}
\usepackage{color}

\usepackage{tikz}
    \usepackage{amssymb,amsfonts,amsmath}
    \usepackage{tkz-euclide}
        \usetikzlibrary{arrows,calc,patterns}
\usepackage{pgfplots}

\definecolor{darkblue}{rgb}{0.1,0.1,.7}
\definecolor{purple}{rgb}{0.6,0,0.6}
\definecolor{orange}{rgb}{0.9,0.6,0}
\usepackage[colorlinks, linkcolor=darkblue, citecolor=darkblue, urlcolor=darkblue, linktocpage]{hyperref} 
\usepackage[square, comma, compress,numbers]{natbib}
\usepackage[]{amsmath}
\usepackage[]{graphicx}
\usepackage[]{latexsym}
\usepackage[utf8]{inputenc}
\usepackage{geometry}
\usepackage{amscd}
\usepackage[all,cmtip]{xy}
\usepackage{mathrsfs}

\usepackage[margin=10pt,font=small,labelfont=bf]{caption}
\usepackage{dsdshorthand}
\usepackage{changepage}
\usepackage{setspace}
\usepackage{tikz}
\usepackage{subcaption}

\usepackage[titles]{tocloft}
\usetikzlibrary{arrows,positioning}

\def\SL2{\widetilde{SL}(2,\mathbb R)}

\def\Index{{\rm Index}}
\def\mC{\mathcal C}

\newcommand\mR{\mathbb{R}}
\newcommand\mZ{\mathbb{Z}}

\numberwithin{equation}{section}

\newcommand{\grp}[1]{\mathrm{#1}}
\newcommand{\str}{\text{Str}\,}

\newcommand {\bes} {\begin {equation*}}
\newcommand {\ees} {\end {equation*}}
\newcommand {\beq} {\begin {equation}}
\newcommand {\eeq} {\end {equation}}
\newcommand {\bea} {\begin {eqnarray}}
\newcommand {\ea} {\end {eqnarray}}
\newcommand {\eea} {\end {eqnarray}}

\newcommand{\Sch}{\text{Sch}}

\numberwithin{equation}{section}

\def\<{\langle}
\def\>{\rangle}

\tikzset{
    >=stealth',
    punkt/.style={
           rectangle,
           rounded corners,
           draw=black, very thick,
           text width=15em,
           minimum height=2em,
           text centered},
    pil/.style={
           ->,
           thick,
           shorten <=2pt,
           shorten >=2pt,}
}

 \def\ie{\begin{equation}\begin{aligned}}
\def\fe{\end{aligned}\end{equation}}

\begin{document}

    \institution{SU}{${}^1$ Stanford Institute for Theoretical Physics, Stanford University, Stanford, CA 94305, USA}
    \institution{OX}{${}^2$ Mathematical Institute, University of Oxford, Oxford, OX2 6GG, UK}
\institution{UCSB}{${}^3$ Physics Department, University of California, Santa Barbara, CA 93106, USA}

\title{
Supersymmetric indices factorize  
}

\authors{Luca V. Iliesiu${}^1$, Murat Kolo\u{g}lu$^2$ and Gustavo J. Turiaci${}^3$ }

\abstract{
The extent to which quantum mechanical features of black holes can be understood from the Euclidean gravity path integral has recently received significant attention. In this paper, we examine this question for the calculation of the supersymmetric index. For concreteness, we focus on the case of charged black holes in asymptotically flat four-dimensional $\mathcal{N}=2$ ungauged supergravity. We show that the gravity path integral with supersymmetric boundary conditions has an infinite family of Kerr-Newman classical saddles with different angular velocities. We argue that fermionic zero-mode fluctuations are present around each of these solutions making their contribution vanish, except for a single saddle that is BPS and gives the expected value of the index. We then turn to non-perturbative corrections involving spacetime wormholes and show that a fermionic zero mode is present in all these geometries, making their contribution vanish once again. This mechanism works for both single- and multi-boundary path integrals. In particular, only disconnected geometries without wormholes contribute to the gravitational path integral which computes the index, and the factorization puzzle that plagues the black hole partition function is resolved for the supersymmetric index. Finally, we classify all other single-centered geometries that yield non-perturbative contributions to the gravitational index of each boundary.
}
\date{}

\maketitle
\tableofcontents

\newpage

\section{Introduction}
According to holography, we expect black holes to be fully described by quantum mechanical degrees of freedom with a discrete spectrum and unitary evolution. However, these quantum mechanical features are not manifest in the gravitational description of a black hole, and understanding how they emerge from geometry is an important problem. Recently, there has been significant progress in understanding these questions by using the Euclidean path integral. For example, the discreteness of the black hole spectrum has an imprint in the late-time behavior of some observables \cite{Maldacena:2001kr}, which in some simple models can be obtained by including spacetime wormholes in the Euclidean path integral \cite{Saad:2019lba}.

While the addition of spacetime wormholes in the gravitational path integral has shed some light on various problems, it has also introduced others. It has been known for quite a while that the sum over spacetime wormholes generates an indeterminacy of coupling constants \cite{Cole, GiStInc}. This issue has gone in the context of holography under the name of the factorization puzzle. It was raised in \cite{Witten:1999xp} and further studied in \cite{Maldacena:2004rf}. The main idea is that spacetime wormholes give contributions to quantities that can only be understood as arising from a disorder average over theories. For example, a connected nonzero contribution to a path integral with multiple boundaries cannot make sense if each boundary is independently described by a single theory. This problem leads to a possible generalization of holography where black holes, and quantum gravity in general, are described by a classical average of quantum systems. Whether this is fundamental or not remains to be seen. Curiously, examples in string theory give unique quantum systems and do not seem to have this ``averaging'' feature.\footnote{Sometimes, it is said that there might not be a factorization problem in higher dimensions because of these examples, but we don't see any justification for this claim from a strictly bulk gravity argument.}

This paper aims to extend the use of Euclidean path integral techniques in gravity. Is it possible to obtain fine-grained information about some special class of states and resolve the factorization puzzle at least for quantities solely involving these states? With this goal in mind, we will focus on the computation of supersymmetry-protected quantities in supergravity. In particular, we will use the gravitational path integral to compute the supersymmetric indices of quantum mechanical systems describing certain black holes. We will clarify the role of fermionic zero modes in this computation and use this understanding to show that spacetimes wormholes do not contribute. 

\begin{figure}[t!]
\begin{center}
\begin{tikzpicture}
    \node[anchor=south west,inner sep=0] at (0,0) {\includegraphics[width=0.4\textwidth]{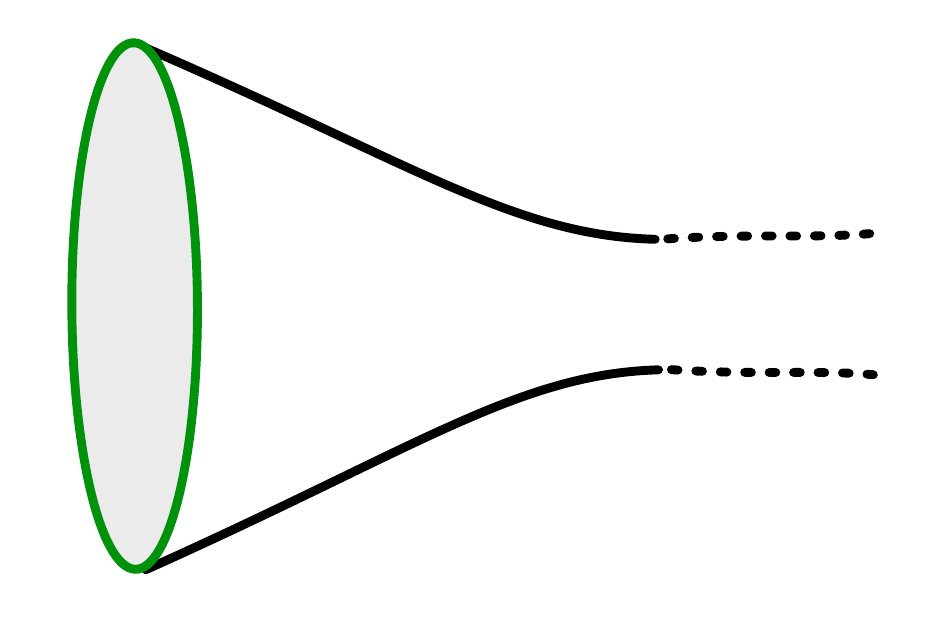}};
     \draw (-3.5,+2.2) node  {$\text{Index}_\text{grav.}(\beta,Q) \,\,\sim  \,\,\, \int Dg_{\mu \nu} \, D\text{[Fields]}$};
      \draw (1.4,-0.1) node  {$F(\tau) = F(\tau+\beta)$};
       \draw (1.4,-0.7) node  {$B(\tau) = B(\tau+\beta)$};
\end{tikzpicture}
\end{center}
\vspace{-0.6cm}
\caption{The gravitational path integral defined for asymptotically flat or AdS geometry with supersymmetric boundary conditions, i.e.~with periodic bosons ($B$) and fermions ($F$). We will first clarify how to explicitly impose such boundary conditions. Then, our goal is to find the geometries that contribute to the gravitational path integral with the above boundary conditions.}
\label{fig:boundaryindexgrav}
\end{figure}

Before we proceed, we want to introduce two notions of what we mean by the `index' of a black hole. The first is the gravitational index, which is given by a gravity calculation that does not rely on the existence of a dual supersymmetric quantum mechanics. For a $d$-dimensional theory, it is defined as a gravitational path integral where the asymptotic boundary is of the form $S^1 \times \cM_{d-2}$ (with $ \cM_{d-2}$ some specified spatial manifold) and where the same periodic boundary conditions are imposed on all bosons and fermions around $S^1$. The proper length of $S^1$ is typically identified as the inverse temperature of the black hole, $\beta$. We will denote such boundaries by a green line as in figure \ref{fig:boundaryindexgrav}.

The second notion of an index is defined in terms of the putative dual supersymmetric quantum field theory describing the black hole \cite{Witten:1982df}. Working in a sector of fixed charge, the index can be generically expressed as
\be \label{eq:QMindex}
\text{Index}_{\rm QFT}(\beta, Q)\equiv \Tr_Q \left[ (-1)^F e^{-\beta H} \right] = (d_\text{boson} - d_\text{fermion})e^{- \beta E_\text{BPS}(Q)}.
\ee
The main property of this quantity is that it only gets contributions from BPS states that preserve some supersymmetry. This is due to a cancellation between all non-BPS bosonic and fermionic states. The energy of BPS states is generically fixed by the supersymmetry algebra to be what we call $E_{\rm BPS}(Q)$, and for this reason, the temperature dependence is extremely simple. The prefactor counts $(d_\text{boson} - d_\text{fermion})$, the number of BPS states weighted with a minus sign for fermions. This quantity should be an integer with no condition on its overall sign.

With these two notions in mind, we want to study to what extent the gravitational index computed from supergravity has the expected properties of a quantum mechanical dual. In the first part of the paper, we focus on perturbative contributions: the classical gravity saddles and the fluctuations around them. We will argue that the index is computed by smooth saddles\footnote{Some previous examples where similar calculations were done are \cite{Dijkgraaf:2000fq} and \cite{Cabo-Bizet:2018ehj}.} and emphasize that the contribution of the one-loop determinant around them is crucial for recovering a sensible answer with the correct temperature dependence in \eqref{eq:QMindex}. In the second part of the paper, we will focus on non-perturbative contributions from spacetime wormholes and will show that they all vanish, implying that supersymmetric protected quantities are not affected by the factorization puzzle.  

To make things concrete, in \textbf{sections \ref{sec:saddle-point-analysis}} and \textbf{\ref{sec:diskN4}} we will discuss the computation of the gravitational index for perhaps the simplest black hole in supergravity. We will consider four-dimensional asymptotically flat $\mathcal{N}=2$ ungauged supergravity. In a sector of fixed $U(1)$ charge $Q$, we expect the index to count---with a sign---the number of extremal Reissner-Nordstr\"om black hole states with an AdS$_2 \times S^2$ throat. At low temperatures, this solution has an approximate $PSU(1,1|2)$ symmetry in the throat, broken by finite temperature effects. As it stands, this theory of $\cN=2$ ungauged supergravity is not expected to be UV complete. Still, our setup will be very helpful to clarify some conceptual questions, and we will indeed find answers consistent with \eqref{eq:QMindex}.\footnote{A better UV complete model might be to study black holes in higher-dimensional AdS, which usually involve throats with approximate $SU(1,1|1)$ symmetry. One could also view the $\cN=2$ ungauged supergravity discussed here as a sector for $\cN=4$ and $\cN=8$ supergravities in asymptotically flat-space which, depending on the matter content, can have a UV completion in string theory.}   

 The procedure we employ to obtain the gravitational index for such black holes is the following. Firstly, we use the fact that $(-1)^F=e^{2\pi i J} $, where $J$ is the angular momentum along any specified spatial direction. We then interpret the gravitational index as the quantity computed by a usual partition function at finite inverse temperature $\beta$ and with an imaginary angular velocity $\Omega = \frac{2 \pi i }{\beta}$. Perhaps against intuition, the gravitational saddles computing the index is a family of Kerr-Newman black holes related by an integer shift $\Omega \to \Omega + \frac{4 \pi i}{\beta}  \cdot \mathbb{Z} $, studied near-extremality in \cite{Heydeman:2020hhw}. All solutions in this infinite family are smooth both in terms of the metric and the spin structure. However, only one of those solutions gives an answer consistent with the quantum mechanical expectation of the index, i.e.~with the gravitational path integral having the same temperature dependence as in \eqref{eq:QMindex}. Therefore, to obtain \eqref{eq:QMindex} one must understand from the bulk perspective why all other saddles do not contribute to the index. The resolution of this puzzle is to include one-loop determinants, especially the contribution from the gravitinos. We will show that all saddles which would give inconsistent contributions---namely, those which are not BPS---come with physical gravitino zero modes. Integrating over these zero modes makes the contributions of those saddles vanish.\footnote{It would be interesting if a similar mechanism is responsible for picking the correct saddles for the index of black holes in AdS$_5\times S^5$ \cite{Aharony:2021zkr}.} These zero modes become gauge modes for the only (BPS) saddle giving the correct answer; accordingly, we should not integrate over them, and the answer does not vanish.\footnote{Since it is rather technical and is not widely discussed in the literature, we will carefully explain the distinction between gauge and physical zero modes in section \ref{sec:diskN4}.} 
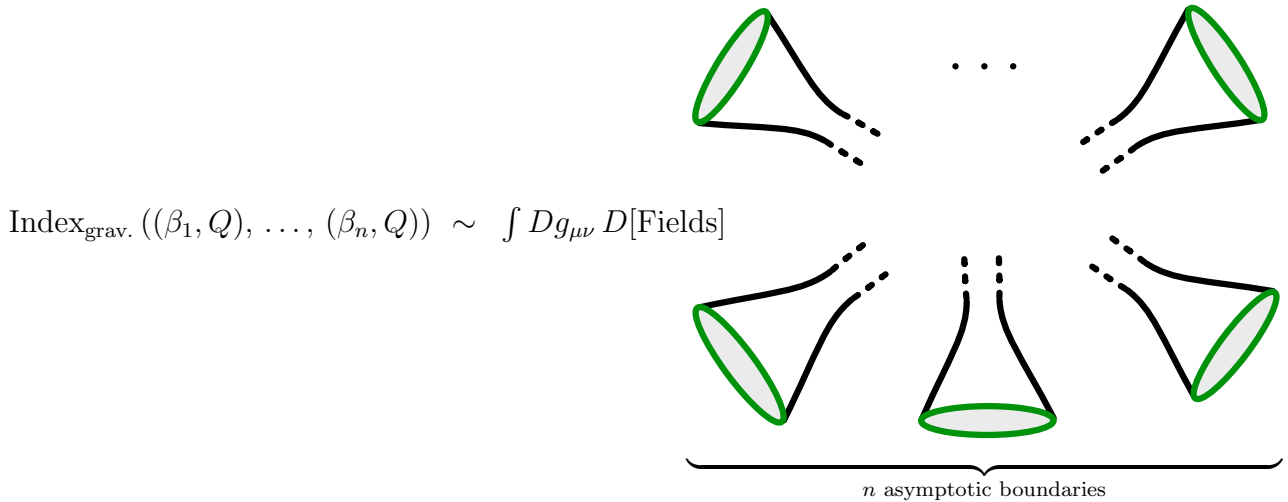
\begin{figure}[t!]
\begin{center}
\begin{tikzpicture}
    \node[anchor=south west,inner sep=0] at (0,0) {\includegraphics[width=0.5\textwidth]{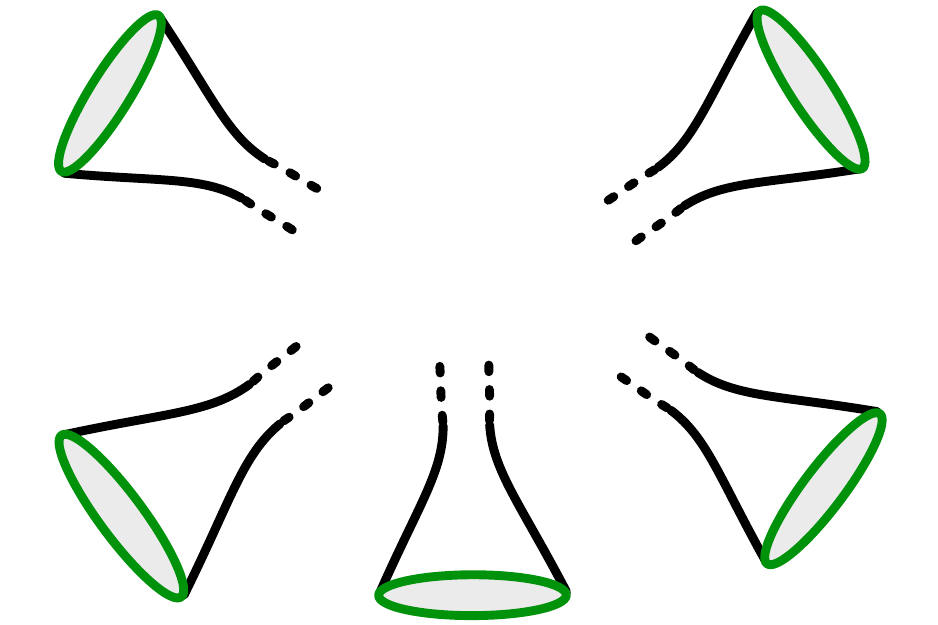}};
     \draw (-3.8,+2.9) node  {$\text{Index}_\text{grav.}\left((\beta_1,Q),\,\dots, \, (\beta_n,Q) \right)\,\,\sim  \,\,\, \int Dg_{\mu \nu} \, D\text{[Fields]}$};
     \draw (4.4,+5.0) node  {\huge ${\dots}$};
    \draw (4.4,-0.5) node  {$\underbrace{\hspace{0.45\textwidth}}_{n \text{ asymptotic boundaries}}$};
\end{tikzpicture}
\end{center}
\vspace{-0.6cm}
\caption{The index with multiple asymptotically-flat boundaries is computed by integrating over geometries and fields 
with index boundary conditions. We will identify which geometries involving topology change near the horizon appear in the gravitational path integral with multiple boundaries, and determine whether they contribute.}
\label{fig:introindexmultiple2}
\end{figure}

Beyond the specific setup discussed in this paper, this clarifies the general procedure for computing the gravitational index in any supergravity theory. Firstly, one identifies a charge $\mathcal{Q}$ in the symmetry algebra such that $(-1)^F =e^{2\pi i \mathcal{Q}}$, which in the case of Reissner-Nordstr\"om is angular momentum. Then, all solutions with an imaginary chemical potential conjugate to this charge should be found, and a careful analysis of fermionic zero modes around each of these solutions should be carried out. To demonstrate a higher-dimensional example, we also work out the case of supergravity in AdS$_3$ in \textbf{section \ref{sec:disk3D}}. 

Understanding the zero modes is an important question, not only because one might find non-BPS saddles as described above, but also because there are cases where all saddles have a vanishing one-loop determinant. In such cases, the answer for the index is zero to leading order (this can happen for black holes with throats with an approximate $SU(1,1|1)$ isometry \cite{Heydeman:2020hhw}). In such cases, there are cancellations in the degeneracy of the index, which is thus different than the degeneracy of all BPS states (given by $d_\text{boson} + d_\text{fermion}$ instead of $d_\text{boson} - d_\text{fermion}$).  For the specific black holes in 4D ungauged supergravity in flatspace, the degeneracy will, however, turn out to purely consist of bosonic states \cite{Sen:2009vz, dabholkar2010no, Heydeman:2020hhw}; thus, we will not witness such cancellations in this paper. 

Having addressed the problem with a single boundary, we can ask what the gravitational index for multiple boundaries is, as in figure \ref{fig:introindexmultiple2}. Our goal is to determine whether connected geometries contribute to the gravitational path integral. Their contribution will determine, from a bulk perspective, whether quantities that are protected by supersymmetry are sensitive in any way to the factorization puzzle. 
\begin{figure}[t!]
\begin{center}
\begin{tikzpicture}
    \node[anchor=south west,inner sep=0] at (0,0) {\includegraphics[width=0.9\textwidth]{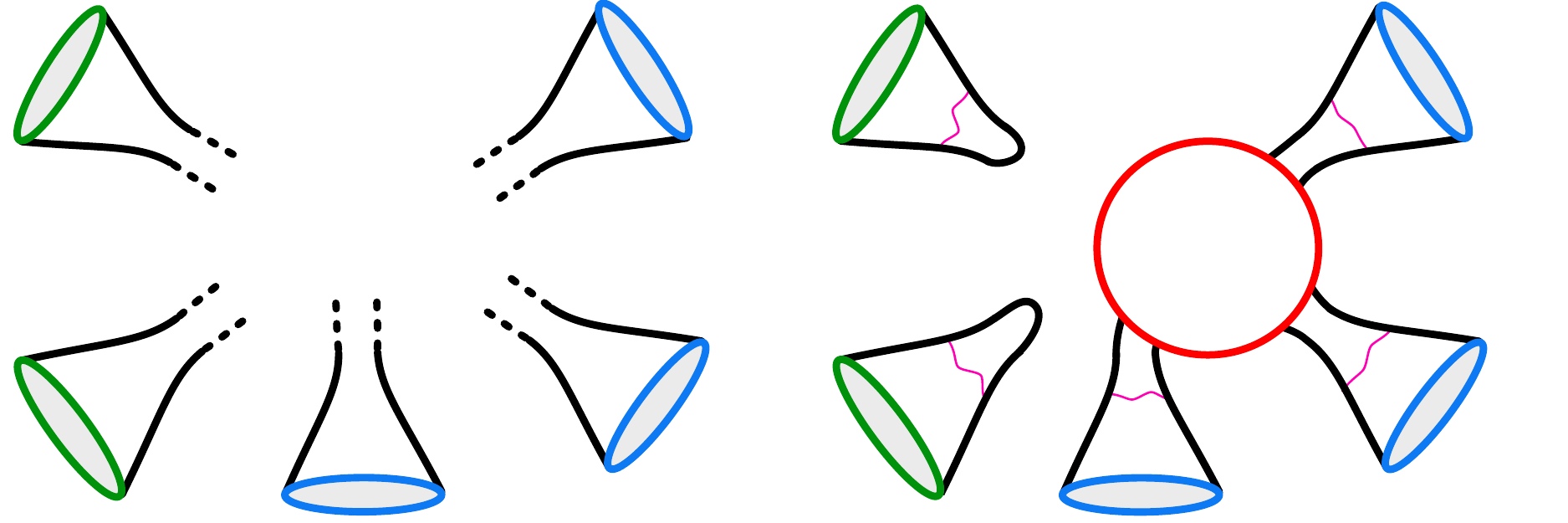}};
     \draw (7.80,+2.7) node  {\Large $=$};
\end{tikzpicture}
\end{center}
\vspace{-0.5cm}
\caption{Cartoon emphasizing the factorization property in gravity with index boundary conditions (on the green boundaries) and non-supersymmetric boundary conditions on other boundaries (represented in blue). The red jagged lines on the right denote the separation between the asymptotic and near-horizon regions. In this paper, we shall prove that, at least in the near-horizon region, the supersymmetric boundaries always factorize while connected geometries with non-supersymmetric boundary conditions could in principle contribute to the gravitational path integral. Consequently, the red blob in the right figure represents a sum over all allowed near-horizon geometries. }
\label{fig:introwomrholes3}
\end{figure}

Before summarizing the results obtained from the gravitational path integral, it is useful to first review how the multi-boundary index could behave in an ensemble average over the putative theories describing the black hole:
\begin{itemize}
    \item The first possibility is that one does not have to consider an ensemble average at all. This is expected from examples of black holes in string theory, all of which involve extended supersymmetry.
    
    \item The second possibility is that one averages over a moduli space of supersymmetric theories without walls of marginal stability, all of which have vacua that preserve supersymmetry.\footnote{This is the case in the $\cN=2$ SYK model \cite{Fu:2016vas}. } In such a case, the value of the index is the same for all members of the ensemble.
    
    \item The third possibility is that one again averages over a moduli space of supersymmetric theories, but the index is not the same in all members of the ensemble due to wall-crossing or supersymmetry breaking. 
    
    \item The final possibility is that one averages over theories that are not necessarily supersymmetric. Yet, supersymmetry can emerge after ensemble averaging.\footnote{As remarked in \cite{Hsin:2020mfa}, this happens for an $O(N)$ internal symmetry of the SYK model. }
    
\end{itemize}
In the first and second scenarios the index factorizes, while for the third and fourth scenarios one finds non-vanishing moments for the index of the ensemble. 

In \textbf{section \ref{sec:factorization}} we will focus on approaching these questions from a bulk computation. We show that all contributions to the gravitational path integral involving a non-trivial topology with spacetime wormholes in the near-horizon region always vanishes due to a gravitino zero mode. 
In particular, this shows that there are no connected contributions with multiple boundaries, thus solving the factorization puzzle for this protected index. This is illustrated in figure \ref{fig:introwomrholes3}. Consequently, our bulk result is consistent with the first two boundary scenarios outlined above and excludes the possibility that ensemble averages from the third and fourth scenarios are dual to bulk supergravity theories. We also show that manifolds with non-trivial topology cannot contribute even to the single-boundary index. It is interesting that this happens within supergravity, without the need to add extra structure to the theory \cite{Blommaert:2019wfy,Saad:2021rcu,Blommaert:2021gha}.

    Having reproduced most of the features summarized by the formula \eqref{eq:QMindex}, the only remaining question is how to see the fact that the degeneracy $d_\text{boson} - d_\text{fermion}$ in the index is an integer number. Explaining this feature from gravity might shed light on the nature of the microstates of extremal black holes. While we will not be able to answer this question fully,\footnote{A concrete computation addressing the exact degeneracy of extremal states in certain theories of gravity whose UV completion is known will appear in forthcoming work.} we will discuss what non-perturbative contributions are present in the gravitational path integral for single-centered black holes in \textbf{section \ref{sec:defects}}.\footnote{Depending on the matter content of the theory and on the charges fixed at the asymptotic boundary, it might be that multi-centered black hole saddles can dominate over the leading saddle, as discussed in~\cite{Denef:2007vg}. This has been referred to as the entropy enigma. However, for $\cN=2$ pure ungauged supergravity  multi-centered solutions are not expected to dominate over the leading saddle. Nevertheless, it would be nice to also understand these subleading contributions to the gravitational index in this theory. In  supergravities with more supersymmetry than $\cN=4$, it was shown that such multi-centered geometries do not contribute to the gravitational index \cite{dabholkar2010no}. Thus, if one treats $\cN=2$ ungauged supergravity as a sector of a theory of supergravity with higher supersymmetry, the classification of geometries that contribute to the index is most likely complete.} These geometries are smooth in the higher-dimensional picture and related upon dimensional reduction to supersymmetric defects in the two-dimensional near-horizon region. There are indications that they are responsible for the integrality of the index, see \cite{Dabholkar:2014ema} and references therein for a concrete example. Curiously, we show that among these defects, one in particular can be identified as the Hawking-Horowitz-Ross solution~\cite{Hawking:1994ii} for extremal black holes; nevertheless, we show that precisely this defect, and, consequently, the solution of \cite{Hawking:1994ii}, also yields a vanishing contribution to the index.  

The new non-perturbative contributions described in the previous paragraph do not spoil the factorization of the index. For example, geometries with one defect and any number of wormholes will vanish. Also, any geometry with multiple defects vanishes as well. The mechanism is once again due to fermionic zero modes.

Finally, we conclude in \textbf{section \ref{sec:discussion}} with a discussion of our results and some open questions. 

\section{The gravitational index of a charged black hole}
\label{sec:saddle-point-analysis}

This section aims to identify the gravitational saddles that contribute to the index of a charged black hole in four-dimensional flat space in the Euclidean path integral. The procedure for identifying the relevant classical saddles is similar to the approach taken for black holes in  AdS$_5$ \cite{Cabo-Bizet:2018ehj}\footnote{A similar approach was suggested in \cite{Dijkgraaf:2000fq} for AdS$_3$ but the details were not spelled out.}. However, an important point we want to stress is the role of quantum effects. We will also review some recent results for the partition function of near-extremal black holes with generic boundary conditions.

\subsection{The grand canonical partition function}

\paragraph{The theory:} For concreteness we will focus on $\mathcal{N}=2$ ungauged supergravity in four dimensions. This theory consists of a metric $g_{\mu\nu}$, a complex spin-3/2 gravitino $\Psi$, and a $U(1)$ graviphoton $A$. We choose units with $G_N=c=\hbar=1$. The quadratic part of the action is given by 
\beq
\label{eq:ungauged-SUGRA-action}
I[g,\Psi,A] = - \frac{1}{16\pi}  \int_M d^4 x \sqrt{g}\left[ R - \frac{1}{4} F^2 - \frac{i}{2} \bar{\Psi}_\mu \Gamma^{\mu\nu\rho} \nabla_\nu \Psi_\rho+\ldots \right]+I_{bdy},
\eeq
where the dots denote terms that are cubic or quartic and are completely fixed by supersymmetry. On closed surfaces the action is invariant under $\mathcal{N}=2$ supersymmetry which act on the gravitino as $\delta_\epsilon \Psi = \hat{\nabla} \epsilon + \mathcal{O}(\Psi^2)$ with $\hat{\nabla}_\mu\equiv \nabla_\mu + \frac{i}{8} F_{\mu\nu}\Gamma^{\mu\nu} \Gamma_\mu$. We are interested in the Euclidean path integral in asymptotically flat space with boundary topology $S^1 \times S^2$,
\beq
ds^2 = d\tau^2 + dr^2 + r^2 (d\theta^2 + \sin^2\theta d\phi^2) + \mathcal{O}(1/r) ,~~~~r\to\infty,
\eeq
with the identifications 
\beq\label{eq:identification4D}
\phi\sim \phi+2\pi \qquad \text{and} \qquad (\tau,\phi) \sim (\tau + \beta, \phi + i \beta \Omega).
\eeq
The asymptotic size of the time circle $\beta=T^{-1}$ is the inverse temperature. The twist parameter $\Omega$ corresponds to angular velocity along a specific direction within $S^2$. We impose $\Psi \sim \mathcal{O}(1/r^2)$ at infinity with periodicity 
\be\label{eq:4Dfermioncontractbdycond}
\Psi(\tau,r,\theta,\phi) &= - \Psi(\tau+\beta,r,\theta,\phi+i\beta \Omega)\\
&= - e^{\beta \Omega J}~ \Psi(\tau+\beta,r,\theta,\phi),
\ee
 where in the second line we defined the angular momentum operator $J$, the generator of rotations along $\phi$. Finally, we fix the graviphoton charge at infinity to be $Q$ magnetic, so $F \to Q \sin \theta d\phi \wedge d\theta$. These choices above require the following boundary terms at fixed radius $r=r_0$ with boundary metric $h_{ab}$ \cite{Hawking:1995ap}:
\be
I_{\rm bdy} =- \frac{1}{8\pi } \int_{\partial M} \sqrt{h} K - \frac{1}{4\pi }\int_{\partial M} \sqrt{h} n_a F^{ab} A_b.  
\ee

\paragraph{The partition function with a rotational chemical potential:} The partition function defined via the Euclidean gravitational path integral \cite{Gibbons:1976ue} is schematically given by 
\be
\label{eq:path-integral-definition-SUGRA}
Z_{\rm grav}(\beta,\Omega,Q) \equiv \int \mathcal{D}g \mathcal{D}\Psi \mathcal{D}A ~e^{-I[g,\Psi,A]},
\ee
with the boundary conditions chosen above. 
For theories with a gravitational dual, when fixing $Q$, $\Omega$ and $\beta$ by imposing the above boundary conditions, the gravitational path integral computes $ {\rm Tr}_Q\left( e^{-\beta H + \beta \Omega J} \right)$, where $ {\rm Tr}_Q$ computes a trace in the sector of charge $Q$. For theories without an explicit gravitational dual, this is not a precise relation since we are not defining a quantum mechanical Hilbert space and operators in a boundary theory which would give an independent calculation of the right-hand side. Nevertheless, we would like to see how to compute the gravitational path integral even in such theories and understand the meaning of the result as a grand canonical partition function.

\paragraph{Classical Saddles:} The classical saddles consistent with the boundary conditions above are the Kerr-Newman geometries. The Lorentzian metric with $\tau=i t$ in terms of the parameters $(a,M,Q)$ is
\bea\label{eq:KerrNewmanEqn}
g_{KN}{}_{\mu\nu}dx^\mu dx^{\nu} &=& - \frac{\Delta}{\rho^2}\left[ dt + a \sin^2\theta d\phi\right]^2 + \frac{\rho^2dr^2}{\Delta}+\rho^2d\theta^2+\frac{ \sin^2\theta}{\rho^2} \left( a dt + (r^2 + a^2)d\phi\right)^2,\nonumber\\
A_{KN}&=&\frac{Q \cos \theta}{\rho^2} ( a dt +(r^2+a^2)d\phi),
\ea
where we define $\rho^2\equiv r^2 + a^2 \cos^2\theta$ and $\Delta \equiv r^2 +a^2-2 M r + Q^2$. The gravitino solution has $\Psi=0$ and the $U(1)$ gauge field corresponds to the magnetic charge $Q$. The metric has an outer and inner event horizon at radii
\beq
r_{\pm} = M \pm \sqrt{M^2 - Q^2 - a^2}.
\eeq
The metric above corresponds to Boyer–Lindquist coordinates, for which the metric goes to a flat one at infinity, and the information of the angular velocity is in the identification \eqref{eq:identification4D}, coming from the smoothness of the Euclidean horizon. Another useful choice of coordinates which we will exploit in section \ref{sec:diskN4} rotate with the black hole and replaces $\phi$ by $\phi'=\phi - i\Omega \tau$. In these coordinates, the identification is $(\tau,\phi')\sim(\tau+\beta,\phi')$, making the time circle contractible. The information of the angular velocity now appears in the asymptotic expression for the metric at large $r$. The corotating coordinate will be useful to make the smoothness of the horizon manifest in the next section and also simplifies the periodicity conditions. 

Going to Euclidean signature, we will focus on the outer horizon $r_+$ and restrict to $r\in[r_+,\infty)$. The area of the horizon in these coordinates is given by $\mathcal{A} = 4 \pi (r_+^2 + a^2)$. The solution depends on $Q$, which is fixed by our boundary conditions. The final step is to use the smoothness of the Euclidean horizon to determine $\beta$ and $\Omega$ in terms of $M$ and $a$, or equivalently $r_+$ and $a$, which gives\footnote{The ADM mass and angular momentum are given by $M=(a^2 + Q^2 + r_+^2)/(2r_+)$ and $J=a M$. Since we are not imposing boundary conditions that fix energy or angular momentum, we will not give these parameters a physical interpretation.}
\beq\label{eq:4DidOmegaT}
\Omega = \frac{a}{r_+^2 + a^2},~~~~~~T=\beta^{-1}= \frac{r_+^2-a^2-Q^2}{4\pi(r_+^2 + a^2) r_+}\, .
\eeq
The classical Euclidean action evaluated at the Kerr-Newman solution $g_{KN}$ and $A_{KN}$ is denoted by $I_{\rm classical}(\beta,\Omega,Q) \equiv I[g_{KN},A_{KN},\Psi=0]$.  Since on a classical solution $R=0$ and $F^2$ is a total derivative, a very simple calculation gives \cite{Gibbons:1976ue}
\beq\label{eq:KNactionclassical}
I_{\rm classical}(\beta,\Omega,Q)= \frac{\beta}{2}\frac{a^2 + Q^2 + r_+^2}{2r_+} + \frac{\beta}{2}\frac{Q^2r_+}{r_+^2 + a^2},
\eeq
where the parameters $r_+$ and $a$ should be thought of as implicit functions of $\beta$, $Q$ and $\Omega$. As usual, this was regularized by subtracting the contribution from empty flat space.

\paragraph{Solving for the partition function:}   The boundary conditions at infinity are not modified under the shift $i \beta \Omega \to i \beta \Omega - 4 \pi n $, with $n\in \mathbb{Z}$ (a theory without fermions would be invariant under $2\pi n$ shifts instead). Therefore, the path integral instructs us to sum over these different solutions which share the same boundary conditions \cite{Heydeman:2020hhw} (see also \cite{Aharony:2021zkr} for a similar observation in another context). The final answer for the semiclassical black hole path integral takes the form,
\beq\label{eq:semiclassicalZgeneral}
Z_{\rm grav}(\beta,\Omega,Q) = \sum_{n\in\mathbb{Z}} e^{ - I_{\rm classical}\left(\beta, \Omega +\frac{4\pi i}{\beta} n, Q\right)}~Z_{\text{1-loop}}(\beta,\Omega,Q;n) (1 + \ldots)\, . 
\eeq
The sum is over classical solutions and the first term gives their classical action contribution $ e^{ - I_{\rm classical}\left(\beta, \Omega +\frac{4\pi i}{\beta} n, Q\right)}$. To give some intuition on how it behaves we can compute it at low temperatures with fixed real $i \beta \Omega$. The answer is 
\bea
\label{eq:ActionNSmallT4D}
-I_{\rm classical}\Big(\beta,\Omega+ \frac{4\pi i}{\beta}n ,Q\Big)&=&-\beta Q + \pi Q^2 +\frac{Q^3  (4\pi^2 - (i \beta \Omega - 4 \pi n)^2)}{2\beta} + \ldots \nonumber\\
&\overset{\Omega \to \frac{4\pi i \alpha}{\beta}}{=}&-\beta Q + \pi Q^2 +\frac{2\pi^2 Q^3  (1-4(\alpha+n)^2)}{\beta} + \ldots,
\ea
where we have introduced the convenient parameter $\a \equiv -\frac{i \beta \Omega}{4\pi}$. From~\eqref{eq:ActionNSmallT4D}, we see that generically the $n=0$ term dominates, although, as we will see next, leading quantum effects can alter this. The dots denote terms subleading at low temperatures. The form of the second line is reproduced by a near-horizon analysis reviewed in section \ref{sec:diskN4}.

The term $Z_\text{1-loop}(\beta,\Omega,Q;n)$ is the one-loop determinant of the metric, graviphoton, and gravitino fluctuations around the $n$-th solution (and possibly also of the matter fields), which distinguishes between bosonic Einstein-Maxwell theory and $\mathcal{N}=2$ ungauged supergravity. At large charge, $Q$, low temperature, and small enough angular velocity (fixed real $\alpha$), \cite{Heydeman:2020hhw} showed that this quantity is approximated by 
\beq\label{eq:1loopsec2}
Z_\text{1-loop}(\beta,\Omega,Q;n) \overset{\Omega \to \frac{4\pi i \alpha}{\beta}}{\sim} \beta \frac{(\alpha+n)\cot(\pi\alpha)}{(1-4(\alpha+n)^2)^2} ~Q^{c_\text{log}} \left[1+ O(Q^{-1})\right]\,.
\eeq
The power of $Q$, which we denote $c_\text{log}$ is captured by Sen's quantum entropy function \cite{Sen:2011ba} and depends on the matter content of the theory, but can be absorbed by a logarithmic correction to the area term in the classical action and will not be very important for our purposes. The temperature-dependent prefactor in \eqref{eq:1loopsec2} comes from a careful treatment of zero modes and is not captured by Sen's entropy function. As explained in \cite{Heydeman:2020hhw}, it is responsible for the well-defined space of extremal BPS black holes as well as the mass gap $E_{\rm gap}=1/8 Q^3$. Supersymmetry is crucial to derive these results since extremal non-BPS black holes do not exhibit a gap and do not have a well-defined space of extremal states \cite{Iliesiu:2020qvm}. Finally, the parenthesis in \eqref{eq:semiclassicalZgeneral} denotes higher-order quantum corrections, which are suppressed at large $Q$.

\subsection{Defining the gravitational index}

Having introduced the relevant gravitational path integral, we can now move on to the main question we want to address: how to compute the Witten index of a black hole from a gravitational perspective. This is schematically defined as
\beq 
{\rm Index}(\beta,Q) = {\rm Tr}_Q\left[ (-1)^F e^{-\beta H } \right]\,.
\eeq 
The procedure we use is very general, but the details depend on the particular theory.  

We will begin by explaining a naive approach that turns out to be misleading. Imagine we begin with the Reissner-Nordstr\"om black hole, which has the metric given in \eqref{eq:KerrNewmanEqn} but with $a=0$ and fixed $T$. In the notation above, this corresponds to computing $Z(\beta,\Omega=0)={\rm Tr}_Q\left( e^{-\beta H } \right)$. The smooth solution requires the boundary conditions on the fermion to be $\Psi(\tau + \beta,r,\theta, \phi) = - \Psi(\tau,r,\theta,\phi)$. The naive approach would be to demand by fiat that $\Psi$ is periodic, which one expects is equivalent to an insertion of $(-1)^F$ while leaving the metric intact. However, the resulting solutions involve singular configurations for the fermions since the time circle is contractible in the bulk. Another option is to keep the extremal metric which does not have a horizon, and pick periodic fermions on the now non-contractible time circle. The issue with this saddle is that it predicts a wrong value for the index \cite{Hawking:1994ii}. Thus, naively, it appears that there are no smooth contributions to this would-be definition of the gravitational index, and, therefore, this is not a desirable approach to the problem.

Instead, we will use the following identity: $(-1)^F = e^{2\pi i J}$. $J$ here is the generator of $\phi$ rotations but a similar formula applies for rotations about other axes (this fact will lead to bosonic zero-modes when studying fluctuations). This identity is obvious due to the fermions having half-integer spins. Accordingly, the proposal to compute the index is 
\be
{\rm Index}(\beta,Q) &= Z_{\rm grav}\Big(\beta, \Omega =\frac{2\pi i}{\beta},Q\Big)\nn \\ &= \sum_{n \geq 0} e^{ - I_{\rm classical}\left(\beta, \frac{2\pi i}{\beta} +\frac{4\pi i}{\beta} n, Q\right)}~Z_\text{1-loop}\left(\beta,\frac{2\pi i}{\beta},Q;n\right) (1 + \ldots)\, .
\ee
In terms of the gravity theory, this definition has a completely different interpretation from the naive approach outlined in the previous paragraph. Instead of changing the boundary conditions of the fermions by hand, we turn on an imaginary angular velocity $\Omega \to \frac{2\pi i}{\beta} +\frac{4\pi i}{\beta} n$. The solutions that enter the path integral computing the index are made smooth thanks to this rotation. This purely imaginary value of $\Omega$ corresponds to real angular velocity in Euclidean space since $\Omega_E = i \Omega$. We will also comment shortly on the reality of the metric in Euclidean signature. The solutions with $\Omega= \pm \left(\frac{2\pi i}\beta +\frac{4\pi i}{\beta} n\right)$ (or equivalently shifting $n\to-1-n$) are equivalent by a parity transformation, and since this is a symmetry of the theory we will restrict ourselves to the positive sign and to $n\geq 0$.\footnote{\label{footnote:moduli space of rotating black holes}When $\Omega=0$, the saddles with $\pm n$ are related by rotation and should not be summed separately. When $\Omega=\frac{2\pi i}{\beta}$ the same is true about $n$ and $-1-n$. Therefore for the index, we can focus on $n\geq 0$. When $\Omega$ has a generic value, all saddles $n\in\mathbb{Z}$ are independent.} 

This choice of $\Omega$, for any value of $n$, implies that the fermions are periodic around the time circle
\bea\label{eq:4Dsmoothfermions}
\Psi(\tau,r,\theta,\phi) &=& - \Psi(\tau+\beta,r,\theta,\phi+2\pi) , \nonumber\\
&=& \Psi(\tau+\beta,r,\theta,\phi) . 
\ea
In the first equality, we used smoothness with respect to the contractible circle at the horizon. In the second line we use that the fermion is always antiperiodic around $\phi \sim \phi+2\pi$. Therefore we found a classical configuration that solves the equations of motion and is completely smooth both for bosonic and fermionic fields. Since the $n=0$ solution will prove important, we will denote the angular velocity of this solution in Lorentzian signature by $\Omega_\star = 2\pi i/\beta$.

\subsection{The saddles of the index}

\paragraph{Supersymmetry:} Now, we will check that the $n=0$ smooth solution, with the Euclidean horizon at a finite distance and at finite temperature (as opposed to the Reissner-Nordstr\"om zero-temperature case), is supersymmetric. We will then come back to the solutions with $n>0$ shortly. We can use \eqref{eq:4DidOmegaT} to write down $\beta(a_\star,r_+,Q)\,  \Omega(a_\star,r_+,Q) = 2\pi i $ and solve for $a_\star$. This gives $a\to i r_+ \pm i Q $ and we pick $a_\star = i r_+ - i Q$ (since the other solution is not physical). Now, we can use the equation that determines $r_+$ and insert this value of $a_\star$, yielding
\be
\Delta(r_+)&=r_+^2 + a_\star^2 -2Mr_+ + Q^2 = 2 r_+ (Q-M)=0,\nonumber\\
&\qquad \Rightarrow ~~ M=Q .
\ee
Then, picking $\Omega \to \Omega_\star= 2\pi i/\beta$ automatically fixes $M=Q$, meaning that the Killing spinor equation $\hat{\nabla}_\mu \epsilon = 0$ is integrable. For example, adapting the results of section V.C of \cite{Caldarelli:1998hg} to flat space, we have ${\rm det}( [\hat{\nabla}_\mu,\hat{\nabla}_\nu] )\propto (M^2-Q^2)^2$ regardless of the angular velocity. Since the equation is integrable and the boundary conditions are compatible with the circle that becomes contractible in the bulk, we can safely assume a solution exists, even though we will not write it down explicitly. 

Even though these solutions have $M=Q$, they still allow a nonzero temperature. To see this, take $a_\star=i \mathfrak{a}$ with real $\mathfrak{a}$ and $M=Q$. This has $r_+ = Q +\mathfrak{a}$ and the temperature is
\beq
T(\mathfrak{a}) = \frac{1}{2\pi} \frac{\mathfrak{a}}{Q(Q + 2 \mathfrak{a})},~~~~M(\mathfrak{a})=Q,~~~~\Omega(\mathfrak{a}) = 2\pi i T(\mathfrak{a}).
\eeq
 As a function of temperature, the size of the horizon is $r_+(T) = Q \frac{1-2\pi Q T}{1-4\pi Q T}$, only valid for $T<1/4\pi Q$. This solution is completely real in Euclidean signature as long as $0<T<1/4\pi Q$. It is also smooth since $\rho>0$ everywhere even though $a$ is purely imaginary. For $T>1/4\pi Q$,  $r_+$ is complex but we can imagine a complex saddle where $r$ goes along a complex direction similar to \cite{Maldacena:2019cbz}. Consequently, even if we restrict ourselves to  saddles with a real Euclidean metric, we can always study the regime with sufficiently low temperatures where this leading saddle is real. 

\paragraph{Action and one-loop effects:} We can now compute the classical action for $\Omega_\star$, first for $n=0$. This is straightforward using \eqref{eq:KNactionclassical}. For any value of $\mathfrak{a}$, and therefore any temperature, the answer is given by 
\beq
- I_{\rm classical}(\beta,\Omega_\star,Q)= -\beta Q + \pi Q^2,~~~~\Rightarrow~~{\rm Index}(\beta,Q) = e^{- \beta Q} ~e^{\pi Q^2}.
\eeq
This is precisely what we need: we obtain the term capturing the ground state energy, $- \beta Q$, and the remaining temperature-independent term is the entropy $\pi Q^2$. \emph{A priori} one might get confused because the horizon area $\mathcal{A}=4 \pi Q(Q+2\mathfrak{a} ) $ is not temperature independent but including the remaining contributions to the action resolves this misdirection. Explicitly, combining the Bekenstein-Hawking entropy term which is given by the area, $S_\text{BH} = \mathcal A/4$, with the chemical potential term, $\Omega J$, present in the grand potential,
\be 
S_\text{BH} + \beta\, \Omega J = \pi Q^2\,,
\ee
one finds the leading value of the entropy associated to the degeneracy in the index. To compare our result to the naive answer one would find when solely imposing the periodicity of fermions without turning on an angular momentum, we can compute the classical action for $\Omega=n=0$ which gives $e^{-I_{\rm classical}(\Omega=0)}\sim e^{-\beta Q + \pi Q^2 + \frac{2\pi^2 Q^3}{\beta}+\,\cdots}$, and clearly does not have the properties of an index.

So far, we have focused on the contribution of the $n=0$ saddle to the index. The Gibbons-Hawking procedure still instructs us to sum over all saddles $\Omega \to \Omega_\star + 4 \pi i n / \beta$, with $n\geq 0$. This is a problem since configurations with $n\neq 0 $ are not supersymmetric: The Killing spinor equation is not integrable and, from \eqref{eq:ActionNSmallT4D}, it can be seen that they have an associated mass $M(\beta,\Omega_\star + \frac{4\pi i n}{\beta} ,Q)=Q+\frac{2\pi^2 Q^3  (1-(1+2n)^2)}{\beta} + \mathcal{O}(\beta^{-2})\neq Q$, unless $n=0$ for which $M\big(\beta,\Omega_\star ,Q\big)=Q$. Relatedly, the classical action on these solutions does not have the temperature dependence expected of an index, as seen from equation \eqref{eq:ActionNSmallT4D}. Thus, including these saddles would incorrectly imply that the index does not solely count the BPS states in some supersymmetric quantum system and, instead, receives contributions from other states. Typically, because they lack supersymmetry these solutions are discarded from the index, however the bulk mechanism responsible for this is usually not discussed.\footnote{For instance, in \cite{Aharony:2021zkr} the sum over all such solutions is discussed but only the BPS solution is kept. } The answer to this puzzle is to remember that we need to include the one-loop determinants in the calculation, and these are crucial. From the low-temperature approximation given in \eqref{eq:1loopsec2}, noting that $\Omega= \Omega_\star$ corresponds to $\a=1/2$, it is immediate to see that 
\beq
Z_\text{1-loop}(\beta,\Omega_\star,Q; n)=0,~~~n>0.
\eeq
Therefore, the contributions of the non-supersymmetric saddles vanish, as expected from an index. 
In section \ref{sec:diskN4} we will explain how the presence of a gravitino zero mode is responsible for the one-loop determinant vanishing. Our explanation will not rely on the details of the one-loop determinant and can be generalized to other cases. 

When $n=0$ and the solution is supersymmetric, the gravitino zero mode is an unphysical gauge mode, equivalent to a superdiffeomorphism, and should not be integrated over; therefore, the determinant for $n=0$ will turn out to be nonzero. This is difficult to see from \eqref{eq:1loopsec2} since naively, there is a divergence that cancels between $n=0$ and $n=-1$. This is due to a bosonic zero mode corresponding to metric perturbations that can be written as \eqref{eq:KerrNewmanEqn} but around a different axis on $S^2$ (this only works for $\Omega=0$ or $\Omega_\star$). In section \ref{sec:diskN4} we will do the calculation exactly at $\Omega = \Omega_\star$, being careful about both fermionic and bosonic zero modes, and obtain a finite, nonzero, and temperature-independent answer for $Z_\text{1-loop}(\beta,\Omega_\star,Q; n=0)$. This result is not trivial, and there are examples with less supersymmetry where the one-loop determinant makes the index vanish to leading order, even though the degeneracy is not vanishing, see appendix A of \cite{Heydeman:2020hhw}.

This combination of classical and one-loop effects would give the full perturbative answer for the index. We have ignored logarithmic terms in the charge, which can be treated by combining our analysis with e.g.~\cite{Sen:2011ba}, but we will not do so here. Instead, in section \ref{sec:factorization} we will analyze corrections from spacetime wormholes and prove that their contribution vanishes due to fermionic zero modes, resolving the factorization puzzle for the index. Finally, in section \ref{sec:defects} we will analyze the only possible non-perturbative corrections to the index.

\section{Quantum effects: Supergravity on the disk}
\label{sec:diskN4}

In this section, we resolve the questions that we posed previously about the computation of the one-loop determinant for the BPS and non-BPS saddles in the calculation of the supersymmetric index without relying on the exact answer. We shall first address these questions for the index for black holes in $\cN=2$ ungauged supergravity in asymptotically flat space by studying the dimensional reduction in the near-horizon region to $\cN=4$ JT supergravity derived in \cite{Heydeman:2020hhw}\footnote{Other recent work relating JT gravity to higher dimensional black holes are \cite{Nayak:2018qej,Castro:2018ffi,Moitra:2019bub,Castro:2019crn,Castro:2021fhc}.}. Finally, we also discuss the one-loop calculation in AdS$_3$ for $(4,4)$ supergravity, which can be directly computed without relying on the AdS$_2$ throat.

\subsection{Setup: Dimensional reduction of 4D $\cN=2$ supergravity around a black hole}

We will focus on fluctuations around the Kerr-Newman metric near extremality, at low temperatures and fixed $i \beta \Omega$. We will follow the presentation in \cite{Heydeman:2020hhw}. In this limit, the dominant contribution from the Euclidean path integral to the temperature dependence of the one-loop determinants comes from a very specific set of modes that live in the AdS$_2$ throat. Moreover, since $\Omega$ is small (since it is $\sim \beta^{-1}$), we can incorporate rotation as a small fluctuation around the Reissner-Nordstr\"om (RN) AdS$_2\times S^2$ throat. As explained in \cite{Iliesiu:2020qvm}, for metric fluctuations we can restrict to the ansatz
\begin{equation}
\label{eq:dim-red-metric-ansatz}
    ds^2_{4D} = \frac{r_0}{\chi^{1/2}}g_{\mu \nu}dx^\mu dx^\nu + \chi \, h_{mn}(dy^m + T^m_{i} B^{i}_{\mu}dx^{\mu})(dy^n + T^n_{j} B^{j}_{\nu}dx^{\nu}) \, ,
\end{equation}
where $x^{\mu,\nu}=\{t,r\}$ and $g_{\mu\nu}(x)$ is an arbitrary metric depending on these coordinates alone. The angles $y^{m,n}=\{\theta,\phi'\}$ parametrize the $S^2$ metric with $h_{mn}dy^mdy^n = d\theta^2 + \sin^2 \theta d(\phi')^2 \, $.\footnote{The difference between $\phi$ (from section \ref{sec:saddle-point-analysis}), and $\phi'$ (from \eqref{eq:dim-red-metric-ansatz}) will be re-emphasized shortly.} We incorporate rotation by an $SU(2)$ gauge field $B^i_\mu(x)$, which transforms as a $\mathbf{3}$ of $SU(2)$ and enters through the $S^2$ Killing vectors $T_i^m(y)$, conventions for which are presented in \cite{Heydeman:2020hhw, Iliesiu:2020qvm}. The dilaton field $\chi(x)$ yields the area of the $S^2$, which we expect to slowly grow radially as we go away from the black hole horizon. We will also be working in the canonical ensemble for the overall $U(1)$ charge of the system by fixing $Q$. The length scale $r_0$ defines the horizon size that an extremal RN black hole with charge $Q$ would have, so $r_0=Q$.

Plugging such an ansatz into the full supergravity action \eqref{eq:ungauged-SUGRA-action} and dimensionally reducing on the internal $S^2$, the bosonic part of the action becomes
\begin{align}
\label{eq:2d-bosoni-action}
I_{\text{2D, bulk}}^{\text{(bosonic)}} = -\frac{1}{4} \int \! \! d^2x \sqrt{g} \left (\chi R + \frac{2r_0}{\chi^{1/2}} -\frac{2r_0 Q^2}{\chi^{3/2}} -\frac{1}{3} \frac{\chi^{5/2}}{r_0} \tr_{SU(2)} H_{\mu\nu}H^{\mu\nu} \right )  \, ,
\end{align}
where we have explicitly integrated out the $U(1)$ gauge field\footnote{It is important that we impose the fixed-charge boundary condition, instead of fixed chemical potential.} and $H_{\mu\nu}$ is the field strength associated to the $SU(2)$ gauge field $B$. If we impose that the $SU(2)$ gauge field is trivial with $B=0$ on all of spacetime, then the saddle point for $g_{\mu \nu}(x)$ and $\chi(x)$ in the action \eqref{eq:2d-bosoni-action} precisely yields the 4D metric profile for a RN black hole. If one analyzes the saddle point with an $SU(2)$ holonomy turned on at the asymptotic boundary, one finds that, as long as the angular momentum in the system is sufficiently small, $g_{\mu \nu}$, $B$, and $\chi(x)$ reproduce the more general Kerr-Newman solution \eqref{eq:KerrNewmanEqn} (up to a simple coordinate transformation whose role we will emphasize shortly) \cite{Iliesiu:2020qvm}.

Our goal is to first include the fermionic fields coming from the gravitino, which we have so far neglected in \eqref{eq:2d-bosoni-action}, and then to quantize the fluctuations of all fields around this background. Towards that end, we will employ the following strategy: 
\begin{itemize}
    \item We will first obtain the near-horizon action for both bosonic and fermionic fields in a  $1/r_0^2$ expansion. This expansion is equivalent to an expansion in terms of the Gibbons-Hawking entropy, which for an extremal black hole with $U(1)$ charge $Q$ is
\be 
S_0  = \pi r_0^2 = \pi Q^2\,.
\ee
Since we are interested in macroscopic black holes, which all have $S_0 \gg 1$, this expansion of the action will nicely organize the fluctuations of all the fields in the 2D action in the near-horizon region.

\item The resulting action after including the  fermions will be equivalent to a topological field theory, namely a $PSU(1,1|2)$ BF theory.  Fluctuations in this theory are extremely simple since they are given solely by  $PSU(1,1|2)$ large gauge transformations, which in the gravitational theory are equivalent to large super-diffeomorphisms  (i.e.,~that can change the boundary value of certain fields). Since the bulk action that we study is invariant under super-diffeomorphisms, only the boundary term---which is present in between the near-horizon region and the asymptotic region---will prove important when studying the determinant from the fluctuations in the theory. 
 
 \item To understand the proper boundary term, we find the appropriate boundary conditions that need to be imposed at the boundary of the near-horizon region. We do this by using the equations of motion in the region separating the boundary of the near-horizon region and the asymptotic boundary at infinity to translate the boundary conditions imposed at the asymptotic boundary to the near-horizon region.

\item Finally, we study the action of the large super-diffeomorphisms on this boundary term to obtain the one-loop determinant from all relevant fluctuations.

\end{itemize}

\subsection{The near-horizon action}

Let's begin executing our strategy by determining the near-horizon action. In the throat region, it is appropriate to expand the dilaton $\chi(x)$ around its value at the horizon,
\begin{equation}
   \chi(x) = r_0^2 + 2  \Phi(x) \,,
\end{equation}
such that the varying part $\Phi\ll r_0^2 \sim S_0$ is suppressed compared to the area of the extremal black hole $S_0$. In this expansion, the bosonic part of the action becomes
\begin{align}
\label{eq:2D-bosonic-action}
I_{\text{2D, bulk}}^{\text{(bosonic)}} = I_{\cN=4 \text{ JT, bulk}}^{\text{(bosonic)}} = S_0 - \frac{1}{2} \int \! \! d^2x \sqrt{g} \, \Phi \left (R + \frac{2}{r_0 ^2}\right ) -  i \int   \tr_{SU(2)}b H \, + \cO\left(1/r_0^2\right),
\end{align}
where we have introduced the Lagrange multiplier zero-form field $b^i(x)$ to rewrite the Yang-Mills term in \eqref{eq:2d-bosoni-action} linearly in the $SU(2)$ field strength $H$ but quadratically in $b$; nevertheless, the quadratic term in $b^2$ is suppressed by a higher power of $1/r_0^2$ and is thus absent from the leading action. The equation of motion of $\Phi$ and $b$ in \eqref{eq:2D-bosonic-action} imposes that $R = -2/r_0^2$ and $H=0$, which is indeed the case for an extremal or near-extremal black hole whose horizon size is close to $r_0$.

Before studying fluctuations in \eqref{eq:2D-bosonic-action}, we have to specify the fermionic part of the supergravity action in the near-horizon region, again in a $ {1/r_0^2}$ expansion. This can be done by either imposing supersymmetry in the resulting 2D dimensionally-reduced theory or by dimensionally reducing the fermionic terms in the original $\cN=2$ ungauged supergravity action \eqref{eq:ungauged-SUGRA-action}, which produces two-dimensional gravitini $\psi^p=(\psi_{+1/2}^p, \psi_{-1/2}^p)^T$ and dilatini $\lambda^p=(\lambda_{+1/2}^p, \lambda_{-1/2}^p)^T$, with indices $p=1,2$ that transform in the fundamental of $SU(2)$. In both cases, one finds that the overall 2D action \cite{Heydeman:2020hhw} can be written in the first-order formalism, introducing frame one-forms $e^{a=1,2}_\mu$ and spin connection $\omega$, as
\begin{align}
\label{eq:final-action-from-dim-red-nhr-2d}
I_{\cN=4\text{ JT, bulk}} &= - \int \Phi \left ( d\omega + \frac{1}{2r_0 ^2} \varepsilon_{ab} e^a \wedge e^b -\frac{2}{r_0 }\bar{\psi}_p  \wedge \psi^p \right ) \nonumber \\  & \qquad \quad +\frac{i}{2} b_{i}\left(H^{i} -\frac{2}{r_0}\bar{\psi}_p (\gamma_3') (\sigma^i)^{p}{}_q \wedge \psi^q \right ) - \left(2i \lambda_p \mathcal{D} \bar \psi^p  + \textrm{ h.c.}\right) + \cO\left(1/r_0^2\right) \,,
\end{align}
where the supercovariant gauge exterior derivative is given by
\begin{equation}
\mathcal{D} \psi^p=  d \psi^p + \frac{1}{2}\omega \gamma_3' \wedge \psi^p - \frac{1}{2 r_0 } \gamma_3' \gamma_a' e^a \wedge \psi^p + i B^i (\sigma_i)^p{}_q \wedge \psi^q \, ,
\label{eq:2dsuperexterior}
\end{equation}
following the spinor conventions discussed in appendix B of \cite{Heydeman:2020hhw}. When working in the first-order formalism, one should also explicitly impose the vanishing of the super-torsion. To that end, we can add to the action the term
\be 
\label{eq:super-tors-Lagr-multiplier}
i \frac{\phi_a}{r_0} \left (d e^a + \varepsilon^{ab} \omega \wedge e_b - 2\,\bar{\psi}_p \gamma'^a \wedge  \psi^{p} \right )\, ,\ee
where the fields  $\phi_a$ serve as Lagrange multipliers which impose the vanishing of the super-torsion in the two-dimensional theory.

Having recovered the full action in the near-horizon region, it is useful to analyze fluctuations in this theory by first rewriting \eqref{eq:final-action-from-dim-red-nhr-2d}, together with \eqref{eq:super-tors-Lagr-multiplier}, in a simpler form where all fluctuations can be rephrased as $PSU(1,1|2)$ gauge transformations. This is given by an $PSU(1,1|2)$ BF theory with  action \cite{Cardenas:2018krd, Heydeman:2020hhw}
\be
\label{eq:JT-N=4-action}
I_{BF} =- i \int \str \phi F\,, ~~~~ F = dA - A\wedge A\,,
\ee
where $A$ is a $\mathfrak{psu}(1, 1|2)$ gauge field and $\phi$ is a $\mathfrak{psu}(1,1|2)$ valued zero-form field. The gauge field can be written in terms of the supermultiplet of the frame $e^a$ and spin connection $\omega$, also consisting of the SU(2) gauge field $B^i$ and the gravitinos $\psi_{p}{}^{ \a=\pm1/2}$,\footnote{Here and in similar notation with spinor fields, $\alpha$ is a spinor index parameter, not to be confused with the angular velocity parameter $\alpha=-i\beta \Omega/(4\pi)$.} as:
\be 
\label{eq:gauge-field-ansatz}
A(x) &= \sqrt{\frac{\Lambda}{2}} \left[e^1(x) L_0 + \frac{e^2(x)}2 \left(L_{1} - L_{-1}\right) \right]- \frac{\omega(x)}2 \left(L_1+L_{-1}\right) + B^i(x) T_i  \nonumber \\ &\quad +   \left(\frac{\Lambda}{2}\right)^{\frac{1}{4}}\left(\bar{\psi}^{p}{}_{ \a}(x) G_{p}{}^{ \a} +  \psi_{p}{}^{ \a}(x) \bar{G}^p{}_{ \a}\right)\,,
\ee
where we will use the same conventions for the $\mathfrak{psu}(1, 1|2)$ generators $L_{0, \pm 1}$, $T_i$, $G_p{}^\alpha$, and $G^p{}_\alpha$ as in \cite{Heydeman:2020hhw}, with $p=1,2$. $\Lambda=1/r_0^2$ denotes the absolute value of the 2D cosmological constant. The generator $T^3$ in this section corresponds to the angular momentum $J$ in the previous section. The zero-form field $\phi$ consists of fields in the same supermultiplet as the dilaton $\phi^0$, which include the Lagrange multipliers for the super-torsion $\phi^{1,2}$ and for the $SU(2)$ gauge field $b^i$, as well as the dilatinos $\lambda_p{}^\alpha$ and $\bar \lambda^p{}_\alpha$.
Putting $A(x)$ and $\phi(x)$ into the action  \eqref{eq:JT-N=4-action}, one can recover the action \eqref{eq:final-action-from-dim-red-nhr-2d} obtained from the gravitational reduction of the 4D theory. Integrating out $\phi$, one finds that the $PSU(1, 1|2)$ gauge field is flat, and therefore the bulk term completely vanishes, leaving only a boundary term which we will analyze in the next subsection.

An important ingredient from the dictionary relating the BF theory \eqref{eq:JT-N=4-action} and JT supergravity is the equivalence between infinitesimal $PSU(1,1|2)$ gauge transformations in the former and infinitesimal super-diffeomorphisms in the latter. In particular, the infinitesimal supersymmetry transformations of the super-JT action \eqref{eq:JT-N=4-action}, which will play an important role in our analysis, can be obtained by considering the infinitesimal fermionic gauge transformation whose gauge parameter takes the form $\Lambda = \epsilon^{p\a} G_{p \a} + \bar \epsilon^{p \a} \bar G_{p \a}$ (not to be confused with the cosmological constant), with $\delta A = d \Lambda + [A, \Lambda]$. Since the bulk term completely vanishes in JT supergravity and we are left with only a boundary term. In order to analyze fluctuations in this theory, we will thus only have to study the action of gauge transformations on this boundary term. 

\subsection{Boundary conditions from 4D to 2D} 
\label{sec:bdy-cond-4d-to-2d}

To compute the gravitational path integral in the near-horizon region it is necessary to specify the boundary conditions for all the fields at the edge of the AdS$_2$ region. We start by imposing that the metric in Fefferman-Graham gauge takes the form 
\be 
\label{eq:FG-gauge-metric}
g_{\mu\nu}dx^\mu dx^\nu= d\tilde r^2 + \left(\frac{1}4 e^{2\tilde r} -\tilde S_b(\tau) + \dots \right) d\tau^2 \,,
\ee
where the boundary is placed at fixed, but large, $\tilde{r}$ (which is a simple function of the radial coordinate $r$ of section \ref{sec:saddle-point-analysis}), with $e^{-\tilde r_\text{bdy}} = \epsilon/2$. To place the theory at finite temperature, the Euclidean time $\tau$ is identified as $\tau \sim \tau + \beta$ from which the proper boundary length is given by $L = \beta/\epsilon$ as $\tilde r \to \infty$. As we approach the boundary, the leading part of the metric is fixed while the sub-leading pieces are allowed to vary, with $\tilde S_b$ denoting the leading varying component, with $\tilde S_b(\tau) = \tilde S_b(\tau+\beta)$. Similarly, the black hole solution dictates that the value of the dilaton field at the AdS$_2$ boundary is given by $\Phi = \Phi_r/\epsilon$, with $\Phi_r = r_0^3=Q^3$. 

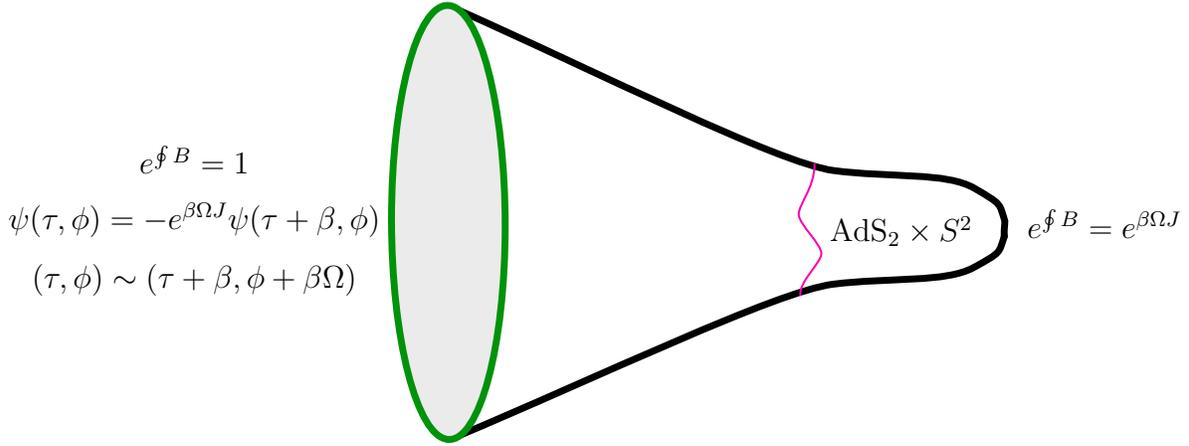
\begin{figure}[h!]
         \begin{center}
\begin{tikzpicture}
    \node[anchor=south west,inner sep=0] at (0,0) {\includegraphics[width=0.55\textwidth]{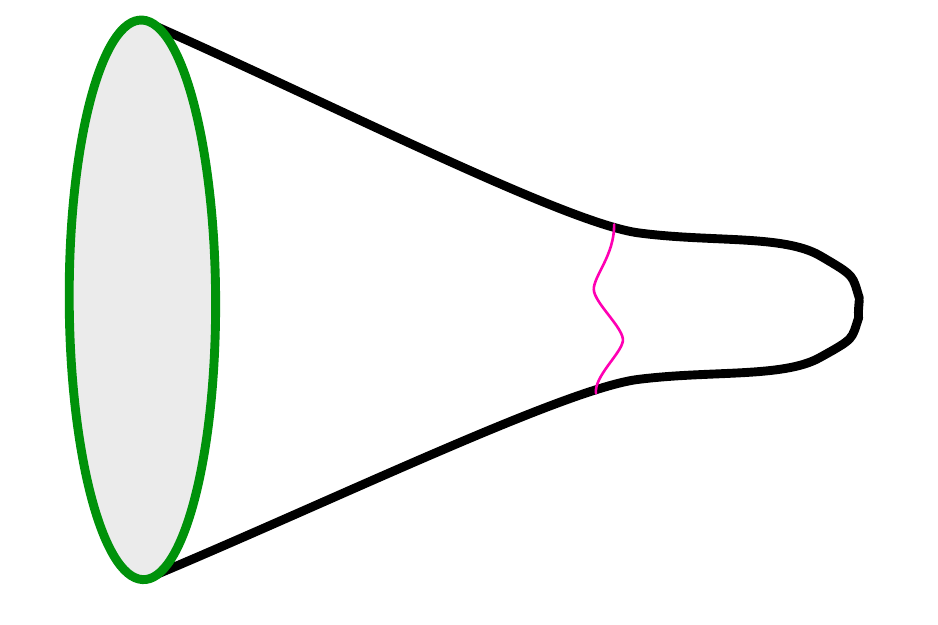}};
     \draw (-1.9,+4.2) node  {$e^{\oint B} = 1$};
      \draw (-1.9,+3.4) node  {$\psi(\tau, \phi) = - e^{\beta \Omega J} \psi(\tau+\beta, \phi)$};
       \draw (-1.9,+2.6) node  {$(\tau, \phi) \sim (\tau+\beta, \phi + \beta \Omega)$};
        \draw (10.2,+3.35) node  {$e^{\oint B} = e^{\beta \Omega J}$};
         \draw (7.5,+3.25) node  {AdS$_2 \times S^2$};
\end{tikzpicture}
\end{center}
     \caption{Summary of boundary conditions and the singular $SU(2)$ gauge field when working in the Boyer-Lindquist coordinates. }
     \label{fig:coord1}
 \end{figure}

The boundary conditions on the gravitino, dilatino and the $SU(2)$ gauge field are more subtle. When performing the dimensional reduction, it is useful if all  curves that are contractible in 2D are also contractible in the initial gravitational theory. This is not the case if we work in the Boyer-Lindquist coordinates from \eqref{eq:KerrNewmanEqn}. Specifically, if we treat $r$ and $t$ from \eqref{eq:KerrNewmanEqn} as the coordinates in the dimensionally reduced theory, then the boundary conditions for the $SU(2)$ gauge field and the fermionic fields at the asymptotic boundary can be summarized through the diagram in figure~\ref{fig:coord1}.
 
 Curves that appear contractible on the two-dimensional manifold, parametrized by $t$ and $r$  of the Boyer-Lindquist coordinates, such as the curves at constant $r$, are only contractible in higher dimensions if paired with an appropriate twist of $\phi$. Consequently, for constant $\phi$, the fermionic field is not antiperiodic, and the holonomy of the $SU(2)$ gauge field is trivial at the asymptotic boundary. If one computes the holonomy of the $SU(2)$ gauge field in these coordinates on a closed curve along Euclidean time arbitrarily close to $r = r_+$, then one finds $e^{\oint B} = e^{\beta \Omega T_3}$. Thus, a consequence of the fact that cycles that are contractible in 2D are not contractible in the original theory is that the $SU(2)$ gauge field seems singular at the horizon.

\begin{figure}[h!]
     \begin{center}
\begin{tikzpicture}
    \node[anchor=south west,inner sep=0] at (0,0) {\includegraphics[width=0.55\textwidth]{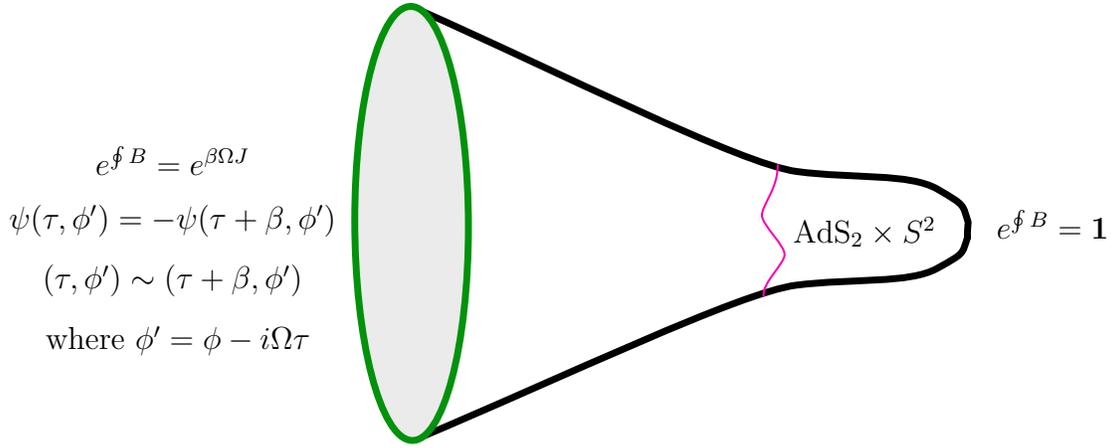}};
     \draw (-1.7,+4.2) node  {$e^{\oint B} = e^{\beta \Omega J}$};
      \draw (-1.7,+3.4) node  {$\psi(\tau, \phi') = -\psi(\tau+\beta, \phi')$};
       \draw (-1.7,+2.6) node  {$(\tau, \phi') \sim (\tau+\beta, \phi')$};
     \draw (-1.7,+1.8) node  {$\text{ where } \phi' = \phi -i \Omega \tau$};
             \draw (10.0,+3.35) node  {$e^{\oint B} = \mathbf{1}$};
         \draw (7.5,+3.25) node  {AdS$_2 \times S^2$};
\end{tikzpicture}
\end{center}
      
     \caption{Summary of boundary conditions in corotating coordinates. The $SU(2)$ gauge field is now smooth.}
     \label{fig:coord2}
 \end{figure}

To resolve this issue and find the correct boundary conditions at the asymptotic boundary, we consider the coordinate change $\phi' = \phi + \Omega t$ mentioned in section \ref{sec:saddle-point-analysis}, where $\phi'$ now parametrizes one of the coordinates of $S^2$ in the metric ansatz \eqref{eq:dim-red-metric-ansatz}. We will refer to these coordinates as the corotating coordinates since an observer placed at constant $\phi'$ rotates together with the black hole horizon. In these corotating coordinates, the boundary conditions for the $SU(2)$ gauge field and the fermionic fields at the asymptotic boundary are summarized in figure \ref{fig:coord2}.
 
 Circles of constant $r$ and $\phi'$ are now contractible in both 2D and in the original 4D solution and, consequently, the gravitino $\psi$ is now antiperiodic around the thermal circle. Thus, we now have to impose a non-trivial holonomy for the $SU(2)$ gauge field at the asymptotic boundary, which, however, resolves the singularity of the $SU(2)$ gauge field at $r=r_+$. Thus, in these coordinates, the equations of motion for the gauge field can be solved on the entire manifold, and one finds that the saddle point of \eqref{eq:2d-bosoni-action} is in fact given by expressing the metric in these coordinates. 
 
 To better understand the relation between the coordinate systems presented in figures \ref{fig:coord1} and  \ref{fig:coord2} in terms of 2D fields, we note that by performing a gauge transformation $h(\tau) = e^{\Omega \tau T_3}$ ---which is illegal since it does not obey the correct periodicity boundary conditions around the thermal circle---one replaces the boundary conditions from those in the corotating coordinates with those in the Boyer-Lindquist coordinates. Since $h(\tau)$ does not obey the correct periodicity for an allowed gauge transformation, this transformation introduces the singularity in the $SU(2)$ gauge field at the horizon that can be observed in figure \ref{fig:coord1}.

Having clarified the boundary conditions at the asymptotic boundary, we can now discuss the boundary conditions at the edge of the near-horizon region. If we impose that the gravitino vanishes at the asymptotic boundary, then at the boundary of AdS$_2$ we solely allow the subleading components of the gravitino to survive. In the Fefferman-Graham coordinate system \eqref{eq:FG-gauge-metric} this implies $\psi_{+1/2}=0$ and 
\be 
\label{eq:bdy-cond-gravitino}
\psi_{-1/2} = \tilde S_f(\tau) e^{-\tilde{r}/2} +\dots\,, \qquad \bar \psi_{-1/2}  =   \bar {\tilde S_f}{}(\tau) e^{-\tilde{r}/2} + \dots\,
\ee
with $\tilde S_f(\tau+\beta) = - \tilde S_f(\tau)$, an $SU(2)$ doublet fermion field proportional to the boundary supercharge of the theory. This component $\tilde{S}_f$ is allowed to fluctuate.  

Next, for the $SU(2)$ gauge field we wish to impose a fixed holonomy $h = \cP \exp \left(\oint_{r\to\infty} B\right) = \exp(2\pi i \alpha\sigma^3) $ at the asymptotic boundary, with $\alpha \sim \alpha +1$, and we adopt the notation $\alpha \equiv - i \beta \Omega/ (4\pi)$. To obtain the boundary conditions at the edge of the AdS$_2$ region we will  follow the strategy in appendix A of \cite{Iliesiu:2020qvm}. We will work in a gauge where $B_r = 0$ to obtain the general solution for the gauge field $B$ and for the zero-form field $b$ (the Lagrange multiplier for $B$ in 2D):
\be 
\label{eq:B-field-solution}
B= i\frac{2\pi \alpha T^3}{\b} \left(1+\frac{\mC}{r^3}\right) d\tau\,, \qquad H_{r\tau} =-i\frac{6\pi \alpha T^3}{\b} \frac{\mC}{r^4}\,, \qquad b = \frac{4\pi \mC \,\alpha\, T^3}{\b}\,,
\ee
where $\mC$ is an undetermined constant. Solving for $\mC$, we find that $B_\tau$ and $b$ are related at the boundary that separates the near-horizon region from the asymptotic region as 
\be
\label{eq:SU(2)-gauge-field-sol}
B_\tau  =\frac{ 2\pi  i\alpha T^3}{\b} + \frac{i b}{2 r_0^3}= \frac{ 2\pi  i \alpha T^3}{\b} + \frac{i b }{2\Phi_r} \, .
\ee
Thus, the boundary condition at the asymptotic boundary reduces to fixing, at the boundary of AdS$_2$, the combination 
\be
\label{eq:bc-SU(2)-gauge-field}
B_\tau - \frac{i b}{2\Phi_r }  = i\frac{2 \pi \alpha}{\beta}T^3\,,
\ee
which implies that $B_\tau$ is varying at the boundary of the AdS$_2$ region. To simplify the notation, it will prove convenient to refer to $B_\tau|_{\tilde r \to \infty}  \equiv \tilde S_T = \tilde S_T^i T_i$, with $\tilde S_T(\tau+\beta) = \tilde S_T(\tau)$. The boundary condition that corresponds to computing the index once again has $\alpha=1/2$ in this notation.

To make better contact with the value of the fields in the Boyer-Lindquist coordinates, we can perform the gauge transformation $h=e^{\Omega \tau T_3}$ mentioned above. We will denote the values of $\psi$ and $B_\tau$ at the AdS$_2$ boundary after performing this transformation by $S_f(\tau)$ and $S_T(\tau)$, respectively. Under such a gauge transformation, the boundary conditions for $S_f$ and $S_T$ are given by, $S_f(\tau+\beta) = - e^{2\pi i \alpha \sigma^3}S_f(\tau)$ and $S_T^i(\tau+\beta) = e^{2\pi i \alpha \sigma^3} S_T^i(\tau)$. In this gauge, the boundary condition for the $SU(2)$ gauge field is the same as \eqref{eq:bc-SU(2)-gauge-field} but with a vanishing right-hand side. Therefore, the information of the chemical potential is encoded in the monodromy of the fields around the thermal circle.

Finally, with these boundary conditions, one can find the correct boundary term in the associated $PSU(1, 1|2)$ BF theory consistent with the variational principle:
\be 
\label{eq:boundary-action-N=4-superJT}
I_{\cN=4 \text{ super-JT},\,\text{bdy.}} = -{\Phi_r} \int_{\partial \cM} d\tau S_b(\tau)\,,
\ee
where $S_b$ is defined as 
\be
\label{eq:def-cL(u)-and-bdy-Lagr}
S_b(\tau)  = \tilde S_b(\tau) + \frac{1}2 \left((S_T^1(\tau))^2+(S^2_T(\tau))^2+ (S^3_T(\tau))^2\right)\,.
\ee 
By studying how $S_b(\tau)$ transforms under infintesimal superdiffeomorphisms one can identify $S_b(\tau)$ as the $\cN=4$ super-Schwarzian derivative and $I_{\cN=4 \text{ super-JT},\,\text{bdy.}}$ as the $\cN=4$ super-Schwarzian action. Similarly, one can identify $S_f(\tau)$ and $S_T(\tau)$ as the other components of the   $\cN=4$ super-Schwarzian action when written in the superspace formalism.  Thus the super-Schwarzian derivative can be parametrized by a  $\text{SDiff}(S^{1|4})$ transformation which can be expressed in terms of a bosonic diffeomorphism $f(\tau)$, a bosonic $SU(2)$ transformation $g(\tau)$, as well as fermionic transformations which we can parametrize using 4 Majorana fermions, $\eta^p(\tau)$ and $\bar \eta_p(\tau)$. In terms of these fields the boundary action \eqref{eq:boundary-action-N=4-superJT} can be written as 
\be 
\label{eq:N=4-super-Schw-bosonic-comp}
I_{\cN=4 \text{ super-JT},\,\text{bdy.}} = -\Phi_r \int_0^\beta d\tau S_b(\tau) =- \Phi_r \int_0^\b d\tau \left[\Sch(f,\tau) + \Tr(g^{-1} \partial_\tau g)^2 + ({\rm fermions})\right]\, ,
\ee 
where 
\be 
\Sch(f,\tau) =\frac{f'''(\tau)}{f'(\tau)} - \frac{3}2 \left(\frac{f''(\tau)}{f'(\tau)}\right)^2\,
\ee
is the usual Schwarzian derivative together with the Lagrangian $ \Tr(g^{-1} \partial_\tau g)^2 $ of a particle moving on an $SU(2)$ manifold. The parenthesis includes all terms which involve $\eta(\tau)$ or $\bar \eta(\tau)$, including the couplings of these fermionic fields to $f(\tau)$ or $g(\tau)$. At the boundary, the gauge field $B_\tau = S_T$ together with the decaying part of the gravitino are both allowed to vary, and they are given by
\be
B_\tau^i = S_T^i = (g^{-1} d g)^i + (\text{fermions})\,, \qquad S_f = (\text{fermions})\,,
\ee
where for brevity we have once again suppressed the $\eta$ and $\bar \eta$ dependence of $S_T$ and $S_f$.  
For the remainder of this section, we will not rely on the super-Schwarzian action \eqref{eq:N=4-super-Schw-bosonic-comp} since we will be able to understand the presence of zero modes by directly looking at the action of super-diffeomorphisms on $S_b$, $S_f$, and $S_T$ in the bulk. We will solely use the saddle-point values of $S_b$, $S_f$, and $S_T$, around which we will compute the effect of the fluctuations. These saddle-point values are given by solving the equations of motion for $f(\tau)$, $g(\tau)$, $\eta(\tau)$, and $\bar \eta(\tau)$ in \eqref{eq:N=4-super-Schw-bosonic-comp}, with the boundary conditions which follow from those for $S_b$, $S_f$, and $S_T$ for the choice of Boyer-Lindquist coordinates described above:\footnote{Note that in the corotating coordinates, which have a smooth $SU(2)$ gauge field configuration at the horizon, the boundary condition for $g$ is modified, and there is an additional term in the action   \eqref{eq:N=4-super-Schw-bosonic-comp} which is due to the ``illegal'' gauge transformation which relates the two coordinate systems. }
\be 
f(\tau+\beta) = f(\tau)\,,\qquad  g(\tau+\beta) = e^{2\pi i \alpha \sigma_3} g(\tau) \,,\qquad   \eta(\tau+\beta) = - e^{2\pi i \alpha \sigma_3} \eta(\tau)\,.
\ee 

The solutions to the equations of motion are given by
\be
\label{eq:classical-solution-N=4-super-Schwarzian}
f(\tau)=\tan(\pi \tau/\b)\,, \qquad g(\tau) = \exp\left(2\pi i \sigma_3 \left(n+\alpha\right) \frac{\tau}\beta\right)\,,\qquad \eta(\tau) = 0\,, \qquad \bar \eta(\tau) = 0\,,
\ee
where $n \in \mZ$ labels the different solutions of the $SU(2)$ gauge field configurations. These solutions precisely correspond to the different saddles found in section \ref{sec:saddle-point-analysis} which represent the redundancy of the holonomy under the shift $\alpha \to \alpha + n$. We can in fact recognize that the saddles in \eqref{eq:N=4-super-Schw-bosonic-comp} match the gravitational Kerr-Newman saddles discussed in section \ref{sec:saddle-point-analysis}. This follows from the saddle-point value of $S_b $ (we also list the saddle-point values of $S_f^p$ and $S_T^i$ for later use)  
\be 
\label{eq:saddle-points-disk-N=4-Schw}
S_b^{\text{(saddle)}} = \frac{2\pi^2}{\beta^2} \left(1 - 4(n+\alpha)^2\right)\,, \qquad S_f^{\text{(saddle)}} =0 \,, \qquad S_T^{i,\text{ (saddle)}} = \frac{4\pi i}{\beta}(n+\alpha) \delta^{i,3}\,.
\ee
where $\alpha = 1/2$ gives the particular values associated to the index. By using the value of $\Phi_r = Q^3$, we get that at the saddle-point 
\be
-I_{\cN=4 \text{ super-JT},\,\text{bdy.}} = \frac{2\pi^2 Q^3}{\beta} \left(1 - 4(n+\alpha)^2\right)\, ,
\ee
which together with the entropy term $S_0$ in \eqref{eq:2D-bosonic-action} and the value of the action in the region between the AdS$_2$ boundary and the asymptotic boundary (which simply yields $-\beta Q$) we recover the saddle-point value of the 4D action obtained in \eqref{eq:ActionNSmallT4D}. Having matched the value of the classical saddle from section \ref{sec:saddle-point-analysis} using the super-Schwarzian technology, we can now proceed to study the existence of zero modes by analyzing the action of bulk super-diffeomorphisms on $S_b$, $S_f$, and $S_T$.

\subsection{Super-diffeomorphisms and zero modes } 

Fluctuations in JT supergravity are given by the action of large bulk super-diffeomorphisms, which preserve the AdS$_2$ boundary conditions discussed above. We are interested in finding the zero modes which lead to the vanishing of non-BPS saddles or, as we shall see in the next section, can lead to the vanishing of contributions from all connected geometries. In order for such zero modes to lead to the vanishing of such saddles, they need to be (i) fermionic, (ii) without a noncompact bosonic zero mode counterpart that might cause a divergence, and (iii) not be gauged away when quotienting the path integral over all metrics and gravitino configurations by super-diffeomorphisms. To make the latter point precise, it is important to distinguish two types of zero modes:
\begin{itemize}
\item \textbf{Gauged zero modes:} In supergravity, fluctuations  obtained from superdiffeomorphisms that leave the boundary values of all fields invariant, 
\be 
\label{eq:gauge-zero-modes-cond}
\delta S_b= 0\,, \qquad \qquad \delta S_f = 0\,, \qquad \qquad \delta S_T = 0\,,
\ee
should not be integrated over. Thus, even though they appear as zero modes of the action, they are not integrated over and they do not make the saddle that we expand around vanish.  
In terms of the BF formulation of the theory, transformations that obey \eqref{eq:gauge-zero-modes-cond} have a vanishing symplectic measure and thus once again should not be integrated over. In terms of the super-Schwarzian theory introduced above, we will see that these zero modes generate a global symmetry which we should quotient by when performing the path integral. This global symmetry is the $PSU(1, 1|2)$ isometry in the near-horizon metric (which in turn is the same as the gauge group in the BF theory).

\item \textbf{Physical zero modes:} Fluctuations obtained from super-diffeomorphisms that leave the action invariant (to quadratic and possibly higher orders in the fluctuation), but change the boundary values of the fields at the boundary, i.e.
\be
\delta^2 S_b = 0\,,\qquad\text{ and }\qquad   \delta S_f \neq 0\, \,\, \text{or}\,\,\,\delta S_T \neq 0\,,
\ee
are physical zero modes. If this is a fermionic fluctuation, then the field configuration around which we have expanded has a vanishing contribution to the path integral. If we have bosonic zero modes, we need to make sure their moduli space is compact such that the final answer is not divergent. 

\end{itemize}

Our task therefore is to find and distinguish these two types of zero modes in the near-horizon region. It is convenient to recast the fluctuations implemented by super-diffeomorphisms in terms of gauge transformations which change the $PSU(1, 1|2)$ gauge field as
\be
\delta A = d\Lambda + [A, \Lambda],
\ee
where we should solely consider gauge transformations that do not affect the boundary conditions discussed above. A general expression for them can be found in section 2.2 of \cite{Heydeman:2020hhw}:
\be
\Lambda(\tau) &= \frac{\xi}2 L_1 - \xi' L_{0} + \left(\frac{1}2 \bar S_f   \epsilon +\frac{1}2 S_f \bar \epsilon - \frac{1}2  \xi (S_T)^2 +S_b \xi + \xi'' \right) L_{-1} - (t^i -\xi S^i_T) T_i \nn \\ &\quad + \frac{1}2  \bar \epsilon^p G_{p,\frac{1}2} - \left((\bar \epsilon^p)' - \frac{1}2 \xi \bar S_f^p  - \frac{1}2 S_T^i (\sigma^i{}^*){}^p_q\, \bar \epsilon^q\right)  G_{p,-\frac{1}2} + \frac{1}2 \epsilon^p \bar G_{p,\frac{1}2} \nn \\ &\quad - \left( (\epsilon^p)' - \frac{1}2 \xi  S_f^p  - \frac{1}2 S_T^i (\sigma^i)^p_q\,  \epsilon^q\right)  \bar G_{p,-\frac{1}2},
\ee
where we see that diffeomorphism, supersymmetry and rotation generators are mixed by the fall-off constrains. It is easy using the results in \cite{Heydeman:2020hhw} to translate these transformations to the second-order formalism, but we will not do that here. Such transformations are parametrized by four bosonic fields $\xi$ and $t_{i=1,2,3}$ and four fermionic fields $\epsilon_a$, and $\bar \epsilon_a$ with $a = 1, 2$. The transformations on $S_b$, $S_f$ and $S_T^i$ are given by
\bea 
\label{eq:Sb-transformation}
\delta S_b&=&\xi S_b'+2S_b \xi'+ \xi'''- S_T^i (t^i)'+\frac{1}{2}\left(3\bar{S}_f\epsilon'+\bar{S}_f'\epsilon-3\bar{\epsilon}'S_f-\bar{\epsilon} S_f'\right), \nonumber\\ 
\delta S_f^p&=& \xi (S_f^{p})'+\frac{3}{2}S_f^p \xi'-\epsilon^p S_b-2(\epsilon^p)''-\frac{1}{2}t^i (\sigma^i){}^p_qS_f^q+(S_T^i)' (\sigma^i{})^p_q\epsilon^q+2 S_T^i (\sigma^i{})^p_q(\epsilon^q)',\nonumber\\ 
\delta S_T^i&=& \left( \xi S_T^i\right)'-(t^i)'+\frac{1}2 \bar{S}_f\sigma^i\epsilon+ \frac{1}2\bar{\epsilon}\sigma^iS_f+i \epsilon^{ijk}t^j S_T^k\, ,
\eea
where $\xi$ implements an infinitesimal bosonic diffeomorphism, $\epsilon$ implements an infinitesimal supersymmetric transformation, and the $t^i$ implement infinitesimal $SU(2)$ gauge transformations. When expanded around the saddle \eqref{eq:saddle-points-disk-N=4-Schw}, the variations become 
\bea
\label{eq:variation-of-Ss-at-saddle}
\delta S_b&=& \frac{4\pi^2}{\beta}(1-4(n+\alpha)^2) \xi'+ \xi'''- \frac{4\pi i}{\beta}  (n+\alpha) (t^3)', \nonumber\\ 
\delta S_f^p&=& -\frac{4\pi^2}{\beta}(1-4(n+\alpha)^2)\epsilon^p -2(\epsilon^p)''+ \frac{8\pi i}{\beta}(n+\alpha) (\sigma^3{})^p_q(\epsilon^q)',\nonumber\\
\delta S_T^i&=& \frac{4\pi i}{\beta}(n+\alpha) \delta^{i,3}\xi' -(t^i)'+i \epsilon^{ijk}t^j  \frac{4\pi i}{\beta}(n+\alpha) \delta^{k,3}.
\eea

\paragraph{Gauged zero modes: } To find the gauged zero modes, all variations in \eqref{eq:variation-of-Ss-at-saddle} have to vanish. From the boundary conditions for $S_b$, $S_f$, and $S_T$, for generic $\alpha$ we have to require that $\xi(\tau+\beta) = \xi(\tau)$, $t_i(\tau+\beta) = (e^{i 4\pi \alpha \sigma_2})_{ij} t_j(\tau) $ for $i, j= 1,2$, $t_3(\tau+\beta) = t_3(\tau)$, and $\epsilon(\tau+\beta) = -e^{2\pi i \alpha \sigma_3}\epsilon(\tau)$.\footnote{The boundary conditions for $\xi$ and $\epsilon$ follow obviously from those for $S_b$ and $S_f$, and $f$ and $\eta$ respectively.  The boundary conditions for $t_i$ follows from that for $g$, since $t$ acts infinitesimally on $g$ as $g \to (1+\frac{1}2 t_i T_i) g$. Thus, 
\be 
g(\beta) = \left(1+\frac{1}2 t_i(\beta) T_i\right) g^\text{sol}(\beta) = \left(1+\frac{1}2 t_i(\beta) T_i\right)e^{2\pi i \alpha \sigma_3} g^\text{sol}(0) = e^{2\pi i \alpha \sigma_3} \left(1+\frac{1}2 t_i(0) T_i\right) g^\text{sol}(0) = e^{2\pi i \alpha \sigma_3} g(0)\,,
\ee
 From the second-to-last equality, the boundary condition for $t_i(\tau)$ follows. }
 For any $\alpha$, the solutions to these equations are as follows. The bosonic diffeomorphisms yield the zero modes
\be 
\label{eq:diffeo-zero-modes}
\xi  &= C_1 + C_2 \cos\frac{2\pi \tau}{\beta}+ C_3 \sin\frac{2\pi \tau}{\beta}\,, 
\ee
while the $SU(2)$ gauge transformations yield
\be
t_3 &= C_4 + 2iC_2\left(n+\alpha\right)\cos(\frac{2\pi \tau}{\beta})+ 2i C_3(n+\alpha) \sin\frac{2\pi \tau}{\beta}\,,\nn\\ t_1& =  C_5 \cos \frac{4\pi (n+\alpha) \tau}{\beta} - C_6  \sin \frac{4\pi (n+\alpha) \tau}{\beta} \,,
\nn \\ 
t_2 &=C_6 \cos \frac{4\pi (n+\alpha) \tau}{\beta} + C_5  \sin \frac{4\pi (n+\alpha) \tau}{\beta}\,,
\ee
and finally for supersymmetry transformations we have the zero modes
\be
\label{eq:fermionic-zero-modes}
\epsilon^1 &= e^{2\pi i (n+\alpha) \frac{\tau}\beta} \left(C_7\, e^{-i\frac{\pi \tau}{\beta}} + C_8 \,e^{i\frac{\pi \tau}{\beta}} \right)\,,\nn \\ 
\epsilon^2 &= e^{-2\pi i (n+\alpha) \frac{\tau}\beta} \left(C_9\, e^{-\frac{i\pi \tau}{\beta}} + C_{10} \,e^{i\frac{\pi \tau}{\beta}} \right)\,.
\ee
Thus, for the disk saddle points in \eqref{eq:saddle-points-disk-N=4-Schw} there are 14 zero modes (six bosonic and eight fermionic, which correspond to the 14 generators of the $PSU(1,1|2)$ supergroup)  that need to be eliminated in order to get a non-degenerate symplectic measure. 

\paragraph{Fermionic physical zero modes: } To study the physical zero modes, we need to study when 
\beq
\delta^2 I_{\cN=4 \text{ super-JT}} \sim \int_0^\beta d\tau\, \delta^2 S_b = 0,
\eeq
to at least quadratic order in the field variations. As mentioned above, we will be particularly interested in fermionic zero modes and therefore we will analyze the expansion of $\delta S_b$ at quadratic order in the supersymmetric variation given by $\epsilon$. Around the saddle point, $S_b$ receives corrections at quadratic orders in the field variations and therefore we can simply set $S_b$ to a constant in the formula for $\delta S_b$. The same is true for $S_T^i$ since it is bosonic and receiving a linear order correction from a fermionic field is impossible. Therefore, the only quadratic contribution to $\delta S_b$  can come from the linear corrections to $S_f$ and $\bar S_f$ in terms of $\epsilon$ and $\bar \epsilon$. In other words, after using \eqref{eq:Sb-transformation}, the quadratic fluctuation of $S_b$ under the appropriate supersymmetric transformation is given by,
\bea 
\label{eq:relevant-susy-transformation}
\delta_\epsilon S_b&=&  \frac{1}{2}\left(3(\delta_\epsilon \bar{S}_f)\epsilon'+(\delta_\epsilon \bar{S}_f)'\epsilon-3\bar{\epsilon}' ( \delta_\epsilon S_f) -\bar{\epsilon} (\delta_\epsilon S_f)'\right), \nonumber\\ 
\delta_\epsilon S_T^i&=& \frac{1}2 (\delta_\epsilon \bar{S}_f)\sigma^i\epsilon+ \frac{1}2\bar{\epsilon}\sigma^i (\delta_\epsilon S_f)\,,
\eea 
with 
\be 
\label{eq:relevant-susy-transformation2}
\delta_\epsilon S_f^p= -\epsilon^p S_b^{\text{(saddle)}}-2(\epsilon^p)''+(S_T^{i, \,\text{(saddle)}})' (\sigma^i{})^p_q\epsilon^q+2 S_T^{i, \text{ (saddle)}} (\sigma^i{})^p_q(\epsilon^q)'\, .
\ee
When reformulating $\cN=4$ JT supergravity as a $PSU(1,1|2)$ BF theory, the superdiffeomorphism \eqref{eq:relevant-susy-transformation} in the second order formalism is equivalent to the gauge transformation 
\be
\Lambda(\tau)=  \left(\frac{1}2 \bar S_f   \epsilon +\frac{1}2 S_f \bar \epsilon  \right) L_{-1} &+ \frac{1}2  \bar \epsilon^p G_{p,\frac{1}2}  - \left((\bar \epsilon^p)' - \frac{1}2 S_T^i (\sigma^i{}^*){}^p_q\, \bar \epsilon^q\right)  G_{p,-\frac{1}2} \nn \\ &+ \frac{1}2 \epsilon^p \bar G_{p,\frac{1}2} - \left( (\epsilon^p)'  - \frac{1}2 S_T^i (\sigma^i)^p_q\,  \epsilon^q\right)  \bar G_{p,-\frac{1}2}\,,
\ee
in the BF description. This equation shows that the mode generated by $\epsilon$ and $\bar{\epsilon}$ also requires a bosonic diffeomorphism (first term above) in order to satisfy the appropriate boundary conditions at infinity.

After integrating by parts, the fermionic variation of $\int_0^\beta d\tau \delta_\epsilon^2 S_b$ is given by
\be 
\label{eq:variation-action-N=4}
\delta_\e^2 I_{\cN=4 \text{ super-JT}} \sim \int_0^\beta d\tau (\delta_\epsilon^2 S_b)\sim \int_0^\beta d\tau \left((\delta_\epsilon \bar S_f) \epsilon' - \bar \epsilon' (\delta_\epsilon S_f)\right) \, .
\ee
The above only vanishes when (a) $\delta_\epsilon \bar S_f = 0$, or (b) $\epsilon = \text{ constant}$, or importantly, when (c) both  $\bar S_f =\epsilon '= 0$. The first case (a) implies that such fluctuations are gauge modes and are already discussed above. The second case (b), corresponds to a genuine zero mode if $\epsilon$ being constant is consistent with the boundary conditions. From the boundary condition for $S_f$ for generic $\alpha$, we have that $\epsilon(\tau +\beta) = -e^{2\pi i \alpha \sigma_3} \epsilon(\tau)$, so we see that this zero mode only exists if $\alpha = 1/2$ (i.e. precisely when we are computing the index of the theory and fermions are periodic). As long as $\delta_\epsilon S_f\neq 0$, this zero mode is physical and its one-loop determinant and contribution to the partition function will vanish. If $\delta_\epsilon \bar S_f = \delta_\epsilon \bar S_f = 0$, then we are in situation (c) and this constant zero mode in the action coincides with one of the gauge zero modes written in \eqref{eq:fermionic-zero-modes}.

From \eqref{eq:relevant-susy-transformation2} we see that for constant $\epsilon$ we have that $\delta_\epsilon \bar S_f^p = -\bar \epsilon^p S_b $. Thus $\delta_\epsilon \bar S_f^p$ only vanishes for constant $\epsilon^p$ if $S_b = 0$. For $n \neq 0$ and $\alpha = 1/2$, $S_b$ is never zero and thus we have found the desired zero mode. For $n = 0$, we have that  $S_b = 0$ and, consequently, $\delta_\epsilon \bar S_f^p  =0$ which implies that this is a gauge zero mode and there are no physical zero modes.\footnote{These are the $C_7$ and $C_9$ zero modes in \eqref{eq:fermionic-zero-modes}, which for $n=0$ and $\alpha= 1/2$ are just constant.} 
Therefore, to summarize, we have proven that when $\a=1/2$ all saddles with $n>0$ have a vanishing fermionic one-loop determinant, while the saddle with $n= 0$ has a non-vanishing, temperature-independent (since $S_b=0$) contribution to the index. We will turn to reviewing the moduli space of saddles and the corresponding bosonic zero modes at $\a=1/2$ in a second to check that the bosonic determinant is well-behaved.

Finally, as a consistency check, note that these boundary fermionic zero modes can be trivially extended in the entire near-horizon region by choosing a spinor $\epsilon(\tilde r, \tau)$ with   $\epsilon|_{\tilde r \to \infty} = \epsilon(\tau)$ from \eqref{eq:relevant-susy-transformation}. Since all solutions in the near-horizon region are related by the action of super-diffeomorphisms, the gravitino solution $ \psi_\mu = \nabla_\mu \epsilon$ is precisely the gravitino fluctuation that corresponds to a physical zero mode. 

\paragraph{Bosonic physical zero modes:} We should also address the physical bosonic $SU(2)$ zero modes.\footnote{We thank J.~Maldacena for suggesting the interpretation given below.} When $\alpha = 1/2$, $e^{2\pi i \alpha \sigma_3} = - 1$ is in the center of $SU(2)$ and there is a moduli space of different solutions for $g(\tau)$ which we explicitly discuss in appendix \ref{sec:SU(2)-BF-appendix}. In the 4D discussion of section \ref{sec:saddle-point-analysis}, this moduli space corresponds to Kerr-Newman solutions rotating along different axes, which for the special cases of $\alpha=0,\tfrac{1}{2}$ are all equivalent, as mentioned in footnote~\ref{footnote:moduli space of rotating black holes}. 
One might be concerned that these bosonic zero modes will eliminate the effect of the physical fermionic zero modes discussed above, or even render the index divergent. This is not the case as explained in appendix \ref{sec:SU(2)-BF-appendix}, because the moduli space is compact, and the integral over the bosonic zero modes gives an integral over this compact moduli space.

Finally, after showing that the one-loop determinant around the saddle that contributes to the index is nonzero and finite we need to argue that it is temperature independent. The only possible temperature dependence is an overall power of $\beta$ which is related to the number of fermionic minus bosonic zero modes. In the case of arbitrary $\alpha$, this prefactor is linear in $\beta$, which would spoil the expected properties of the index. The resolution of this issue is that, as explained in appendix \ref{sec:SU(2)-BF-appendix}, for the case of the index with $\alpha=1/2$ the (compact) bosonic zero modes contribute with an extra factor of $1/\beta$, making the one-loop determinant temperature-independent.

\subsection{Comparison to the exact answer}
Finally, for completeness we are going to show how to reproduce the result of this section starting from the exact $\cN=4$ JT disk partition function for any angular velocity, $\Omega= 4\pi i \alpha/\beta$ \cite{Heydeman:2020hhw}:
\beq\label{eq:diskN4partitionfunction}
Z_\text{Disk}=e^{S_0} \sum_{n\in\mathbb{Z}} \frac{\beta}{\Phi_r} \frac{2 (\alpha+n) \cot(\pi \alpha)}{\pi^3(1-4(n+\alpha)^2)^2}  e^{\frac{2\pi^2\Phi_r}{\beta}(1-4(n+\alpha)^2)} \,.
\eeq
We take the limit $\alpha\to1/2$ of equation \eqref{eq:diskN4partitionfunction}. The contribution from each saddle is 
\beq
\frac{\beta}{\Phi_r}\frac{2 \cot(\pi \alpha)(\alpha+n)}{\pi^3(1-4(n+\alpha)^2)^2} e^{\frac{2 \pi^2 \Phi_r}{\beta}(1-4(n+\alpha)^2)} \Big|_{\alpha \to 1/2}=0 ,~~~\text{unless}~~n=-1,0,
\eeq
due to the $\cot(\pi \alpha) \to 0$ as $\alpha \to 1/2$. The physical fermionic zero modes are responsible for this. Then, only the terms with vanishing classical action contribute. But these are also the terms where the denominator in \eqref{eq:diskN4partitionfunction} vanishes, and therefore the limit has to be taken with some care. The denominator vanishes for each term due to the bosonic $SU(2)$ zero modes arising from the enhanced symmetry at $\a=1/2$, as explained above and in appendix~\ref{sec:SU(2)-BF-appendix}. The resolution is simple; remembering that the saddles with $n=-1,0$ are joined together by the enhanced moduli space at $\a=1/2$, we should add their contributions and take the limit accordingly. 
Expanding each of these two saddles in small $\e$ with $\alpha=1/2+\epsilon$, we see that
\bea
n=0:~~~~~~~~~\frac{\beta}{\Phi_r}\frac{2 \cot(\pi \alpha)(\alpha+0)}{\pi^3(1-4(0+\alpha)^2)^2} e^{\frac{2 \pi^2 \Phi_r}{\beta}(1-4(0+\alpha)^2)}  &=& -\frac{\beta}{\Phi_r}\frac{1}{16\pi^2 \epsilon} + \frac{1}{2} + \mathcal{O}(\epsilon),\nn\\
n=-1:~~~~~\frac{\beta}{\Phi_r}\frac{2 \cot(\pi \alpha)(\alpha-1)}{\pi^3(1-4(-1+\alpha)^2)^2} e^{\frac{2 \pi^2 \Phi_r}{\beta}(1-4(-1+\alpha)^2)} &=& \frac{\beta}{\Phi_r}\frac{1}{16\pi^2 \epsilon} + \frac{1}{2} + \mathcal{O}(\epsilon).
\ea
After summing the contributions of these two saddles, the divergent pieces as $\e\to 0$ cancel, and we obtain a finite limit. The result is simply ${\rm Index}(\beta,Q) = e^{-\beta Q}Z_{\rm Disk}(\beta, \alpha = 1/2) = e^{S_0 - \beta Q}$, which is temperature-independent as expected. 
This reproduces the result of the analysis made in the previous section.

\subsection{An example from $(4,4)$ supergravity in AdS$_3$}\label{sec:disk3D}

We conclude this section by showing a concrete example in higher dimensions for which an explicit holographic dual is known: black holes in asymptotically AdS$_3$ for $\mathcal{N}=(4,4)$ supergravity. We will once again see, this time in AdS, that a gravity one-loop calculation with supersymmetric boundary conditions produces an answer consistent with field theory expectations. 

We will be brief and follow the conventions in \cite{Kraus:2006nb}. $\mathcal{N}=(4,4)$ supergravity comes with a left- and right-moving $SU(2)_L\times SU(2)_R$ symmetry and corresponding Chern-Simons fields in the bulk with integer level $k$. In string theory constructions, they come from isometries of $S^3$ factors appearing in the compactification from ten to three dimensions. In units with unit AdS$_3$ radius, supersymmetry fixes the Newton constant, or equivalently the central charge, to be $c=3/(2G_N)=6k$.

We want to perform a gravity path integral with fixed boundary inverse temperature $\beta$, fixed angular velocity $\theta$ along AdS$_3$ and fixed $SU(2)_L \times SU(2)_R$ chemical potentials $z$ and $\bar{z}$, respectively \cite{Kraus:2006nb}. This corresponds to a boundary torus with complex structure $\tau = (\theta + i \beta)/2\pi$. The chemical potentials $z/\bar{z}$ are conjugate to the $SU(2)_{L/R}$ charges  $J_{L/R}^3$ along the Cartan direction, and in string theory can be interpreted as angular velocity along $S^3$.

Since we are interested in extracting the black hole spectrum, we restrict to the path integral calculation around a BTZ background in AdS$_3$, but we will sum over all saddles corresponding to the $SU(2)$ Chern-Simons fields. We will focus on the Ramond sector with periodic fermions along space, but antiperiodic in time.\footnote{Other sectors and other metric saddles like vacuum AdS$_3$ can be obtained by a combination of half-integer spectral flow and modular transformations, see for example \cite{Dijkgraaf:2000fq,Maloney:2007ud}.} The one-loop gravitational path integral can be computed by the methods of \cite{Maloney:2007ud} and gives \cite{Eguchi:1988af,Heydeman:2020hhw}
\beq\label{eq:oneloop3DA3}
Z_{BTZ} = \sum_{n,\bar{n}\in\mathbb{Z}}
  Z_\text{1-loop}(n,\bar{n}) \exp{\left(\frac{i \pi k}{2\tau}(1-4(z+n)^2)-\frac{i \pi k}{2\bar{\tau}}(1-4(\bar{z}+\bar{n})^2)\right)}
\eeq
with
\beq\label{eq:oneloop3D3}
Z_{\text{1-loop}}(n,\bar{n})= \left|\frac{(1-q')}{\eta(q')}~ \frac{y'{}^{2n}(q'{}^{(\frac{1}{2}+n)^2} y' - q'{}^{(\frac{1}{2}-n)^2}y'{}^{-1})}{\theta_{1}(q,-y'{}^2)} ~\frac{\theta_{3}^2(q',-y')}{\eta(q')^2 (1-y' q'{}^{n+\frac{1}{2}})^2(1-y'{}^{-1} q'{}^{-n+\frac{1}{2}})^2}\right|^2
\eeq
written in terms of $q'=e^{2\pi i(-1/\tau)}$ and $y'=e^{2\pi i (z/\tau)}$. The $\theta_{1,3}(q,y)$ are the Jacobi theta functions and $\eta(q)$ is the Dedekind eta function. Next, we explain the origin of each term in \eqref{eq:oneloop3DA3}: 
\begin{itemize}
\item Fixing the $SU(2)_L$ ($SU(2)_R$) chemical potential corresponds to fixing Dirichlet boundary conditions for the holomorphic (antiholomorphic) component of the corresponding Chern-Simons field. The component unspecified by the boundary condition is fixed by imposing the $SU(2)_{L,R}$ holonomy around the contractible cycle to be the identity, making the configuration smooth. These solutions are labeled by an integer $n$ for $SU(2)_L$ and $\bar{n}$ for $SU(2)_R$ as explained in \cite{Kraus:2006nb}. Then, the overall sum in \eqref{eq:oneloop3DA3} is a sum over distinct Chern-Simons classical configurations and plays the exact same role as the sum over $n$ in \eqref{eq:semiclassicalZgeneral}. 

\item The argument of the exponential in \eqref{eq:oneloop3DA3} is the classical action for the bosonic sector of the $(4,4)$ supergravity action evaluated on each solution. It includes both the Einstein-Hilbert and the Chern-Simons contributions.

\item Finally, \eqref{eq:oneloop3D3} is the supergravity one-loop determinant around each solution. The first term in \eqref{eq:oneloop3D3} is the graviton one-loop determinant \cite{Maloney:2007ud}. The second term is the $SU(2)_k$ Chern-Simons one-loop determinant, which can be easily extracted from the exact CS answer \cite{Elitzur:1989nr}. The third term is the gravitino one-loop determinant, given by a generalization of \cite{David:2009xg}. 
\end{itemize}

The analog of the index is now the elliptic genus \cite{Witten:1993jg}. Computing the elliptic genus corresponds to inserting a factor of $(-1)^{F_R}$ in the path integral, where $F_R$ is a right-moving fermion number. A field theory analysis implies that the answer should be holomorphic and depend only on $\tau$ and $z$. We will see how this emerges from a gravity calculation \cite{Dijkgraaf:2000fq}. 

Firstly, notice that $(-1)^{F_R} = e^{2 \pi i J_R^3}$, where $J_R^3$ is the $SU(2)_R$ charge along the Cartan. Therefore the elliptic genus is computed by a gravitational path integral with, in our conventions, $\bar{z}_\star=1/2$. We would like to stress that this solution is completely smooth in the bulk for the same reasons as in the Kerr-Newman case studied in the previous sections, both for gravity and the Chern-Simons fields. Secondly, the classical action in \eqref{eq:oneloop3DA3} is only holomorphic when $\bar{z}=1/2$ and either $\bar{n}=0$ or $-1$ since for these values $\frac{i \pi k}{2\bar{\tau}} (1-4(\bar{z}_\star+\bar{n})^2)\to 0$.

The contributions from other values of $\bar{n}\neq 0,-1$ are not holomorphic and therefore inconsistent with field theory expectations. This problem is again resolved by remembering to include the one-loop determinant. For any $\bar{n}\neq 0,-1$, we have the relation $\bar{y}'(\bar{z}=1/2)=\bar{q}'{}^{1/2}$, and the theta function appearing in the gravitino one-loop determinant vanishes since $\theta_{3}(\bar{q}',-\bar{q}'{}^{1/2})=0$. This is due to physical fermion zero modes of the gravitino. Only for $\bar{n}=0,-1$ the denominator which explicitly comes from removing gauge zero modes also vanishes, giving a finite result. 

Just like in our previous four-dimensional case, new bosonic zero modes appear in the Chern-Simons sector. They manifest in the fact that the factor $\theta_{1}(\bar{q},-\bar{y}'{}^2)$ in the denominator of \eqref{eq:oneloop3DA3} vanishes when $\bar{z}=1/2$. The resolution is the same as in section \ref{sec:diskN4}, this divergence is canceled between $\bar{n}=0$ and $-1$ and can be reproduced by a one-loop calculation at $\bar{n}=0$ multiplied by the volume of the compact moduli space $S^2$.\footnote{The fact that the final answer is finite was verified in \cite{Eguchi:1988af}. Our one-loop gravity partition function is a product of their $\mathcal{N}=4$ super-Virasoro vacuum character. The Witten index of the vacuum character is one.} Therefore, the final answer is finite and holomorphic, as expected from field theory reasons.

\section{Wormholes and factorization in supergravity} \label{sec:factorization}

So far, our work has laid out essentially two ideas; establishing a suitable definition of a gravitational index and its computation in the disk topology. The focus of our analysis thus far arguably falls within the ``old-school'' holography (or AdS/CFT) paradigm, where one computes a well-studied quantity counting BPS states in a gravitational theory. Having clarified these points and made contact with established literature, we are now in a prime position to foray into the wilder aspects of quantum gravity. 

\subsection{Motivation and setup}

Topology-changing processes in the gravitational path integral bring up some puzzles and open questions. One such puzzle is the issue of so-called factorization. Performing the gravitational path integral with multiple disjoint boundaries leads to contributions where spacetime wormholes connect the boundary regions. These contributions are in addition to the disconnected contributions, and in particular, imply that the partition function $Z_\text{grav.}(\beta)$ fails to factorize into factors coming from each boundary. 
Rather, $Z_\text{grav.}(\b)$ is an averaged observable with higher moments  \cite{Maldacena:2004rf}:
\be
Z_\text{grav.}(\b_1,\, \cdots,\,\b_n) &\equiv 
\begin{tikzpicture}[baseline={([yshift=-.3ex]current bounding box.center)}, scale=0.6]
 \pgftext{\includegraphics[scale=0.5]{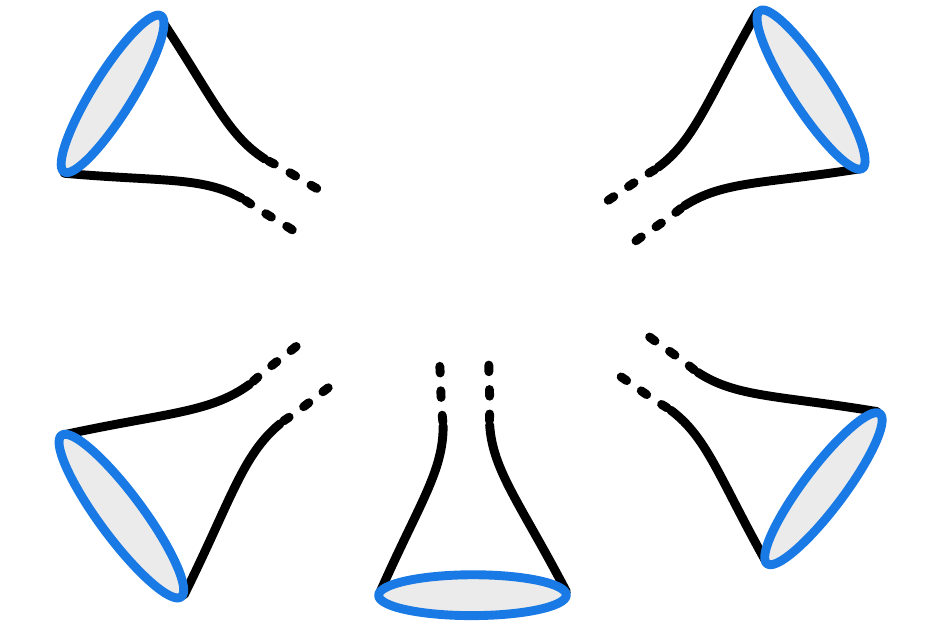}} at (0,0);
   \draw (-6,0) node  {$ \int Dg_{\mu \nu} \, D\text{[Fields]}$};
\end{tikzpicture}\nn \\ &\stackrel{\Large !}{\ne} \prod_{i=1}^n  Z_\text{grav.}(\b_i) \equiv \begin{tikzpicture}[baseline={([yshift=-.3ex]current bounding box.center)}, scale=0.6]
 \pgftext{\includegraphics[scale=0.5]{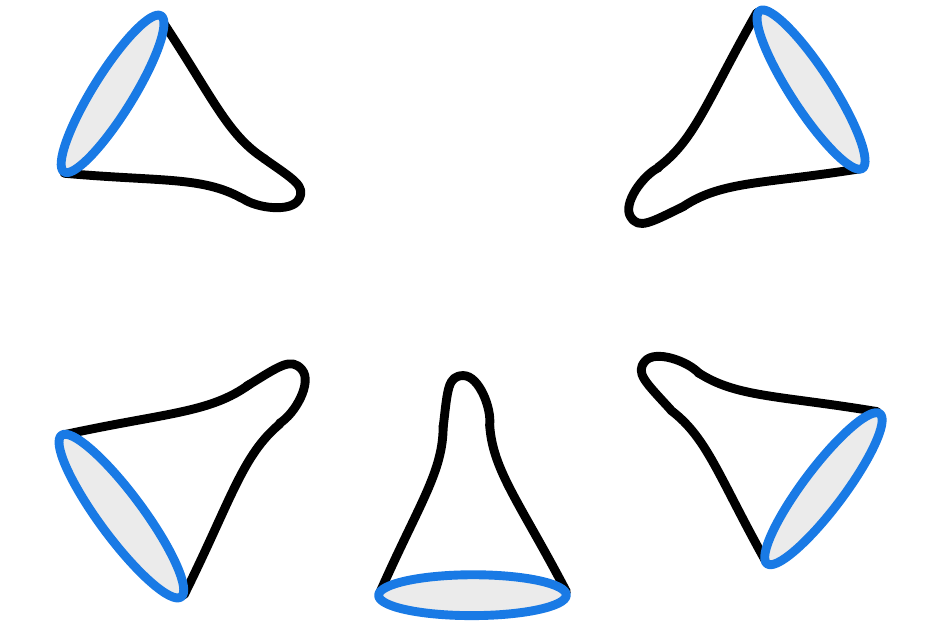}} at (0,0);
  \draw (-6,0) node  {$ \int Dg_{\mu \nu} \, D\text{[Fields]}$};
\end{tikzpicture}\, ,
\ee 
where $Z_\text{grav.}(\b_1,\, \cdots,\,\b_n)$ is the result of the gravitational path integral for $n$ boundaries with temperatures $\beta_1$, $\dots$, $\beta_n$.
The lack of factorization suggests generalizing the interpretation of AdS/CFT to include ensemble averages of dual boundary theories \cite{Saad:2019lba}. 

Curiously, this seems to run against known examples of AdS/CFT coming from string theory, wherein the holographic dual CFT is believed to be unique and not an ensemble. Notably, these constructions are also supersymmetric. This begs the question of whether we can sharpen the factorization puzzle by focusing on supersymmetric observables or at least find consistency between the two camps. Specifically, consider computing the supersymmetric index instead of the partition function. If the index fails to factorize due to the presence of spacetime wormholes in the sum over all geometries, we are faced with a dramatic scenario where we would be forced to reconsider or reinterpret the known supersymmetric AdS/CFT examples. A possibility is to achieve factorization by accounting for all UV-sensitive corrections. Nevertheless, one might hope to resolve the factorization puzzle for these quantities without studying the UV completion of the theory. In this section, we will show that it is indeed possible to resolve the factorization puzzle in this case by solely relying on the gravitational path integral of the low-energy effective theory.

Another open question regarding the contribution of topology-changing processes in the bulk is whether geometries that have a spacetime wormhole connecting different regions of the geometry contribute to the index even in the case of a single boundary. Such geometries are naturally included in the gravitational path integral; however, string theory examples where the index is explicitly known (such as cases in which the low-energy theory is given by 4D $\cN=4$ and $\cN=8$ supergravity) seemingly suggest that they either do not contribute or that their contribution is eliminated by the aforementioned UV-sensitive corrections \cite{dabholkar2010no}.

In this section, we will find resolutions for all these puzzles for the case of the gravitational index whenever the topology change occurs in the near-horizon region. Concretely, we will study the Euclidean gravitational path integral with multiple disconnected boundaries as in figure~\ref{fig:introindexmultiple2}, where on each boundary we will be interested in imposing the supersymmetry-preserving boundary conditions discussed in section \ref{sec:diskN4}. We will show that physical gravitino zero modes are always present on geometries connecting separate boundaries when such supersymmetric boundary conditions are imposed. Integrating over these fermionic zero-mode fluctuations makes the contribution of the connected geometry to the path integral vanish. Similarly, in the case of a single boundary with a near-horizon region that has a complicated topology, gravitino zero modes once again set the contribution to zero. A different way to rephrase these results is as follows. The near-horizon region of a boundary component with supersymmetric boundary conditions generically introduces gravitino zero modes; it is only in the particular case when the bulk geometry is BPS, such as the case of the BPS saddle in the disk topology considered in section~\ref{sec:diskN4}, that these zero modes are gauged, and therefore the contribution to the path integral is nonzero. Therefore, we conclude that the supersymmetric index factorizes and that topology-changing corrections are not present in the index.

Before focusing on the near-horizon AdS$_2$ case, let us comment on the analogous questions in AdS$_{d+1}$ regions with $d>1$, generalizing to higher-dimensional topology change, i.e.~where the topology change does not occur in the near-horizon AdS$_2$ region. To investigate the factorization of the index with AdS$_{d+1}$ regions, and the status of putative dual ensembles of CFT$_d$ for higher $d$, we would need to investigate the zero modes in supergravity with AdS$_{d+1}$ asymptotic regions, with boundary components of topology $S^1\times M_{d-1}$. We will not explicitly analyze these cases, but we expect that when supersymmetric boundary conditions are imposed on the boundary $S^1$, the gravitino zero modes are there and make the connected contributions to the gravitational index vanish. We will comment on the higher-dimensional case at more length in section~\ref{sec:factorization in higher dim}.

Having oriented ourselves, and armed with a road map, let us now lay out the setup of the main calculation of this section. We consider processes captured by AdS$_2$ throat regions, where the low-energy dynamics include a sector given by $\cN=4$ JT supergravity. The observable we want to study is the gravitational path integral with boundaries given by $n$ disjoint thermal circles, each with circumference $\b_i$ and $SU(2)$ holonomy specified by the parameter $\a_i$ as $e^{2\pi i \a_i \sigma_3}$ (as explained in section \ref{sec:saddle-point-analysis} this parameter is related to the angular velocity of the 4D black hole by $\Omega = 4 \pi i \alpha/\beta$). Since the low-energy dynamics are captured by $\cN=4$ JT supergravity, we have a good handle on how to perform the path integral. Firstly, the path integral is computed by summing over bulk topologies, and performing the path integral for fluctuations around each topology, as in~\cite{Saad:2019lba}. Working within the low-energy description, the integral over the dilaton sets the bulk geometry to have constant negative curvature $R=-2/r_0^2$, therefore the geometries in the near-horizon region can be dimensionally reduced to hyperbolic 2-manifolds with $n$ boundaries~\cite{Saad:2019lba}. Therefore, at small temperatures, the gravitational path integral is reproduced by the fluctuations in the $\cN=4$ JT supergravity sector around the hyperbolic geometries, with high-energy fluctuations suppressed by the temperature. We thus expect that the overall gravitational partition function in the theory of 4D  $\cN=2$ supergravity with $n$ boundaries and genus $g$ in the near-horizon region (denoted by $Z_{\text{near-horizon: }g,n}^{\text{4D } \cN=2 \text{ SUGRA}}((\beta_1,\Omega_1) , \ldots, (\beta_n,\Omega_n))$) can be related to the partition function of   $\cN=4$ super-JT gravity with the same topological characteristics (denoted by $Z_{g,n}^{\cN = 4 \text{ JT}}((\beta_1,\alpha_1) , \ldots, (\beta_n,\alpha_n))$) as
\be
\label{eq:low energy multi boundary connected contribution}
Z_{\text{near-horizon: }g,n}^{\text{4D } \cN=2 \text{ SUGRA}}((\beta_1,\Omega_1) , \ldots, (\beta_n,\Omega_n)) &\propto Z_{g,n}^{\cN = 4 \text{ JT}}((\beta_1,\alpha_1) , \ldots, (\beta_n,\alpha_n)),
\ee 
where we expect the contribution from other fields or from corrections of the $\cN = 4 \text{ JT}$ theory to appear as a nonzero proportionality constant that is independent of $\beta_1$, \dots, $\beta_n$ to leading order. The total path integral is then given as a sum over $Z_{g,n}$ for each bulk configuration, including disconnected ones. What we will explicitly show shortly using a zero-mode analysis similar to that in section \ref{sec:diskN4}, is that when computing the index (by setting all $\alpha_i = 1/2$ or equivalently $\Omega_i = 2\pi i/\beta_i$) all connected geometries and all $g\neq 0$ geometries in the RHS of \eqref{eq:low energy multi boundary connected contribution} have a vanishing contribution. Finally, we emphasize that our zero-mode analysis persists in any dilaton-type two-dimensional supergravity for which there is the same amount of supersymmetry, so we are not reliant on the precise form of the low-energy supergravity for this result. This will be related to our discussion in section \ref{sec:defects} of the possible non-perturbative contributions that the index can exhibit. 

\subsection{Superdiffeomorphisms and zero-mode analysis in the near-horizon region} 

When multiple near-horizon geometries are connected, we have several changes of the boundary conditions discussed in section \ref{sec:bdy-cond-4d-to-2d}. We will once again start our analysis in the corotating coordinates on each asymptotic boundary. To begin with, we will study the effect that a Euclidean wormhole has on the boundary conditions at the edge of the near-horizon wormhole. Firstly, we note that while in section \ref{sec:bdy-cond-4d-to-2d} the holonomies for the $PSL(2, \mR)$ and $SU(2)$ connections were trivial when evaluating them on a cycle contractible to the Euclidean horizon, this is no longer the case when studying connected geometries. Because the boundary of the near-horizon region is no longer on a contractible cycle, there could be an arbitrary $PSL(2, \mR)$ and $SU(2)$ holonomy along this cycle. We will study the boundary conditions of the fields described in section \ref{sec:bdy-cond-4d-to-2d} for fixed values of the $PSL(2, \mR)$ and  $SU(2)$ holonomy. We will start by completely turning off all fermionic fields to simplify our analysis. 

In order for the geometry to not be singular, one has to impose that the conjugacy class of the $PSL(2, \mR)$ holonomy is hyperbolic \cite{Witten:2020wvy, Goel:2020yxl},
\be 
\label{eq:holonomy-PSL(2, R)}
\cP \exp{\oint } \left(\sqrt{\frac{\Lambda}{2}} \left[e^1(x) L_0 + \frac{e^2(x)}2 \left(L_{1} - L_{-1}\right) \right]- \frac{\omega(x)}2 \left(L_1+L_{-1}\right) \right) \sim \left(\begin{matrix}
e^{b/2} & \\  & e^{-b/2} 
\end{matrix}\right) \, ,
\ee
with $b \in \mR$ and where the closed integral is along a cycle homotopic to the boundary of the near-horizon region.\footnote{Once again, since in the near-horizon region the bulk is well approximated by a BF theory, the holonomy is the same along any two curves that are homotopic to each other.}   In the near-horizon region, $b$ represents the length of the closed geodesic which is homotopic to the asymptotic boundary. Writing the spin-connection and frame fields in terms of $f(\tau)$, one finds that imposing \eqref{eq:holonomy-PSL(2, R)} implies that around the thermal circle, $f(\tau+\beta) = e^b f(\tau)$ \cite{Saad:2019lba}.

Since we assume that the topology change does not affect the region far away from the horizon, the boundary condition for $B_\tau$ in corotating coordinates remains the same as in \eqref{eq:SU(2)-gauge-field-sol}. However, while in corotating coordinates the holonomy around any cycle contractible to the Euclidean horizon was trivial, we can now impose that the holonomy for the $SU(2)$ gauge field along the same cycle is given by 
\be 
\label{eq:holonomy-for-B}
\cP \exp{\oint } B = e^{2\pi i \varphi \sigma_3}\,.
\ee

\begin{figure}[h!]
\begin{center}
\begin{tikzpicture}
    \node[anchor=south west,inner sep=0] at (0,0) {\includegraphics[width=0.65\textwidth]{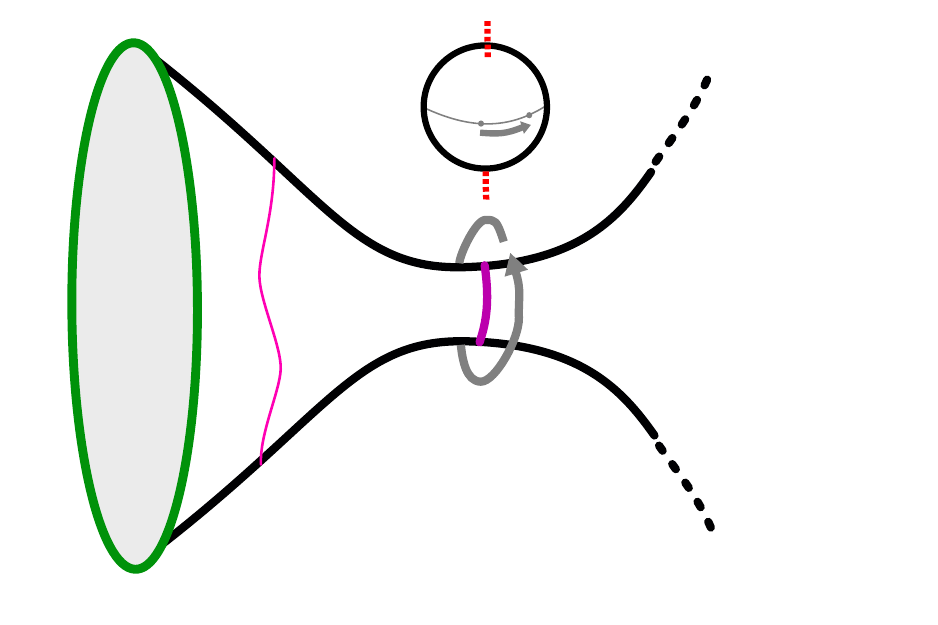}};
     \draw (5.5,+3.0) node  {$b$};
     \draw (6.15,+6.5) node  {$\varphi$};
\end{tikzpicture}
\end{center}
\caption{The geometry of the near-horizon region when the geometry is connected to another asymptotic boundary (located past the dotted curves). The purple curve is the closed geodesic of length $b$. As one goes around that cycle the internal space $S^2$ rotates by an angle $\phi$ due to the $SU(2)$ holonomy.  }
\label{fig:trumpet-bc}
\end{figure}

The physical interpretation of \eqref{eq:holonomy-for-B}, is that as one goes around the wormhole along the thermal circle $\tau$, the internal $S^2$ space experiences a rotation by an angle $\varphi$ around a given axis.
As explained in appendix \ref{sec:SU(2)-BF-appendix}, we can once again perform an illegal gauge transformation which takes us to the Boyer-Lindquist coordinates for which we get 
\be
B_\tau^i = S_T^i = (g^{-1} d g)^i\,,  
\ee
with the boundary conditions for $g$ around the thermal circle given by $ g(\tau+\beta) = e^{-2\pi i \alpha \sigma_3} g(\tau) e^{2\pi i \varphi \sigma_3}$. In these coordinates, the boundary condition for the fermionic fields remain the same as in section \ref{sec:bdy-cond-4d-to-2d}.

Thus, to perform our zero-mode analysis in the presence of the new boundary conditions, we will once again determine the saddle-point values of $S_b$, $S_f$, and $S_T$, around which we will compute the effect of the fluctuations. We will once again discuss the solutions for an arbitrary value of $\alpha$ and later specialize to the case of the index, which has $\alpha  = 1/2$. 
The solution to the equations of motion are given by\footnote{The saddles for the $SU(2)$ field $g(\tau)$ are explained more generally in appendix~\ref{sec:appBFtrumpet}. We simply write the diagonalized saddle here.}
\be
\label{eq:classical-solution-N=4-super-Schwarzian-trumpet}
f(\tau)= e^{\frac{b\tau}{\beta}}\,, \qquad g(\tau) =  \exp\left( 2\pi i \frac{\tau}{\beta} (\alpha \pm \varphi+n)\cdot \sigma_3\right)\,,\qquad \eta(\tau) = 0\,, \qquad \bar \eta(\tau) = 0\,,
\ee
where once again $n \in \mZ$ labels the different solutions of the $SU(2)$ gauge field configurations. The saddle point values of $S_b $, $S_f^p$, and $S_T^i$ on a given boundary component of a geometry with multiple connected near-horizon regions are given by
\be 
\label{eq:saddle-points-disk-N=4-Schw-trumpet}
S_b^{\text{(saddle)}} =  \frac{2\pi^2}{\b^2} \left(\left(\frac{b}{2\pi}\right)^2 + 4  (n+\a\pm \varphi) ^2 \right)\,, \qquad S_f^{\text{(saddle)}} =0 \,, \qquad S_T^{i,\text{ (saddle)}} = \frac{4\pi i}{\beta}(n+\alpha\pm \varphi) \delta^{i,3}\,,
\ee
where $\alpha = 1/2$ gives the particular values associated to the index. By using the value of $\Phi_r = r_0^3$, we get that at the saddle point 
\be
-I_{\cN=4 \text{ super-JT},\,\text{each bdy.}}^\text{connected} = \frac{2\pi^2 Q^3}{\beta} \left(\left(\frac{b}{2\pi}\right)^2+ 4(n+\alpha\pm \varphi)^2\right)\, ,
\ee
which together with the value of the action in the two regions between the AdS$_2$ boundary and the asymptotic boundary (which simply yields $-\beta Q$) we would get 
\be
-I_\text{2-connected BH} = -2\beta Q + \frac{2\pi^2 Q^3}{\beta} \left[2\left(\frac{b}{2\pi}\right)^2+ 4(n+\frac{1}2\pm \varphi)^2 + 4(m+\frac{1}2 \pm' \varphi)^2\right] \, .
\ee
We again proceed to study the zero modes of the action by studying the action of bulk super-diffeomorphisms on $S_b$, $S_f$, and $S_T$. 

\paragraph{Gauged zero modes: } To find the gauged zero modes, we demand that all variations in~\eqref{eq:Sb-transformation} around the saddles~\eqref{eq:saddle-points-disk-N=4-Schw-trumpet} have to vanish. From the boundary conditions for $S_b$, $S_f$, and $S_T$, for generic $\alpha$ we have to require that $\xi(\tau+\beta) = \xi(\tau)$, $t_i(\tau+\beta) = (e^{i 4\pi \alpha \sigma_2})_{ij} t_j(\tau)$ for $i, j= 1,2$, $t_3(\tau+\beta) = t_3(\tau)$, and $\epsilon(\tau+\beta) = -e^{2\pi i \alpha \sigma_3}\epsilon(\tau)$.\footnote{The boundary conditions for $\xi$ follows from the action of infinitesimal diffeomorphisms on $f(\tau) \to f(\tau) + \xi(\tau) f'(\tau) $, from which $\xi(\tau) = \xi(\tau+\beta)$. The boundary condition for $\epsilon$ again immediately follows from that of $S_f$. Again, the boundary conditions for $t_i$ follows from that for $g$:
\be 
\left(1+\frac{1}2 t_i(\beta) T_i\right) g(\beta) = \left(1+\frac{1}2 t_i(\beta) T_i\right)e^{2\pi i \alpha \sigma_3} g(0) e^{-2\pi i \varphi \sigma_3} = e^{2\pi i \alpha \sigma_3} \left(1+\frac{1}2 t_i(0) T_i\right) g(0) e^{-2\pi i \varphi \sigma_3}
\ee
and, from the last equality, the b.c.~for $t_i(\tau)$ follows. }
 For all $\alpha$ the only solution to these equations are
\be 
\label{eq:list-of-zero-modes-trumpet}
\xi  &= C_1, \qquad  t_3 = C_2\,.
\ee
Thus on each boundary of a connected geometry, we only have two solely-bosonic zero modes; one which was part of the $SL(2,\mR)$ generators on the disk, and the other was part of the $SU(2)$ generators of rotations around $S^2$. 

\paragraph{Fermionic physical zero modes: }
Once again, we seek to find modes for which on each boundary
\beq
\delta^2 I_{\cN=4 \text{ super-JT, each bdy.}}^\text{connected} \sim \int_0^\beta d\tau\, \delta^2 S_b = 0,
\eeq
to at least quadratic order in the field variations. Once again, the only quadratic contribution to $\delta S_b$  can come from the linear corrections to $S_f$ and $\bar S_f$ in terms of $\epsilon$ and $\bar \epsilon$. The variations for $S_b$ and $S_T^i$ remain unchanged and are given by \eqref{eq:relevant-susy-transformation}, while the variation of $S_f$ around the saddle is now given by 
\be 
\label{eq:relevant-susy-transformation22}
\delta_\epsilon S_f^p= - \frac{2\pi^2 \epsilon^p}{\b^2} \left(\left(\frac{b}{2\pi}\right)^2 + 4  (n+\a\pm \varphi) ^2 \right) -2(\epsilon^p)''+ \frac{8\pi i}{\beta}(n+\alpha\pm \varphi) (\sigma^3)^p_q(\epsilon^q)'.
\ee
Once again, after integrating by parts, the fermionic variation of $\int_0^\beta du \delta_\epsilon^2 S_b$ is given by \eqref{eq:variation-action-N=4} where $\delta_\epsilon S_f^p$ is given by \eqref{eq:relevant-susy-transformation22}. 
As before, \eqref{eq:variation-action-N=4} only vanishes and the fermionic mode is physical if $\epsilon = \text{ constant}$. Once again, since $\epsilon(\tau +\beta) = -e^{2\pi i \alpha \sigma_3} \epsilon(\tau)$, this zero mode only exists if $\alpha = 1/2$, i.e. when computing the gravitational index. As long as $\delta_\epsilon S_f\neq 0$, this zero mode is physical and its one-loop determinant and contribution to the partition function will vanish. In contrast to our analysis in section \ref{sec:diskN4}, for $\epsilon = \text{ constant}$,  $\delta_\epsilon S_f=- \frac{2\pi^2 \epsilon^p}{\b^2} \left[\left(\frac{b}{2\pi}\right)^2 + 4 \left(n+ \frac{1}2\pm \varphi\right) ^2 \right]$ is nonzero for all values of $b \neq 0$, $n \in \mZ$ and $\varphi$. Therefore, the mode with  $\epsilon = \text{ constant}$ is always a zero mode of the action and is not equivalent to a pure super-diffeomorphism. Consequently, we arrive at the main result of this section which is that the fermionic one-loop determinant on each connected boundary always vanishes due to this zero mode. Thus, all connected geometries vanish in the path integral.  
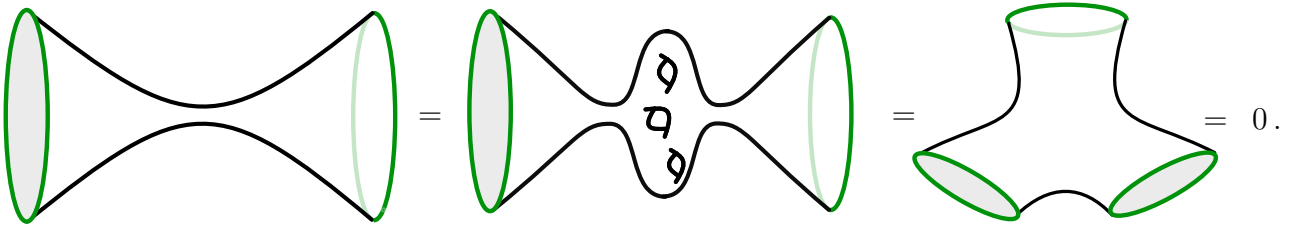
\begin{figure}[h!]
\begin{center}
\begin{tikzpicture}
    \node[anchor=south west,inner sep=0] at (0,0) {\includegraphics[width=0.92\textwidth]{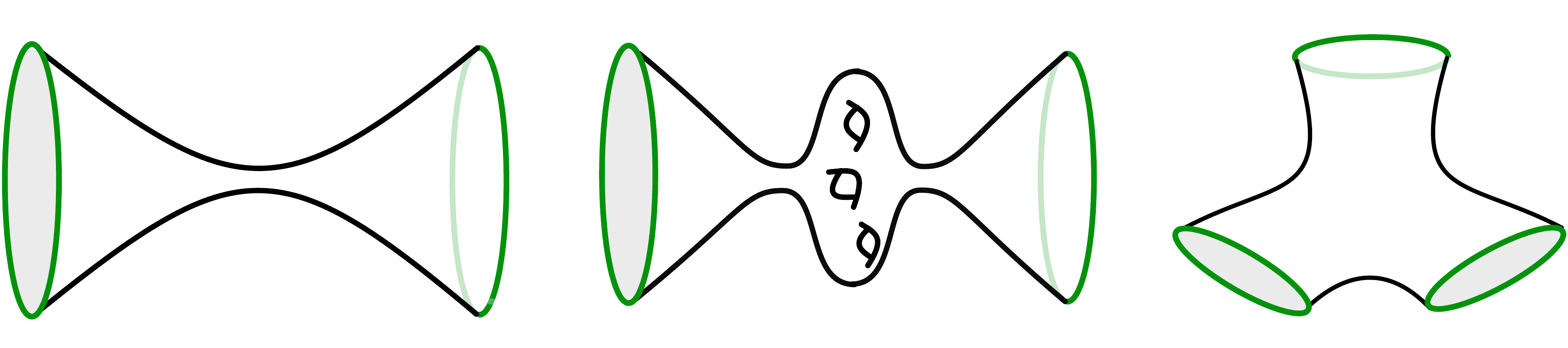}};
     \draw (5.70,+1.7) node  {$=$};
     \draw (12,+1.7) node  {$=$};
      \draw (16.5,+1.7) node  {$= \,\,\,0\,.$};
\end{tikzpicture}
\end{center}
\caption{Examples of  path integral contributions with multiple boundaries vanishing due to fermionic zero modes. }
\label{fig:example-vanishing-1}
\end{figure}

\begin{figure}[h!]
\begin{center}
\begin{tikzpicture}
    \node[anchor=south west,inner sep=0] at (0,0) {\includegraphics[width=0.7\textwidth]{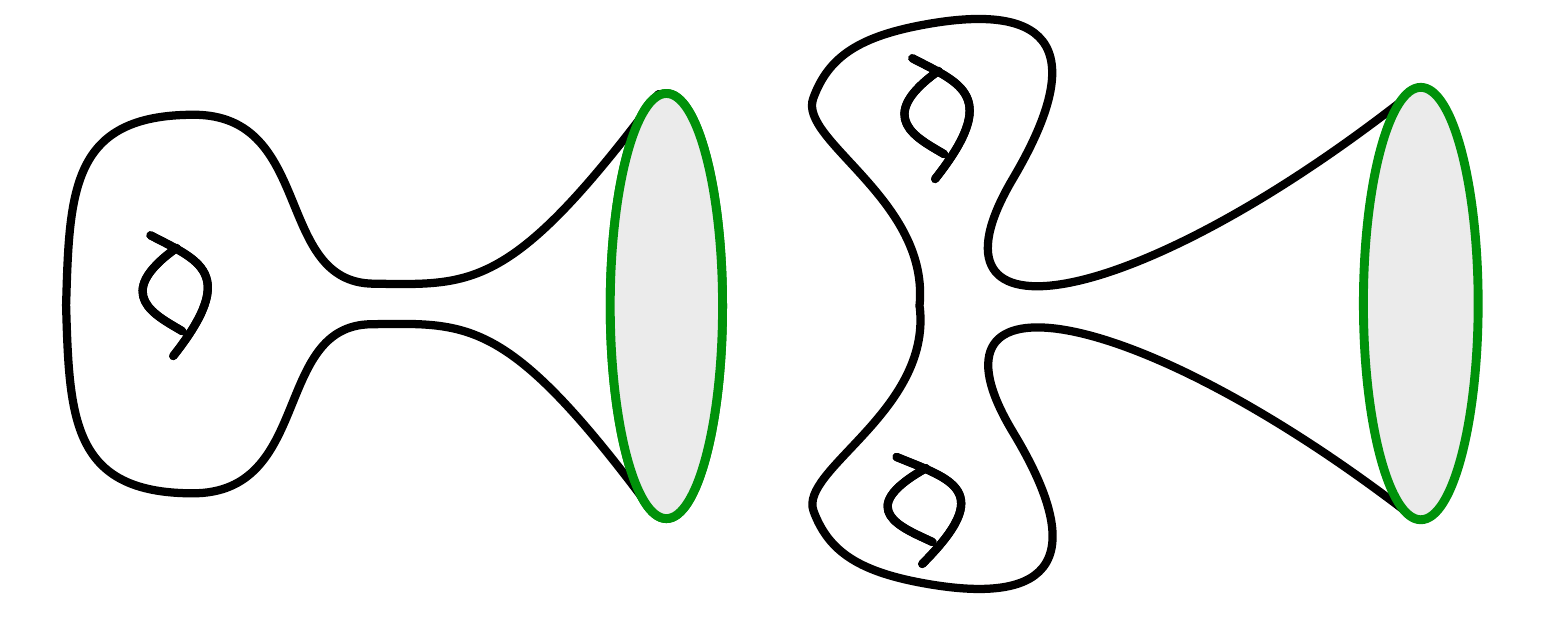}};
     \draw (6.40,+2.65) node  {$=$};
     \draw (12.5,+2.65) node  {$=\,\,\, 0\,.$};
\end{tikzpicture}
\end{center}
\caption{Examples of  path integral contributions with a higher-genus near-horizon region which once again vanish from the index due to fermionic zero modes. }
\label{fig:example-vanishing-2}
\end{figure}

Consequently, at least when only accounting for topology change in the near-horizon region, there is no factorization problem for the gravitational index. Following the notation in figure \ref{fig:introindexmultiple2} we have that
\be 
 \Index_\text{grav.}\left((\beta_1, Q_1), \,\dots,\,(\beta_k, Q_k)\right) = \Index_\text{grav.} (\beta_1, Q_1)\cdots \Index(\beta_k, Q_k)_{\text{grav.}}
\ee
for all $k$. Our result also indicates that the index factorizes from observables involving an arbitrary number of boundaries which do not necessarily have supersymmetry-preserving boundary conditions. As remarked in figure \ref{fig:introwomrholes3}, we have that 
\be 
Z_{\text{grav.}} &\bigg[(\beta_1, \Omega_1 = 2\pi i/\beta_1, Q_1), \,\dots,\, (\beta_k, \Omega_k = 2\pi i/\beta_k, Q_k),\,(\beta_{k+1}, \Omega_{k+1}, Q_{k+1}),\, \dots,\,(\beta_n, \Omega_n, Q_n)\bigg]  \nn \\ &=  \Index_{\text{grav.}}(\beta_1, Q_1) \cdots \Index_{\text{grav.}}(\beta_k, Q_k) Z_{\text{grav.}}\bigg[(\beta_{k+1}, \Omega_{k+1}, Q_{k+1}),\, \dots,\,(\beta_n, \Omega_n, Q_n)\bigg]\, .
\ee
Finally, we note that the presence of a zero mode above has implications that go beyond factorization. Our analysis above holds for any near-horizon region which contains a closed geodesic. Since all higher-genus hyperbolic surfaces contain a closed geodesic in the same homotopy class as the asymptotic boundary, the presence of the zero mode implies that all higher-genus surfaces vanish from the path integral of the gravitational index (see figure \ref{fig:example-vanishing-2}). 

Note that our analysis is applicable even though all surfaces that are under consideration (for instance, those in figure \ref{fig:example-vanishing-1} and \ref{fig:example-vanishing-2}) do not correspond to on-shell gravitational solutions. This is because while we have found the saddle points for $S_b$, $S_f$, and $S_T$, we have not explicitly found them for the dilaton or dilatino. In fact, on any surface that has a closed geodesic, the dilaton equation of motion cannot be satisfied \cite{Saad:2019lba} which implies that such surfaces are not on-shell gravitational solutions in $\cN=4$ super-JT gravity. 

As a consistency check, one can explicitly verify that these zero modes, whose boundary value we have just described, can be extended in the bulk for any connected geometry. For such connected geometries, one can find a spinor $\epsilon(r,\tau)$ such that  $\psi_\mu = \nabla_\mu \epsilon$ with $\epsilon(r, \tau)$ constant and where the gravitino is decaying as in \eqref{eq:bdy-cond-gravitino}  on each one of the boundaries.

\subsection{Check from pure 2D gravity} 
We will explicitly verify the claim that all contributions with wormholes vanish for the case of pure $\mathcal{N}=4$ JT supergravity. This theory is one-loop exact and easily solvable. 

We follow the logic of \cite{Saad:2019lba}. In the case of bosonic JT we first integrate over the dilaton and localize on hyperbolic metrics, which allows for a separation into trumpets and hyperbolic metrics with geodesic boundaries. 

In the case of $\mathcal{N}=4$ JT supergravity we can apply an analogous procedure, integrate out the $PSU(1,1|2)$ adjoint dilaton and localize over flat $PSU(1,1|2)$ connections. This, in particular, restricts the metric to be hyperbolic (and the $SU(2)$ connection to be flat), and we can still cut and glue by trumpets. The only difference now is that when gluing we need to integrate not only over the $SL(2)$ holonomy labeled by the geodesic length $b$ but also over the $SU(2)$ holonomy labeled by the parameter $\varphi$. The reason is that one can reduce the integral over $PSU(1,1|2)$ holonomies on the geodesic boundaries to an integral over diagonal matrices, eliminating the integral over fermionic directions (although the diagonalization can affect the bosonic integration measure). A simpler example is the gluing procedure for $\mathcal{N}=1$ JT supergravity \cite{Stanford:2019vob}.
 \begin{figure}[h!]
     \centering
     \includegraphics[scale=0.25]{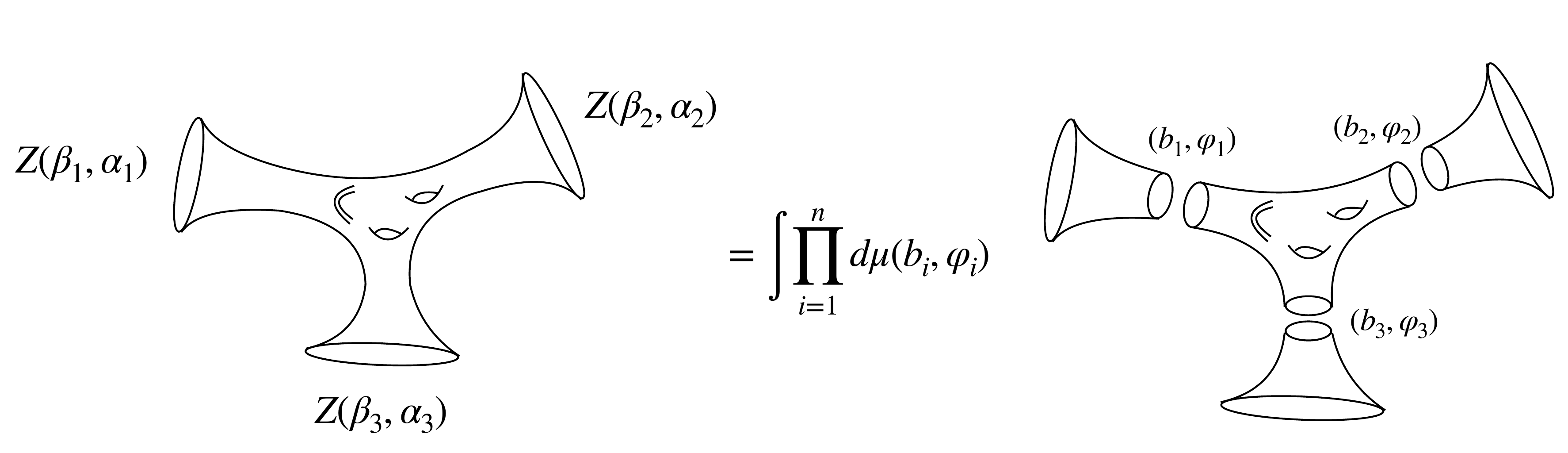}
     \caption{Example of a genus three contribution to the connected three-boundary partition function. The right-hand side illustrates the decomposition \eqref{eq:SSSdecompositionN4} into trumpets and surfaces with geodesic boundaries.}
     \label{fig:purejtwormhole}
 \end{figure}
 
We can write the final answer for the connected partition function of pure $\mathcal{N}=4$ JT supergravity with $n$ boundaries with holographic boundary conditions, renormalized length $\beta_i$ and $SU(2)$ chemical potentials $\alpha_i$ for $i=1,\ldots, n$. For simplicity we restrict to connected contributions with fixed $n$ and fixed genus $g$. The answer is 
\be\label{eq:SSSdecompositionN4}
Z_{ g,n}((\beta_1,\alpha_1) , \ldots, (\beta_n,\alpha_n))_{\rm conn.} &=  \int \left[\prod_{i=1}^n d\mu(b_i,\varphi_i) ~Z_{\rm Trumpet}(\beta_i, \alpha_i ; b_i, \varphi_i) \right] \nonumber\\
& \qquad \qquad \times V_{g,n}((b_1,\varphi_1), \ldots,(b_n,\varphi_n)).
\ee
This is graphically depicted in figure \ref{fig:purejtwormhole}.  The full answer for the gravitational path integral can be easily written as a sum over all disconnected and connected diagrams and also as a sum over genera \cite{Saad:2019lba,Stanford:2019vob}. We leave the gluing measure $d\mu(b,\varphi)\propto dbd\varphi$ arbitrary since its form will not be important. $Z_{\rm Trumpet}(\beta_i, \alpha_i ; b_i, \varphi_i)$ denotes the path integral on the trumpet geometry interpolating between a holographic boundary with boundary conditions $(\beta_i,\alpha_i)$ and a geodesic boundary with length $b_i$ and $SU(2)$ holonomy parametrized by $\varphi_i$ conjugate to $e^{2\pi i \varphi_i \sigma^3}$. This quantity is derived in appendix \ref{app:N4trumpetlocalization} and gives\footnote{We would like to thank E.~Witten for help with this calculation.} 
\beq
\label{eq:maintext:trumpetfull}
Z_\text{Trumpet}(\beta,\alpha) \propto \sum_{n\in \mathbb{Z}} \frac{\Phi_r}{\beta} \frac{ \cot(\pi \alpha)}{\pi \sin(2\pi \varphi)}  \left( e^{-\frac{2\pi^2\Phi_r}{\beta}((\frac{b}{2\pi})^2+4(n-\alpha+\varphi)^2)}-e^{-\frac{2\pi^2\Phi_r}{\beta}((\frac{b}{2\pi})^2+4(n+\alpha+\varphi)^2)}\right) \,.
\eeq
In the second line of \eqref{eq:SSSdecompositionN4} we have $V_{g,n}((b_1,\varphi_1), \ldots,(b_n,\varphi_n))$,  the volume of moduli space of $\mathcal{N}=4$ hyperbolic surfaces equivalent to flat $PSU(1,1|2)$ connections, modulo the supersymmetric mapping class group, with geodesic boundaries with fixed length and holonomies. This is a generalization of the Weil-Petersson volumes of Riemann surfaces, and as far as we know, this quantity has not been computed in the mathematical literature. Fortunately, even though we will assume they exist and are finite, we will not need their precise expression.
 
When we set at least one boundary to be supersymmetric, meaning it computes the gravitational prediction for the index, we need to set $\alpha=1/2$. Then there are two cases. The first is the configuration where this boundary is disconnected and filled by a disk. This can contribute to the index as explained in section \ref{sec:diskN4}. The second are configurations where one either adds spacetime wormholes to the disk or connects the supersymmetric boundary with any other boundary. All these latter cases involve gluing with trumpets, and since
\beq
Z_\text{Trumpet}(\beta,\alpha=1/2)=0,
\eeq
which we check in appendix \ref{app:N4trumpetlocalization}, their contribution vanishes. This cancellation is of course due to the zero modes we have found above.

Finally, in bosonic JT gravity there are doubly non-perturbative effects from D-branes analyzed in \cite{Saad:2019lba}. These effects are only understood from the matrix integral description of these theories. Since we do not know of an analogous duality for $\mathcal{N}=4$ super-JT, we will not analyze these corrections, and we cannot claim that they do not contribute.

\subsection{Higher-dimensional topology change}
\label{sec:factorization in higher dim}
We have analyzed spacetime wormhole configurations in higher dimensions that consist of near-extremal states where the wormholes appear in the asymptotically AdS$_2\times S^2$ sector of the geometry. We can extend this discussion to near-extremal black holes with AdS$_2$ throats and less supersymmetry. Another generalization is to consider more general higher-dimensional configurations which are not described by an effective two-dimensional theory (e.g. by being very far from extremality). 

Although the generalization is conceptually obvious, details depend on the particular case but we can outline the general procedure that emerges from the previous section. Firstly, we argued quite generally in section \ref{sec:saddle-point-analysis} that computing the index with supergravity involves thinking of boundary conditions with fixed (imaginary) chemical potential for a charge $\mathcal{Q}$ that satisfies that $(-1)^F=e^{2\pi i \mathcal{Q}}$. These boundary conditions are supersymmetric and we need to find solutions satisfying them at finite temperature. Importantly, as stressed in section \ref{sec:saddle-point-analysis}, the geometries contributing at finite temperature might look very different to the black holes we expect the index to count (for example, in the case of 4D they involve rotation). 

Then, we need to analyze the one-loop determinants. In the bulk, all these theories have a gravitino $\Psi_\mu$ with a local supersymmetry transformation of the form $\delta_\epsilon \Psi_\mu = D_\mu \epsilon + \ldots$, where $D_\mu$ a first-order differential operator determined by the super-algebra and $\epsilon(x)$ is the local transformation parameter (an example is given by 4D $\mathcal{N}=2$ ungauged supergravity in section \ref{sec:saddle-point-analysis}). When computing the one-loop determinants we can decompose $\Psi_\mu = D_\mu\epsilon + \Psi_\mu^{\perp}$, where $\Psi_\mu^{\perp}$ is orthogonal to any profile written as a local supersymmetry transformation. Then, we can independently integrate over $\epsilon$ and $\Psi_\mu^\perp$ and we will focus on the former. In the particular examples that we studied, we checked that any configuration $\Psi_\mu \sim D_\mu\epsilon$ is a fermionic zero mode (i.e. at least to quadratic order has zero action) as long as $\epsilon$ goes to a constant at infinity independent of boundary coordinates. This is obviously not trivial even if the bulk action is invariant, due to boundary terms that have to be treated carefully. These zero modes make the gravitino one-loop determinant vanish in general. Still, we have seen that when the metric we are expanding around has a Killing spinor, namely there is a solution to the equation $D_\mu\epsilon = 0$, then this fermion zero mode is not physical (i.e. produces a gravitino profile that decays too fast\footnote{By `too fast' we mean a gravitino profile that does not affect the boundary supercharges, which therefore is a gauge mode, as described in section \ref{sec:diskN4}.} 
at infinity) and should not be integrated over, allowing the answer to be nonzero. 

This leads to the very reasonable rule that only supersymmetric geometries contribute to the gravitational path integral with supersymmetric boundary conditions. This already rules out wormholes, since they are described by trumpets near the boundaries. Therefore, our arguments show that spacetime wormholes do not contribute to the value of the index (probed by single-boundary path integrals) nor to its statistics (probed by multi-boundary path integrals). This might give the idea that supersymmetric states decouple from generic states, although we will see an example in the next section that this cannot be so simple.

We would like to comment on a subtlety in higher dimensions. In general, the contributions from spacetime wormholes are not saddle points and therefore the whole idea of performing a one-loop calculation is not well-defined. We can go around this issue in two dimensions, where the dilaton is simply a Lagrange multiplier, but the situation in higher dimensions is more complicated. In a more careful treatement, the procedure outlined above can be combined with the constrained instanton approach of \cite{Cotler:2020lxj,Cotler:2021cqa}. This procedure consists of rewriting the exact path integral as a sum over constrained instantons with a well-defined metric (which are not precisely solutions of the equations of motion). Then, our discussion applies for the calculation of one-loop determinants around these constrained solutions. With this caveat in mind, we expect our conclusion that spacetime wormholes do not affect supersymmetric quantities to hold true.  

We finish this section with a final concrete example. Consider the calculation of the partition function of three-dimensional pure gravity with a negative cosmological constant and torus boundary conditions, and focus on the $T^2 \times ({\rm interval})$ topology. This has been computed in \cite{Cotler:2020ugk} and it is an example of a path integral without saddle points. The extension of the results of \cite{Cotler:2020ugk} to $(4,4)$ supergravity are straightforward, although we leave the details for future work. Just like in JT gravity, we can separate the path integral into two three-dimensional versions of the trumpet geometries. The exact partition function around these geometries can be reduced to a supersymmetric version of the coadjoint orbit action, whose partition function is expected to be a generic super-Virasoro character with a dimension related to the size of the wormhole `bottleneck'. These characters correspond to long representations with zero index \cite{Eguchi:1988af}, making the final answer vanish when imposing supersymmetric boundary conditions.

\section{Other contributions: Supersymmetric defects}\label{sec:defects}

In the previous sections we have argued that one can compute the black hole index using the Euclidean path integral. We also showed that spacetime wormholes do not contribute either to the index or to its statistic. This raises the question: which geometries do contribute to the index? In particular, we know that the disk (Euclidean black hole) cannot be the whole answer. For example, in the case of pure JT supergravity the calculation gives ${\rm Index} \sim e^{S_0}$ which, if we impose that the black hole charge is some integer $Q$, does not necessarily imply that the degeneracy of the index is an integer.

The purpose of this section is to study which other asymptotically AdS$_2$ geometries might affect the index. Just like in the analysis in section \ref{sec:diskN4}, the geometries we will study can be embedded as corrections near the horizon of higher-dimensional black holes with AdS$_2$ throats. We will not attempt to do this here, and simply focus on two-dimensional configurations that can contribute to the index. We leave their embedding in specific higher-dimensional theories for future work.

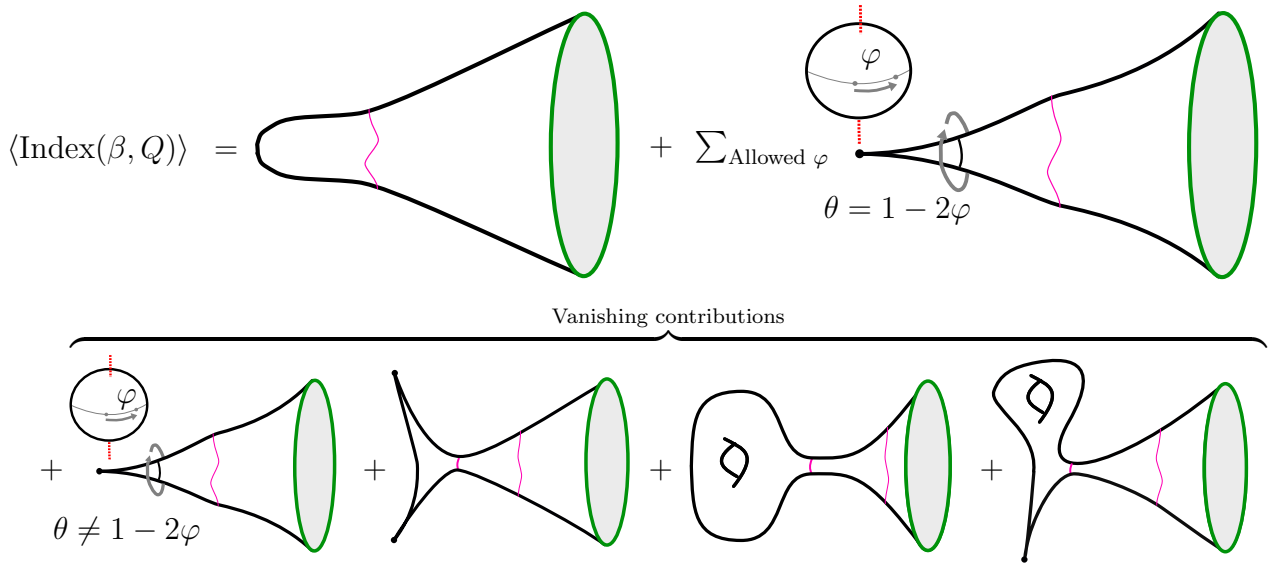
\begin{figure}[t!]
\begin{center}
\begin{tikzpicture}
    \node[anchor=south west,inner sep=0] at (0,0) {\includegraphics[width=0.8\textwidth]{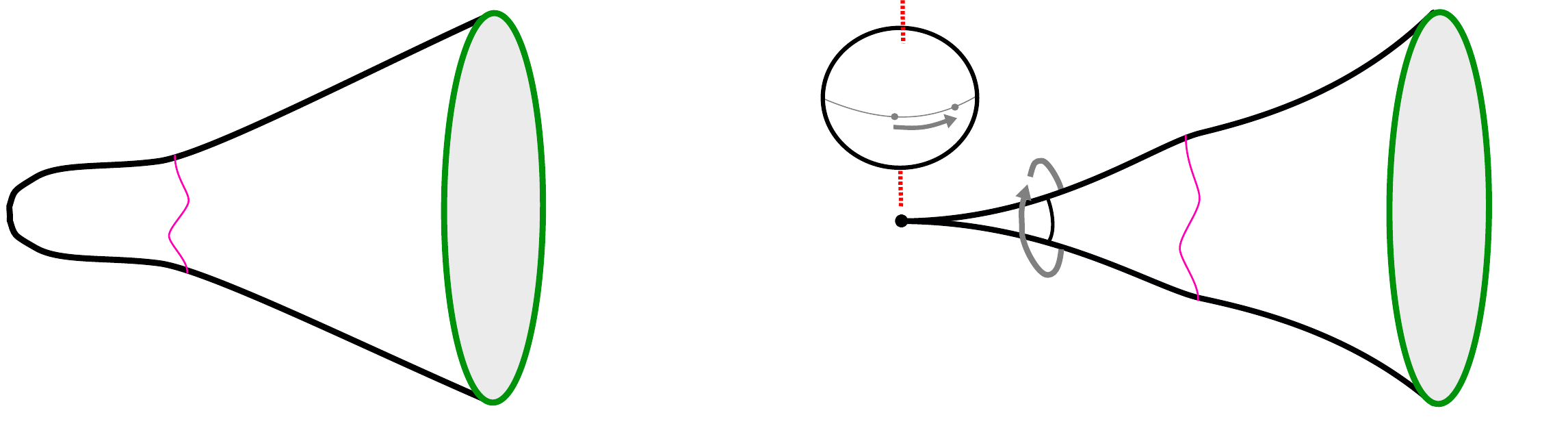}};
     \draw (-1.7, +1.9) node {$\langle \Index(\beta, Q)\rangle \,\,\, = $};
     \draw (6.5,+1.9) node  {$+ \,\,\,\sum_{\text{Allowed } \varphi}$};
     \draw (8.25,+3.1) node  {$\varphi$};
     \draw (8.6,+1.1) node  {$\theta=1-2\varphi$};
\end{tikzpicture}
\vspace{0.2cm}
\begin{tikzpicture}
    \node[anchor=south west,inner sep=0] at (0,0) {\includegraphics[width=0.9\textwidth]{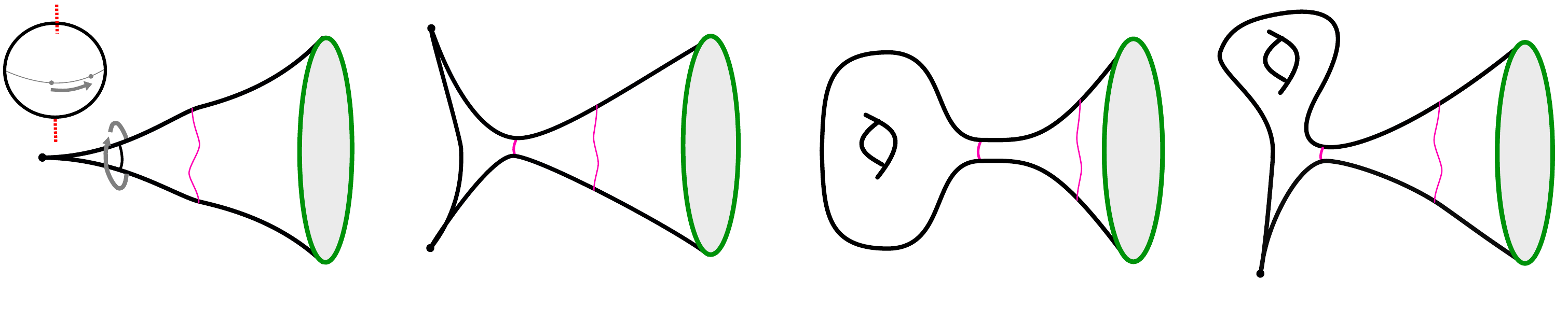}};
     \draw (-0.2, +1.7) node {$+$};
       \draw (0.8, +2.65) node {$\varphi$};
      \draw (0.8, +0.9) node {$\theta\neq 1-2\varphi$};
     \draw (4.1,+1.7) node  {$+$};
     \draw (7.9,+1.7) node  {$+$};
     \draw (12.3,+1.7) node  {$+$};
      \draw (8.0,+3.6) node  {$\overbrace{\hspace{0.9\textwidth}\,}^{\text{Vanishing contributions}}$};
\end{tikzpicture}
\end{center}
   \caption{The possible kinds of geometries appearing in a single-boundary path integral computing the index. In the first line, we show the only geometries that contribute: the disk and the supersymmetric defects (with opening angle $2\pi \theta = 2\pi(1-2\varphi)$ and $SU(2)$ holonomy $\varphi$). In the second line, we summarize contributions that vanish: wormholes, non-supersymmetric defects, and combinations thereof.}
     \label{fig:indexdefect}
\end{figure}

From the perspective of two dimensions, there are three objects we can add to the Euclidean disk to see whether they contribute to the index. The first is to add spacetime wormholes, which we have shown does not affect the index. The second thing we can do is to add end-of-the-world branes as geodesic holes in the disk. These configurations are never supersymmetric, which we explain through the zero-mode analysis of section \ref{sec:diskN4} and through the exact answer from appendix \ref{app:N4trumpetlocalization}, and therefore also do not affect the index. Finally, the third possibility that we will explore here is to add defects. A defect in $\mathcal{N}=4$ JT is specified by two numbers: the deficit angle in the geometry and the $SU(2)$ holonomy around the defect. When a special relation is satisfied between these two parameters the defect preserves enough supersymmetry to cancel fermion zero modes which otherwise would make their contribution to the index vanish. Our results are neatly summarized in figure \ref{fig:indexdefect}.

\begin{figure}[t!]
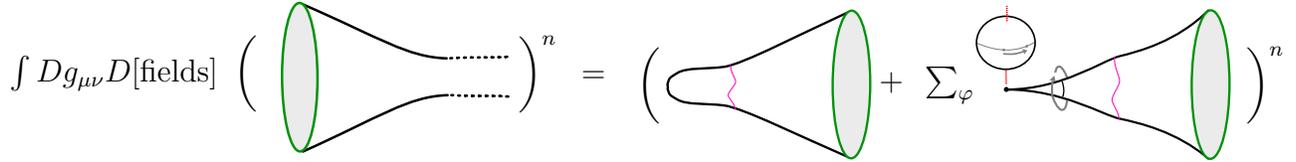

\begin{center}
\begin{tikzpicture}
    \node[anchor=south west,inner sep=0] at (0,0) {\includegraphics[width=0.2\textwidth]{intro1.pdf}};
     \draw (-1.7, +1.25) node {$\int Dg_{\mu \nu} D\text{[fields]} \,\,\, \bigg( $};
     \draw (4.0,+1.25) node  {$\bigg)^n\,\,\, = $};
\end{tikzpicture}
\begin{tikzpicture}
    \node[anchor=south west,inner sep=0] at (0,0) {\includegraphics[width=0.45\textwidth]{defect1.pdf}};
     \draw (3.5,+1.1) node  {$+ \,\,\,\sum_{ \varphi}$};
      \draw (-0.2,+1.1) node  {$\bigg($};
         \draw (8.0,+1.1) node  {$\bigg)^n$};
\end{tikzpicture}
\end{center}
   \caption{Cartoon illustrating the factorization of the index even when allowing defects. }
     \label{fig:factorization-w-defects}
\end{figure}

Even if the defects are supersymmetric and contribute to the index, we would like to emphasize they still do not contribute to connected contributions with multiple boundaries. This is summarized in figure \ref{fig:factorization-w-defects}. This is evidence that the zero-mode analysis of section \ref{sec:factorization} is robust under deformations of JT gravity.

\subsection{The defect geometry}
We start with the metric with a single defect at $r=0$. This is given by
\beq
ds^2 = dr^2 + \sinh^2 r d \tau^2,~~~~~\tau \sim \tau + 2 \pi  \theta\,, 
\eeq
where $\theta$ is the parameter that labels the angular deficit.\footnote{This is called $\alpha$ in \cite{Maxfield:2020ale} but here we use $\theta_\text{here}=\alpha_\text{there}$ to avoid confusion with the $SU(2)$ chemical potential.} We will also allow for the monodromy of the $SU(2)$ gauge field to be non-trivial. We can use the $SU(2)$ symmetry to fix $e^{\oint_\text{def} B} = e^{2\pi i \varphi \sigma^3} $, integrated around the defect, just like we did in the previous section for the holonomy around trumpets. Finally, to determine the insertion of defects in the path integral we need to specify the weight with which each defect contributes \cite{Maxfield:2020ale,Witten:2020wvy}. From the 2D perspective, this is a free parameter and we will denote it by $w$. The contribution from geometries with $k$ defects is then weighted by $w^k$. Defects are therefore parametrized by a geometrical parameter $\theta$, an $SU(2)$ parameter $\varphi$, and the weight $w$.

One might argue that it is not correct to include singular geometries in the gravitational path integral. From a 2D perspective, this has to be interpreted as defining a deformation of JT gravity, with the addition of a gas of defects \cite{Maxfield:2020ale,Witten:2020wvy}. Nevertheless, there are smooth higher-dimensional non-perturbative corrections, which dimensionally reduce in 2D to defects, for example see \cite{Maxfield:2020ale} and also \cite{Dabholkar:2014ema}. In these cases, the higher-dimensional picture determines the value of the angles and weights.

\subsection{Correction to the index}

Using the methods presented in \ref{app:N4trumpetlocalization}, it is straightforward to compute the exact partition function of $\mathcal{N}=4$ JT gravity in the presence of a single defect parametrized by $(\theta,\varphi)$. For arbitrary parameters, the answer is given by 
\beq
\label{eq:generic-defect-part-function}
Z_{(\theta,\varphi)}(\beta,\alpha) = w \sum_{n\in \mathbb{Z}} \frac{\Phi_r}{\beta} \frac{ \cot(\pi \alpha)}{2\pi }  \left( e^{\frac{2\pi^2\Phi_r}{\beta}(\theta^2-4(n-\alpha+\varphi)^2)}-e^{\frac{2\pi^2\Phi_r}{\beta}(\theta^2-4(n+\alpha+\varphi)^2)}\right) \,.
\eeq
The origin of each term in this expression are hopefully clear. The sum is over saddles, while the two terms in parenthesis are saddles related by the Weyl group of $SU(2)$. The prefactors give the one-loop determinants around these saddles. The contribution to the index is $Z_{(\theta,\varphi)}(\beta,\alpha=1/2)=0$. This vanishing can be understood as coming from physical fermion zero modes in the exact same way as in section \ref{sec:diskN4}, and we will not repeat the analysis here.

Defects become supersymmetric when the deficit angle is related to the $SU(2)$ holonomy around them in a specific way. These geometries preserve four Killing spinors, which are enough to absorb the physical zero modes studied in section \ref{sec:diskN4}, for example. More concretely, this phenomenon happens when the defect angle is $\theta_\text{SUSY} = 1 -2 \varphi \neq 0$. We can always pick the $SU(2)$ holonomy parameter such that $0<\varphi<1/2$, so with these conventions $0<\theta_\text{SUSY}<1$. In appendix \ref{app:N4trumpetlocalization} we obtain the one-loop partition function as
\beq
\label{eq:super-defect-part-function}
Z_{(1-2\varphi, \varphi)}(\beta,\alpha)= w \sum_{n\in\mathbb{Z}} \frac{\beta}{\Phi_r} \frac{\cot(\pi \alpha)}{4\pi^3 } \left( \frac{e^{\frac{2\pi^2\Phi_r}{\beta}((1-2\varphi)^2-4(n-\alpha+\varphi)^2)}}{(1-2\alpha+2n)^2} -\frac{e^{\frac{2\pi^2\Phi_r}{\beta}((1-2\varphi)^2-4(n+\alpha+\varphi)^2)}}{(1+2\alpha+2n)^2}\right) \, .
\eeq
The sum involves the same saddles as the generic defects, but due to the presence of gauged fermion zero modes the calculation of the gravitino one-loop determinant is different, and the new terms in the denominator appear. These new terms are responsible for the contribution to the index not vanishing:
\beq
Z_{(1-2\varphi, \varphi)}(\beta,\alpha=1/2)=w~(1-2\varphi).
\eeq
Again, this can be argued using the same methods as in section \ref{sec:diskN4}, as physical zero modes being absorbed by gauge zero modes, even without full knowledge of the details of the partition function. It takes more effort to justify how the one-loop determinant around the saddle computing the index depends on $\varphi$ nevertheless as presented in appendix \ref{app:N4defectlocalization} this can be done by fermionic localization. This can be thought of as a $\mathcal{N}=4$ version of the Ramond punctures in \cite{Stanford:2019vob}.

The leading contribution to the index from the disk is given by ${\rm Index}(\beta) = e^{S_0}$. Including the corrections from supersymmetric defects, and allowing for the presence of a set of $N_F$ different flavors labeled by $\varphi_i$ and $w_i$, with $i=1,\ldots, N_F$, the total answer for the index is  
\beq
{\rm Index}(\beta) = e^{S_0} + \sum_{i=1}^{N_F} w_i~ (1-2\varphi_i)\, .
\eeq
In examples such as \cite{Dabholkar:2014ema}, there are an infinite number of defects that have $1-2\varphi_i = 1/i$ where $i\in \mathbb{Z}$ and $i\geq 2$. Moreover $w(\varphi_i) \sim e^{S_0 \phi_i} \sim e^{S_0/i}$ is exponentially suppressed in $S_0$ compared to the disk, and therefore they are non-perturbative.  

Finally, it is easy to see using the same methods as in section \ref{sec:factorization} that geometries with more than one defect do not contribute to the index. This happens because even if locally the defect is supersymmetric, geometries with more than one defect do not preserve any supersymmetry. In all these cases there are physical gravitino zero modes and therefore for the purpose of computing the index we can restrict to a single defect. We summarize this in figure \ref{fig:indexdefect}. For the same reason, the presence of defects (supersymmetric or not) does not spoil the factorization of the index.\footnote{We can justify this for sharp defects with $0<\theta<1/2$. It would be nice to check whether the same is true for defects with $1/2<\theta<1$ studied in \cite{Turiaci:2020fjj}.}

\subsection{Correction to the mass gap}
There is an interesting feature in the black hole spectrum of non-supersymmetric states, implied by the inclusion of the defects considered above.

Analyzing the non-supersymmetric sector of the spectrum is a difficult problem since, as opposed to the BPS states, this sector is affected by wormholes, non-supersymmetric defects, as well as multiple defects. In particular, it was argued in \cite{Maxfield:2020ale, Witten:2020wvy} that the sum over defects results in a shift of the extremal energy for the case of bosonic dilaton-gravity. This cannot be correct in supergravity since the BPS bound is protected by supersymmetry. Instead, we will argue here that the presence of defects in supergravity results in a correction to the black hole mass gap between the BPS states and the first non-extremal black hole, in a sector of fixed charge.

For simplicity we will focus on the case of $\mathcal{N}=4$ JT supergravity with a single supersymmetric defect. We can write a density of states for the continuum part of \eqref{eq:super-defect-part-function} that corrects the disk density of states. The answer is given by a rewriting of the partition function as
\beq
Z_{(1-2\varphi, \varphi)}(\beta,\alpha) =w~(1-2\varphi) +  \sum_{J} \int_{E_{\rm gap}(J^2)}^\infty dE e^{-\beta E } \rho_{(1-2\varphi, \varphi)}(E,J) \left( \chi_J(\alpha)+2 \chi_{J-\frac{1}{2}}(\alpha)+\chi_{J-1}(\alpha)\right).
\eeq
The first term is the contribution from $E=0$ BPS states, which are only made of $J=0$ states since it is independent of the $SU(2)$ chemical potential $\alpha$. This term captures precisely the corrections to the index mentioned in the previous section. The continuum part is expanded in supermultiplets $J\oplus2(J-1/2)\oplus (J-1)$ which start at an energy gap above extremality $E_{\rm gap}(J)=J^2/(2\Phi_r)$, similar to what happens in the disk without defects considered in \cite{Heydeman:2020hhw}. Decomposing \eqref{eq:super-defect-part-function} gives
\be
\rho_{(1-2\varphi, \varphi)}(E,J) &= w \cos(4J \pi \varphi) \frac{J \sinh\left(2\pi(1-2\varphi)\sqrt{2\Phi_rE-J^2}\right)}{2\pi^2 \Phi_r E^2} \nonumber\\
& \qquad + w \sin (4J \pi \varphi) \frac{(2\Phi_r E-2J^2)\cosh \left( 2\pi(1-2\varphi) \sqrt{2\Phi_rE-J^2}\right)}{4\pi^2\Phi_r E^2 \sqrt{2\Phi_rE-J^2}}\, .
\ee
A similar formula for non-supersymmetric states, without a contribution to BPS states, can be derived from equation \eqref{eq:N4trumpetlocalizationrho}. 
 \begin{figure}[t!]
     \centering
     \includegraphics[scale=0.4]{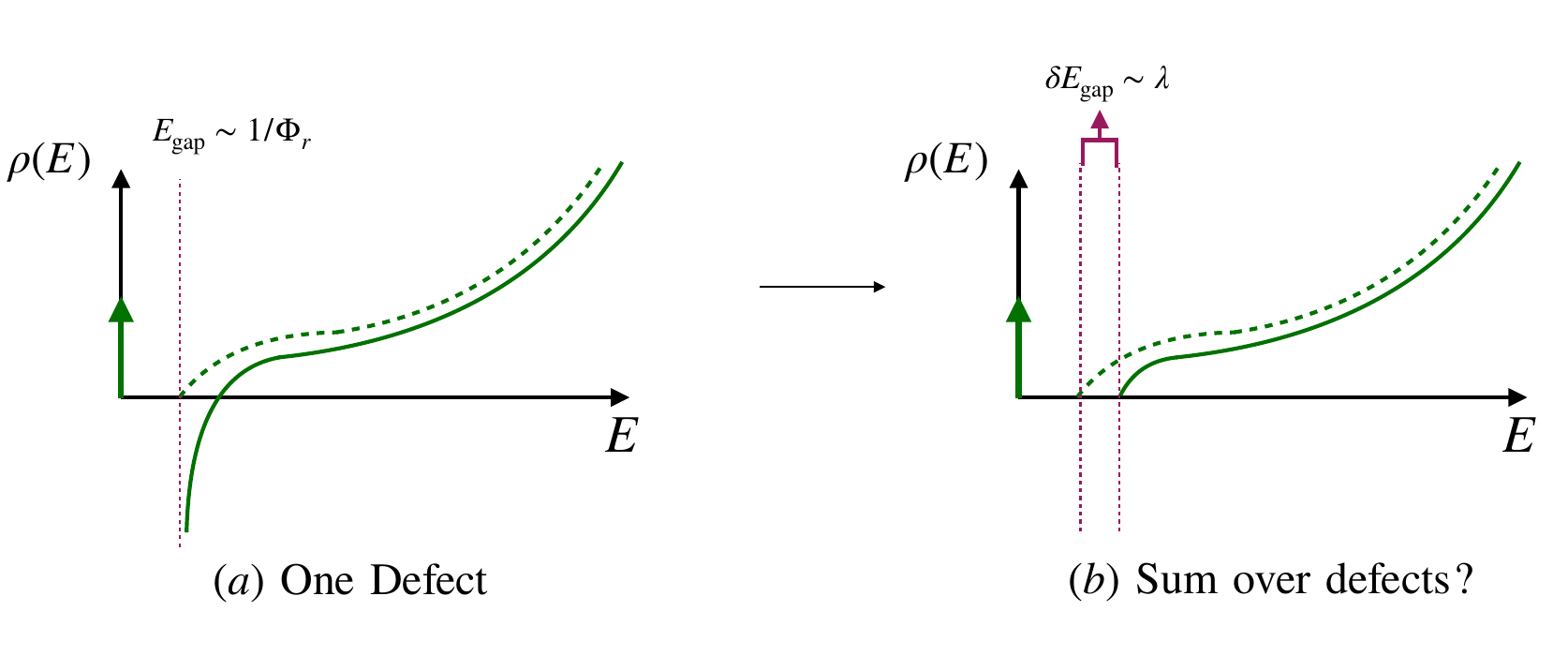}
     \caption{(a) Spectrum of black holes corrected by one defect (solid line) compared to case without defect (dashed line). The total answer diverges at $E_{\rm gap}$. (b) We expect a sum over defects to resum into a smooth curve with square-root edge and a corrected value $E_{\rm gap} + \delta E_{\rm gap}$ as explained in the text.}
     \label{fig:gapcorrection}
 \end{figure}
 
In the case of bosonic JT, the presence of a single defect causes a divergence at extremality as $\delta \rho\sim w /\sqrt{E}$ \cite{Maxfield:2020ale,Witten:2020wvy}. Instead, for $\mathcal{N}=4$ supergravity with a single defect, the divergence appears in a fixed $J$ sector as we approach the mass gap $E\to E_{\rm gap}(J)\neq 0$ since the non-supersymmetric sector starts at $J=1/2$. To see this, expand 
\beq
\rho_{(1-2\varphi, \varphi)}(J,E_{\rm gap}(J)+\varepsilon) \sim   w  \frac{\sin (4J \pi \varphi)}{\sqrt{\varepsilon}} ,\qquad \text{with } \varepsilon\ll 1,
\eeq
while $\rho_{\rm disk} \sim e^{S_0}\sqrt{\varepsilon}\to0$ \cite{Heydeman:2020hhw}. This behavior is problematic especially when the prefactor is negative since this can make the whole genus zero density of states negative for $\varepsilon$ small enough. This situation is analogous to \cite{Maxfield:2020ale,Witten:2020wvy} and we propose a similar solution. We conjecture that  summing over a gas of defects should produce a non-perturbative modification of the position of the edge
\beq
\delta E_{\rm gap}(J) \propto  e^{-S_0}w \sin (4J \pi \varphi).
\eeq
 The divergence from the one-defect contribution then only appears from expanding to leading order $\rho \sim e^{S_0}\sqrt{\varepsilon + \delta E_{\rm gap}} \sim  e^{S_0} \sqrt{\varepsilon} +(e^{S_0} \delta E_{\rm gap} ) \varepsilon^{-1/2} + \cdots$, with the dots corresponding to higher powers of $\varepsilon$ that correspond to multiple defects. In figure \ref{fig:gapcorrection} we show these effects for the case of the spectrum of zero-angular-momentum states. In this case, the gap is controlled by the $J=1/2$ supermultiplet.

We can apply the same analysis to the addition of a gas of multiple supersymmetric defects, and also a gas of non-supersymmetric defects. The answer is the same, a divergence appears near the edge of the spectrum that can be resolved by a non-perturbative shift of the black hole mass gap. As it stands, this is a conjecture, since we do not know how to calculate the contribution from multiple defects.

\subsection{Why the Hawking-Horowitz-Ross solution does not contribute}

\begin{figure}[h!]
\centering
\includegraphics[width=0.9\textwidth]{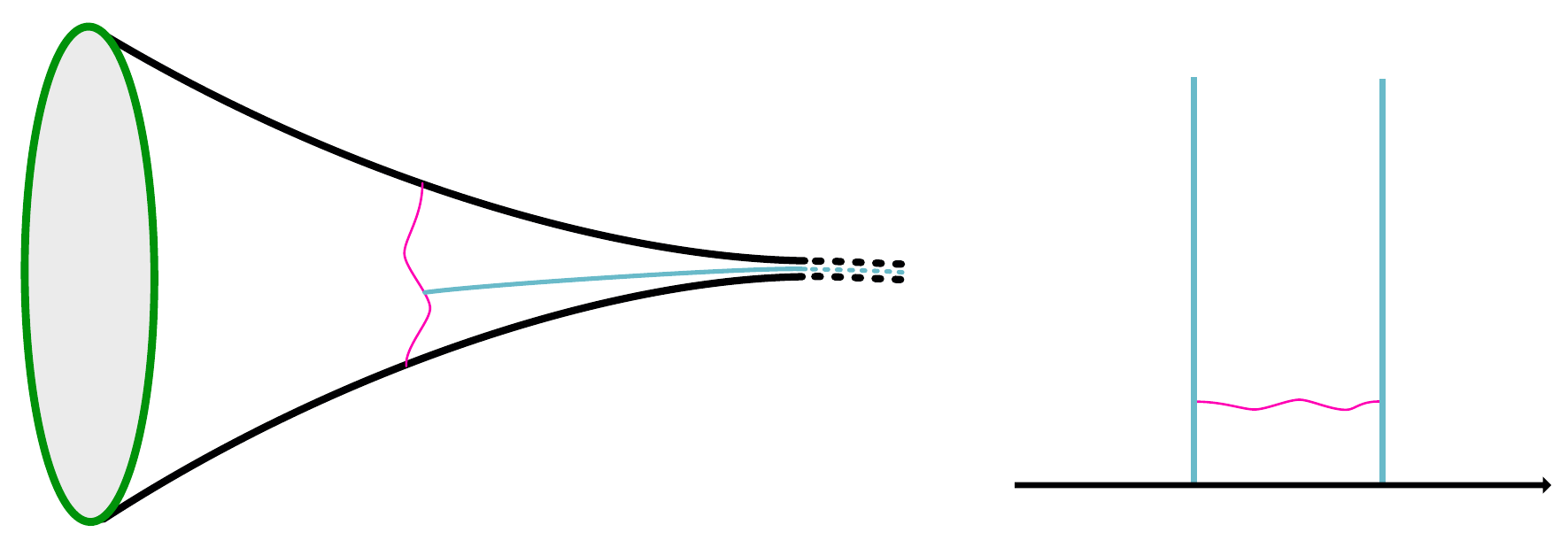} 
\caption{A cartoon of the Hawking-Horowitz-Ross solution. The AdS$_2$ region is located to the right of the purple curve in the left figure. In the right figure, we show the embedding of this region in the Poincar\'e upper half-plane.  Since we identify the thermal circle to be periodic, $\tau \sim \tau +\beta$, the AdS$_2$ region of the left geometry is a $\mZ$ quoteint of the Poincar\'e upper half-plane where the two blue geodesic are identified. This is equivalent to a trumpet with $b=0$ or to a defect with $1-2\varphi = 0$.   }
\label{fig:Hawking-Horowitz-Ross}
\end{figure}

Having understood what kind of surfaces contribute to the gravitational index, we can now revisit the problem of whether the Hawking-Horowitz-Ross solution \cite{Hawking:1995ap}, briefly mentioned in section \ref{sec:saddle-point-analysis}, also contributes to the index. While this solution has very small entropy, it could in principle yield a non-perturbative correction to the index which can compete with the other surfaces shown in figure \ref{fig:indexdefect}. Since we have classified all the surfaces that contribute to the index we simply have to understand where this solution fits in our classification of surfaces. In the throat of the extremal black hole, the metric of the Hawking-Horowitz-Ross solution can be expressed as 
\be 
\label{eq:throat-extrmal-BH}
ds^2 = \frac{(r-r_0)^2}{r_0^2}{d\tau^2} + \frac{r_0^2}{(r-r_0)^2}{dr^2} + r^2 d\Omega_2^2 \, ,
\ee
where $d\Omega_2^2$ yields the metric on the internal $S^2$. Taking $z = r_0/(r-r_0)$ we recover the standard form of the Poincar\'e metric. The horizon is located at $r = r_0$ and $z \to \infty$. Thus this solution does not have a horizon, and we are free to choose whatever periodicity $\tau \sim \tau+\beta$ we desire, and, as usual, this will fix the inverse temperature at the asymptotic boundary. The identification  $\tau \sim \tau+\beta$ is a $\mathbb Z$ quotient of the Poincar\'e upper half-plane, which thus yields either a trumpet or a defect in the hyperbolic plane depending on the quotient. In our case, as shown in figure \ref{fig:Hawking-Horowitz-Ross} this identifies two vertical geodesics in the Poincar\'e upper half-plane. The geodesic distance between any two vertical geodesics is $b=0$ and therefore we conclude that the AdS$_2$ region of the manifold is a special case of the trumpet with $b=0$ (this also happens to be equivalent to the particular defect with $\theta = 0$). Since this geometry has no horizon, we are also free to choose the $SU(2)$ holonomy around the non-contractible cycle on this surface to be $e^{2\pi i \varphi \sigma_3}$, with $\varphi \in [0, 1/2]$. This is equivalent to fixing the boundary conditions for the $SU(2)$ field or equivalently, for the $S^2$ component of the metric, as $z \to \infty$. 

For $\varphi\neq 1/2$ the analysis in section \ref{sec:factorization} shows that such surfaces always have physical fermionic zero modes which means that the one-loop determinant of the fermionic fluctuations around these saddles always vanishes. For $\varphi = 1/2$  one finds that the physical zero-mode could become a gauge zero mode; however, the exact answer from \eqref{eq:maintext:trumpetfull} for $\varphi = 1/2$ still explicitly shows that the one-loop determinant is vanishing. 

In summary, we have thus shown that the Hawking-Horowitz-Ross saddle does not contribute to the index, even as a non-perturbative correction, since it can be identified with a limit of the trumpet solution discussed in section \ref{sec:factorization}.

\section{Discussion} \label{sec:discussion}
We have specified a gravitational prescription for computing supersymmetric protected quantities such as the index of black holes states. We focused on the case of four-dimensional black holes in $\mathcal{N}=2$ supergravity. At the perturbative level, the answer is given by a saddle-point geometry that is smooth, but including one-loop effects is important to get sensible answers. Moreover, we argued that geometries with spacetime wormholes do not contribute. Our computation thus shows that to distinguish between a gravity theory with or without disorder, one cannot restrict to protected quantities. We finish with some open questions that we leave for future work.

\begin{itemize}
    \item Our arguments for the vanishing of contributions with spacetime wormholes rely on the presence of an asymptotically AdS$_2$ throat. Still, we expect the presence of physical fermionic zero modes to be quite robust and apply in more generic situations. Nevertheless, it would be great to show this explicitly.
    
    \item If a theory of supergravity is dual to an average over an ensemble of boundary duals, each member has to have the same index. Nevertheless, in AdS/CFT, we know of examples of quantum theories where the index can jump over some walls in moduli space. It would be interesting to understand whether averaging over theories with different indices can correspond to wormhole contributions that might not fall under the type assumed in this paper. Alternatively, one can show whether averaging over a moduli space that contains jumps of the index is always disallowed. 
    
    \item It would be interesting to extend our analysis to black holes with an approximate $SU(1,1|1)$ symmetry in the throat. These are the canonical examples of supersymmetric black holes in AdS$_4$ and AdS$_5$ \cite{Gutowski:2004ez, Cacciatori:2009iz}. In this case, based on $\mathcal{N}=2$ JT supergravity theory, not all BPS states are bosonic, and we expect the index to vanish to leading order in some cases, see appendix A of  \cite{Heydeman:2020hhw}. Making this connection precise is work in progress \cite{wip}.   
    
    \item Understanding how to add non-perturbative corrections to obtain an integer index is an interesting challenge. Some progress, using localization in supergravity, was done in \cite{Dabholkar:2014ema} and references therein. Nevertheless, these work focus primarily on contribution from matter fluctuations in different geometries. We hope our methods can help include fluctuations of gravity in these calculations. In order to do this analysis, going slightly away from extremality is crucial to treat bosonic zero modes appropriately \cite{Iliesiu:2020qvm}. A similar question in cases with a smaller amount of supersymmetry appears in \cite{Aharony:2021zkr}.
    
    \item Finally, it would be interesting to extend our observations regarding factorization of the index into other protected quantities such as Wilson lines or certain boundary correlators. This is especially interesting since such quantities could have a more explicit dependence on the coupling constants of the boundary theory and could shed an even stronger light on the allowed space of couplings for a putative ensemble average.
\end{itemize}

\subsection*{Acknowledgements} 
We thank J.~Cotler, A.~Dabholkar, T.~Hartman, G.~Horowitz, K.~Jensen, J.~Maldacena, S.~Murthy, S.~Pufu, S.~Shenker, D.~Stanford, B.~Rayhaun, L.~Tizzano and E.~Witten for valuable discussions. LVI and GJT would like to thank M. Heydeman and W. Zhao for collaboration on related matters. GJT is supported by a Fundamental Physics Fellowship. LVI is supported by the Simons Collaboration on Ultra-Quantum Matter, a Simons Foundation Grant with No. 651440. The work of MK is supported by funding from the European Research Council (ERC) under the European Union’s Horizon 2020 research and innovation programme (grant agreement No 787185).

\appendix 

\section{SU(2) BF theory on the disk and trumpet} 
\label{sec:SU(2)-BF-appendix}

In this appendix, we will consider some salient aspects of 2D BF theory, focusing on the case with gauge group $G=SU(2)$. We will review the computations of the partition functions on a disk and a trumpet. Our emphasis will be on the localization computation, as the method and results for when $G=SU(2)$ will be relevant for computations in $\cN=4$ JT supergravity. 

For the computation of the gravitational index in $\cN=4$ JT supergravity, one needs to turn on holonomies for the R-symmetry gauge field corresponding to the center element $-1\in SU(2)$. There are subtleties associated to turning on such special values of holonomies, which we will investigate in the simpler case of the BF theory. We will see that special care needs to be taken with the classical saddles of the BF theory path integral when the holonomies take values in the center $Z(G)$ of the gauge group $G$. For such holonomies, we will show that the space of classical solutions is enhanced from isolated points to disjoint unions of finite dimensional moduli spaces. Accordingly, care must be taken when quantizing the theory and performing the path integral due to the presence of bosonic zero-modes. In particular, one must separate the integral over the zero modes along the moduli space of solutions from the rest of the fluctuations. We will also see that localization via the Duistermaat-Heckman formula needs to be amended appropriately to account for such degenerate cases.

\subsection{Disk}

We start with considering 2D BF theory on the disk. A nice review of BF theory is given in
~\cite{Kapec:2019ecr}; we will focus on presenting complementary details that are relevant for our purposes. On a two-manifold $\Sigma$ with boundary, the action of BF theory in the notation of section \ref{sec:diskN4} is~\cite{Blommaert:2018oro}
\be -i \int_\Sigma \Tr ~b H - \Phi_r \int_{\partial \Sigma} d\tau \Tr \left[ (B_\tau +\mu)^2\right],\ee where $H$ is the field strength of the gauge field $B$ and $b$ is an adjoint scalar. One also has the boundary condition labeled by a number $\Phi_r$ and $\mu\in \mathfrak{g}\equiv {\rm Lie}(G)$, given by
\be b= 2i \Phi_r(B_\tau+\mu)\big|_{\partial\Sigma}. \ee
With this choice $\mu$ plays the role of boundary chemical potential. 
After performing the path integral for the $b$ field in the bulk, one obtains the constraint $H=0$, so we are left with a path integral over flat connections. On the disk, a flat connection takes the form 
\be 
\label{eq:BF bulk connection}
B = g d g^{-1}, 
\ee where $g(x)$ is a map from the disk to the gauge group $G$. The gauge theory identifies solutions related by gauge transformations, so there are no physical modes in the bulk. However, allowed gauge transformations vanish on the boundary, so the boundary modes of $g$ are physical. On the boundary, we have $B_\tau = g\partial_\tau g^{-1}$, with action 
\be
I = - 
\Phi_r \int_0^\beta d\tau \Tr \left[ (g \partial_\tau g^{-1} + \mu )^2 \right].
\ee Now $g(\tau)$ is a map $S^1_\beta \to G$, and the action is that of a 1d sigma model into the group manifold $G$~\cite{Blommaert:2018oue}.
The parameter $\Phi_r$ controls the size of quantum effects. In $\cN=4$ super-JT, due to supersymmetry the BF theory part of the action has this parameter equal to the coupling of the Schwarzian theory set by the boundary value of the dilaton, $\Phi_r$, and in order to make direct contact with the rest of this article we denote it as such here. 

The action with $\mu=0$ possesses two copies of $G$ as global symmetries, $G_L\times G_R$, each copy acting on the field $g$ by constant left and right multiplications, respectively. The chemical potential $\mu$ is for the left-multiplication symmetry $G_L$. For generic values of $\mu$, $G_L$ is broken to a Cartan torus $T_L$, one for which $ \mu \in {\rm Lie}( T_L)$. Furthermore, right-multiplication by a constant group element leaves the bulk connection~\eqref{eq:BF bulk connection} invariant. Therefore, the symmetry $G_R$ is redundant, and we should quotient the space of fields ${\rm Loop}(G)$ by the right-acting $G_R$, which identifies the correct integration space as ${\rm Loop}(G)/G$.\footnote{The path integral over ${\rm Loop}(G)$ is related to the path integral over ${\rm Loop}(G)/G$ by the volume of the group $\vol(G)$ (as measured by an appropriate metric), coming from integrating over the right $G$ action. This was explained in \cite{Kapec:2019ecr} as the difference between the ultralocal and the symplectic measures on the respective field spaces.} 

In the disk, the group valued field has to be single valued $g(\beta) = g(0)$. In order to incorporate the effect of the chemical potential $\mu$, we can redefine the field $g(\tau)$ as $\tilde g (\tau) = e^{-\mu \tau} g(\tau)$ and the action takes the simpler form
\be
\label{eq:particle on group simple action}
I = -\Phi_r \int_0^\beta d\tau \Tr \left[ (\tilde g \partial_\tau \tilde g^{-1} )^2 \right],
\ee but the new field has the modified boundary condition 
\be \label{eq:g boundary condition disk}
\tilde g(\beta) = e^{-\beta \mu} \tilde g(0).
\ee In the context of BF theory coupled to JT gravity this change of variables also has the advantage of decoupling the particle on a group action with the Schwarzian mode \cite{Mertens:2019tcm}. The equation of motion is
\be
\partial_\tau \Tr(\tilde g \partial_\tau \tilde g^{-1})^2 =0.
\ee
We begin considering the case of an arbitrary group $G$. The classical solutions are given by $\tilde g= \exp(i \epsilon_i t^i \tau)$, where $t^i$ are a basis of generators for $\mathfrak{g}$
and the constants $\epsilon^i$ depend on the boundary conditions. For generic $\mu$, there are a discrete set of classical solutions. To describe them, we use part of the $G_L$ symmetry to diagonalize $\mu$ to lie along a Cartan subalgebra $\mathfrak{g}_0= {\rm Lie}(T_L)$, breaking the symmetry group to $T_L$. For what follows, it will be convenient to write $-\beta \mu = 2 \pi i \alpha$. We pick generators $t_0^i$ for $\mathfrak{g}_0$ such that $\exp(2\pi i t_0^i)=1$ for each $i$. In terms of these generators, we can diagonalize $\alpha = \alpha_i t_0^i$. Recalling the boundary condition~(\ref{eq:g boundary condition disk}), we can now determine the classical solutions to be
\be
\label{eq:classical solutions generic}
\tilde g_{n_i}(\tau) = \exp \left(2\pi i t_0^i (n_i+\alpha_i) \frac{\tau}{\beta} \right),
\ee where $n_i \in \Z^r$ with $r={\rm rank} (G)$. Since the solutions lie along the Cartan subalgebra, we can compensate for the action of $T_L$ (by constant left-multiplication) by an appropriate right-action in $T_R\subset G_R$, which was gauged. Therefore, the classical solutions are non-degenerate points labeled by $n_i$. Specializing to the case of $G=SU(2)$, we have the simple expression
\be
\label{eq:su2 saddles}
\tilde g_n(\tau ) =  \exp \left(2\pi i \sigma_3 (n+\alpha) \frac{\tau}{\beta} \right),
\ee where $n\in \Z$ and $\sigma_3$ is the usual Pauli matrix generating the chosen Cartan subalgebra of $\mathfrak{su}(2)$.

Before carrying on with the computation of the disk partition function, let us analyze the space of classical solutions at special values of $\mu$, or equivalently $\alpha$. The special values are when $\exp(2\pi i \alpha)$ commutes with more than just a Cartan torus $T_L$, and thus $G_L$ is either unbroken or broken to a subgroup strictly larger than $T_L$. The prototypical case for such a scenario is when $\exp(2\pi i \alpha)\in Z(G)$, in which case $G_L$ is unbroken. This turns out to be exactly the relevant case for our purposes since computing the index in section \ref{sec:diskN4} involves picking $e^{2\pi i \alpha}=-1\in Z(SU(2))$.

Before launching into the details, it is instructive to give a heuristic explanation for why such values of $\alpha$ are special. A generic $\alpha$ picks out a particular direction within the Lie algebra $\mathfrak{g}$, which can be thought of as the direction of the shortest path from the identity to the group element $\exp(2\pi i \alpha)$. Indeed, the classical solutions~(\ref{eq:classical solutions generic}) are the geodesics on $G$ connecting $1$ and $\exp(2\pi i \alpha)$, with $n_i$ counting the windings around the Cartan torus $T$. When $\exp(2\pi i \alpha)\in Z(G)$, there are many equivalent paths related by rotations in $G/T$. In other words, given a (nontrivial) classical solution $\tilde g_{n_i}(\tau)$ with windings along a given Cartan torus $T$, we can generate new solutions using the restored $G_L$ action as
\be
\tilde g_{n_i,h}(\tau) = h \tilde g_{n_i}(\tau) h^{-1}, \qquad h\in G/T\, .
\ee Therefore, except when $\tilde g_{n_i}(\tau)$ is the constant solution with $n_i=0$, $\a=0$, the non-degenerate set of classical solutions is enhanced to a disjoint union of moduli spaces isomorphic to $G/T$.\footnote{Furthermore, classical solutions with $n_i$ that are related by the action of the Weyl group combine into the same connected component, as we will see below explicitly for $G=SU(2)$.}

In the case of $G=SU(2)$, we can be very explicit (the generalization to arbitrary $G$ is straightforward). The special values are when $\exp(2\pi i \alpha) = \pm  1$. Thinking of $SU(2)$ as $S^3$, we need to count geodesics between a point chosen as the origin and itself, or between the origin and the antipodal point. Except for the geodesic of zero length between the origin and itself, there are an $S^2\cong SU(2)/U(1)$ worth of directions for the geodesics to lie along. We can write the corresponding moduli spaces of classical solutions for given winding $n$ as 
\be
\label{eq:disk degenerate saddles su2}
\cM_n^\a = \left\{ \tilde g_{n,\hat y}(\tau ) =  \exp \left(2\pi i \, \hat y \cdot \sigma \, (n+\alpha) \frac{\tau}{\beta} \right), \quad \hat y \in S^2\cong SU(2)/U(1)\right\}, \quad \text{for }\alpha =0,\tfrac{1}{2}.
\ee We have introduced a unit vector $\hat y = (y^1,y^2,y^3)\in \mathfrak{g}^*$ to parametrize the direction $\hat y \cdot \sigma = \hat y^i \sigma_i$ chosen within $\mathfrak{g}=\mathfrak{su}(2)$. Finally, we note that the solutions are now labeled by $n=0,1,\cdots$, since windings with negative $n$ are mapped into this set by $\hat y \mapsto -\hat y$. In summary, when $\exp(2\pi i \alpha)\in Z(G)$, there is a moduli space of classical solutions which is enhanced from a collection of isolated points to include continuous sets. For $G=SU(2)$, the moduli spaces $\cM^\a$ are given by the disjoint unions
\be
\label{eq:disk moduli spaces}
\alpha = 0: &\qquad \cM^0 = \bigsqcup_{n\ge0} \cM_n^0 \cong  \bullet  \, \sqcup \bigsqcup_{n\ge 1} S^2 \nn
\\ \alpha = \tfrac{1}{2}: &\qquad \cM^{\frac{1}{2}} \! = \bigsqcup_{n\ge0} \cM_n^{\frac{1}{2}} \cong \bigsqcup_{n\ge 0} S^2 \nn\\ 
\alpha \ne 0,\tfrac{1}{2}: &\qquad \cM^{\a} \!= \bigsqcup_{n\in \Z} \cM_n^\a \cong \bigsqcup_{n\in \Z} \bullet \cong \Z \, ,
\ee
where $\bullet$ denotes a (non-degenerate) point.

When evaluating the path integral, care must be taken with the enhanced moduli space of classical solutions, as the directions along the moduli spaces are zero modes. Practically speaking, the integral along the zero mode directions are pulled out from the path integral. One first performs the path integral over the oscillator modes for each point on the moduli space, and finally integrates over the compact moduli space. A similar discussion for 2D Yang-Mills can be found in \cite{Witten:1992xu}.

We now turn to the computation of the partition function, focusing on the case of $G=SU(2)$. The path integral can be evaluated via localization \cite{picken1989propagator,Witten:1992xu} or via canonical quantization~\cite{Witten:1991we}. We will begin by briefly reviewing the calculation via localization. 

When the moduli space consists of isolated non-degenerate points, as in the case for generic $\alpha$ above, the localization result is one-loop exact \cite{picken1989propagator}. Accordingly, the answer takes the form of a sum over saddles with one-loop fluctuations, schematically
\be
Z= \sum_{n\in \Z} Z_{\text{1-loop}} e^{-I^{\text{on-shell}}}\, .
\ee For the saddles~(\ref{eq:su2 saddles}), the action is 
\be
I^{\text{on-shell}} = \frac{8 \pi ^2 (n+ \alpha)^2 \Phi_r}{\beta}\, .
\ee To compute the one-loop fluctuations around the saddles, we can parametrize the fluctuating field as
\be
\tilde g(\tau) = \exp\left( \left[ 2\pi i (n+\a) \frac{\tau}{\beta}+i \epsilon_3(\tau)\right] \sigma_3 \right) \exp\left( i \epsilon_2(\tau) \sigma_2 \right) \exp\left( i \epsilon_1(\tau) \sigma_1\right)\, .
\ee The action to quadratic order is given by \cite{picken1989propagator,Heydeman:2020hhw}
\be
I_\text{quad} &= \frac{8\pi^2 \Phi_r}{\beta} (n+\a)^2 + 2\Phi_r \int_0^\beta d\tau \left( (\e_1(\tau)')^2+(\e_2(\tau)')^2+(\e_3(\tau)')^2  + \frac{8\pi(n+\a)}{\beta} \e_2(\tau) \e_1(\tau)'\right)\, ,
\ee and the corresponding regularized one-loop determinant around each saddle $n$ gives~\cite{picken1989propagator}
\be
Z_{\text{1-loop}} = C \frac{\Phi_r^{3/2}}{\beta^{3/2}} \frac{(n+\a)}{\sin(2\pi \a)},
\ee where $C$ is some regularization-dependent constant. We can understand the factor $\Phi_r^{3/2}/\beta^{3/2}$ as coming from removing the three zero modes of constant $\e_i$. These zero modes generate precisely the $G_R\cong SU(2)$ symmetry which is gauged, and modding them out produces this factor in the one-loop determinant.  

Putting all these ingredients together, we obtain the disk partition function
\be
\label{eq:BF Z_disk at generic holonomy from localization}
Z_\text{disk} = C \sum_{n\in \Z} \frac{\Phi_r^{3/2}}{\beta^{3/2}}  \frac{(n+\a)}{ \sin(2\pi \a)} e^{-\frac{8\pi^2 \Phi_r}{\beta} (n+\a)^2} \, .
\ee

The prefactor $C$ can be obtained by picking a particular regulator, for example the zeta function, or by comparing with the result from canonical quantization of BF theory
\be
Z_\text{disk} &= \sum_{J=0,\frac{1}{2},\cdots} (2J+1) \chi_J(\alpha) e^{-\frac{\beta}{2\Phi_r} \left(J(J+1)+\frac{1}{4}\right)} \nn
\\ &= \sum_{n\in \Z} \frac{\Phi_r^{3/2}}{\beta^{3/2}}\frac{4(2\pi)^{3/2}(n+\alpha)}{\sin \left(2\pi \alpha\right)} e^{-\frac{8\pi^2\Phi_r}{\beta}(n+ \alpha)^2} \, ,
\ee
with $\chi_J(\alpha) = \frac{\sin (2\pi \alpha(2J+1))}{\sin(2\pi \alpha)}$ the character of the spin-$J$ $SU(2)$ representation. This gives $C=4(2\pi)^{3/2}$, which we adopt.

Finally, we turn to analyzing the partition function at the special holonomies $\a =0,\frac{1}{2}$ where the solutions are not isolated but have moduli spaces of positive dimension. The easiest way to obtain the partition function at these holonomies is to take the limits as $\a\to 0,\frac{1}{2}$ of the expression~\eqref{eq:BF Z_disk at generic holonomy from localization} at generic $\a$, yielding
\be
\label{eq:Z disk at alpha=0}
\lim_{\a\to 0}Z_{\text{disk}} &= \frac{C}{2\pi} \frac{\Phi_r^{3/2}}{ \beta^{3/2}} +  \sum_{n\ge 1} e^{-\frac{8 \pi ^2 n^2 \Phi_r}{\beta }} C\left(-16 \pi  n^2 \frac{\Phi_r^{5/2}}{\beta^{5/2} }+ \frac{1}{\pi}\frac{\Phi_r^{3/2}}{\beta^{3/2} }   \right) ,
\\ 
\label{eq:Z disk at alpha=1/2}
\lim_{\a\to \frac{1}{2}}Z_{\text{disk}} &=  \sum_{n\ge 0} e^{-\frac{8 \pi ^2 \left(n+\frac{1}{2}\right)^2 \Phi_r }{\beta }} C \left(16 \pi  \left(n+\tfrac{1}{2}\right)^2 \frac{\Phi_r ^{5/2} }{\beta ^{5/2}} -\frac{1}{\pi} \frac{\Phi_r ^{3/2} }{\beta ^{3/2}} \right) .
\ee
Such limits yield the correct expressions due to the validity of the canonical quantization computation. However, the localization computation has some added subtleties at these special holonomies, which we will now comment on.

We observe that the result has the expected properties due to the enhanced moduli spaces~\eqref{eq:disk moduli spaces}. The $n=0$ saddle for the $\a=0$ case remains non-degenerate, and correspondingly the localization around this saddle is still one-loop exact. This can be seen from the corresponding factor $\Phi_r^{3/2}/ \beta^{3/2}$, meaning there are the same three zero modes as in the one-loop determinant around the non-degenerate saddles at generic holonomy. The absence of higher-order terms in the coupling $\beta/\Phi_r$ for this saddle imply that the localization around the saddle is one-loop exact. 

The rest of the saddles are degenerate, with 2-dimensional moduli spaces $\cM^\a_n\cong S^2$. Accordingly, localization around these degenerate saddles are two-loop exact \cite{Witten:1992xu}. Finally, we note that although the degenerate moduli spaces introduce bosonic zero modes around the saddles, the path integral is indeed finite. This is due to the fact that after correctly treating the bosonic zero-modes, one obtains integrals over the compact, finite-dimensional moduli spaces $\cM^\a_n$ which yield finite results, as we now explain.

Firstly, let's argue that the action for fluctuations around the degenerate saddles is independent of the coordinate on the moduli space. We can parametrize fluctuations around the degenerate saddles~\eqref{eq:disk degenerate saddles su2} analogously to the non-degenerate saddles,
\be
\tilde g_{n,\hat y} (\tau) = \exp\left(2\pi i (n+\a) \frac{\tau}{\beta} \hat y\cdot \sigma\right) \exp(i \e_3(\tau) \sigma_3)\exp(i \e_2(\tau) \sigma_2)\exp(i \e_2(\tau) \sigma_2), \qquad \a=0,\tfrac{1}{2}\, .
\ee We can relate fluctuations around different points $\hat y$ by translating along the moduli space via conjugations by $h\in G/T$. Since $\exp\left(2\pi i (n+\a) \frac{\tau}{\beta} \hat y\cdot \sigma\right) = h \exp\left(2\pi i (n+\a) \frac{\tau}{\beta} \sigma_3 \right) h^{-1} $ at $\a=0,\tfrac{1}{2}$ for some suitable $h\in G/T$, we can redefine the fluctuations $\e_h= h \e h^{-1}$ around each point $h$ of the moduli space. Recalling that the action is invariant under $G_L\times G_R$ at the special holonomies of interest, we see that the action $I_{n,h}[\e_i]$ for fluctuations around the degenerate saddles is independent of $h$. Therefore, the path integral factorizes: 
\be
\label{eq:degenerate path integral invariance on saddle}
Z_\text{disk} \supset \sum_n \int_{\cM_n^\a \cong G/T} \mu_n^\a(h) \int \cD\e_i e^{-I_{n,h}[\e_i]} = \sum_n \vol(\cM_n^\a) \int \cD\e_i e^{-I_{n}[\e_i]} \, .
\ee The integrals over the moduli spaces $\cM_n^\a$ are done with a suitable measure $\mu$, which is inherited from the symplectic form $\omega$ as
\be
\mu_n^\a(h) = \int \left(\prod_i d\psi_i\right) \exp({\omega|_{\cM_n^\a}(h)} )\, .
\ee The volume $\vol(\cM_n^\a)$ is as measured by $\mu_n^\a$.

We are left with evaluating the one-loop path integral and integrating over the moduli space. It suffices to evaluate the one-loop determinant around a single saddle, which we pick to be $h=1$ ($\vec y = (0,0,1)$). Expanding the fluctuations in modes, $\e_i = \sum_{m\in \Z} \e_{i,m} e^{2\pi i m \frac{\tau}{\beta}}$, and writing $\e_{i,\pm m} =\e_{i,m}^R \pm i \e_{i,m}^I$ for $m>0$
, the quadratic action is 
\be
I_\text{quad} &= \frac{16 \pi^2 \Phi_r}{\beta} \sum_{m > 0} \bigg[  m^2 ((\epsilon_{1,m}^R)^2 + (\epsilon_{2,m}^R)^2 + (\epsilon_{3,m}^R)^2+ (\epsilon_{1,m}^I)^2+ (\epsilon_{2,m}^I)^2+ (\epsilon_{3,m}^I)^2) \nn \\ & \qquad \qquad \qquad \qquad  - 4 m(n+\alpha)(\epsilon_{1, m}^R \epsilon_{2, m}^I - \epsilon_{1, m}^I \epsilon_{2, m}^R)
\bigg]\,. 
\ee
Considering generic holonomies for a second, the regularized one-loop determinant gives 
\be
\det_{\text{one-loop}} \propto \frac{\Phi_r^{3/2}}{\beta^{3/2}} \prod_{m\ge 1} \frac{m^2}{m^2 -4(n+\a)^2} = \frac{\Phi_r^{3/2}}{\beta^{3/2}} \frac{2\pi (n+\a)}{\sin(2\pi \a)} \, ,
\ee where we have removed the three constant-$\e_i$ zero modes as before. Now, we see that at the special holonomies $\a=0,\tfrac{1}{2}$, the modes $\e_{i,m}^R$ with $i=1,2$ and $m=2(n+\a)$ give rise to zero modes with a divergent determinant. This is indeed expected, since exactly these modes correspond to perturbing the saddle $g_n(\tau)= e^{2\pi i (n+\a)\frac{\tau}{\beta} \s_3}$ along the moduli space. We should remove the integral over these infinitesimal modes, and instead integrate over the full compact moduli space, as anticipated in~\eqref{eq:degenerate path integral invariance on saddle}. Remembering that the symplectic form 
\be
\omega = -c \sum_i \sum_{m\ge 1} (8\pi m) d\e_{i,m}^R\wedge d\e_{i,m}^I
\ee
gives the symplectic measure for the integral over fluctuations \cite{Kapec:2019ecr}
\be
\mu_\text{symplectic} = \prod_{i}\prod_{m\ge 1} (8\pi c m) d\e_{i,m}^R d\e_{i,m}^I,
\ee removing these zero modes from the functional determinant gives the regularized answer 
\be
\label{eq:degenerate moduli space one loop det}
\det_{\text{one-loop}} \propto \frac{1}{16 \pi c (n+\a)} \frac{\Phi_r^{5/2}}{\beta^{5/2}}  \frac{\prod_{m\ge 1} m^2}{\prod_{\substack{m\ge 1\\ m\ne 2(n+\a)}} (m^2 -4(n+\a)^2)} =  (-1)^{2\a+1} \frac{(n+\a)}{2 \pi c } \frac{\Phi_r^{5/2}}{\beta^{5/2}}\, ,  \qquad \a=0,\tfrac{1}{2}.
\ee
Finally, we have to integrate over the moduli space $G/T$, with respect to the measure induced by the symplectic form. Parametrizing the moduli space as $g_{n,\hat y}(\tau) = \exp(i  f_n(\tau) \hat y\cdot \sigma)$ with $f_n(\tau) = 2\pi(n+\a)\frac{\tau}{\beta}$ and coordinates $\hat y = (\sin \theta\sin\phi,\sin\theta\cos\phi,\cos\theta)$, the symplectic form on the moduli space is
\be
\omega|_{\cM_n^\a} &= c \int_0^\beta d\tau \Tr\left[ dg^{-1} g \wedge \partial_\tau(dg^{-1} g) \right] \nn
\\ &= 4c \psi_2 \psi_1 \sin\theta \int_0^\beta d\tau f'(\tau) \sin^2f(\tau)  = 4\pi c (n+\a) \psi_2 \psi_1 \sin\theta\, ,
\ee where $\psi_1 = \delta \theta$ and $\psi_2=\delta \phi$. Therefore the symplectic form gives the volume
\be
\vol\left(\cM_n^{\a=0,\frac{1}{2}}\right) = \int_{\cM_n^\a} \int d\psi_1 d\psi_2 e^{\omega} = 16\pi^2 c (n+\a) \, .
\ee Combining with the determinant~\eqref{eq:degenerate moduli space one loop det}, we obtain the one-loop factor
\be
Z_{\text{1-loop}} \propto (-1)^{2\a+1} 8\pi (n+\a)^2 \frac{\Phi_r^{5/2}}{\beta^{5/2}}\, ,
\ee 
which agrees with the 1-loop pieces obtained by taking the $\a \to 0,\frac{1}{2}$ limits in~\eqref{eq:Z disk at alpha=0} and~\eqref{eq:Z disk at alpha=1/2} up to a nonzero proportionality constant that is independent of $n$ and $\alpha$. The two-loop contribution can be computed \emph{\`a la} \cite{Stanford:2017thb}, but we will omit doing so here.

\subsection{Trumpet}\label{sec:appBFtrumpet}

Next, we turn to the partition function of BF theory on a trumpet. Topologically, the space is a cylinder, but with different boundary conditions on each end. One end has the same boundary conditions as the disk example above, while the other end has ``topological'' boundary conditions without any fluctuating fields (see \cite{Kapec:2019ecr}). Correspondingly, the theory can be reduced to the particle-on-a-group theory on the non-topological boundary as with the disk example, but with a suitable generalization accounting for the holonomy one can now turn on for the bulk gauge field around the topological boundary. 
We once again begin by considering a generic simple Lie group $G$, and we will later specialize to $G=SU(2)$. 
The general flat gauge field in the bulk is given by
\be
B = g(a d\tau) g^{-1} + g d g^{-1} ,
\ee 
with corresponding holonomy $e^{-\beta a}$ around the periodic direction $\tau$. The resulting action on the boundary is~\cite{Kapec:2019ecr}
\be
I = -\Phi_r \int_0^\beta d\tau \Tr\left[ \left( g \partial_\tau g^{-1} + g a g^{-1} + \mu\right)^2 \right]\, .
\ee Once again, we can redefine the group-valued field as 
\be
\tilde g (\tau) = e^{-\mu \tau} g(\tau) e^{-a\tau}
\ee to obtain the simpler action~\eqref{eq:particle on group simple action}, 
where now the field satisfies the periodicity condition
\be
\tilde g(\beta) = V^{-1} \tilde g (0) U, \qquad U= e^{-\b a}, \quad V= e^{\beta \mu}.
\ee
For generic $U,V$, the global symmetry of the system is broken from $G_L\times G_R$ down to $T_L\times T_R$ due to these boundary conditions. 
To solve for the classical configurations, we diagonalize $U,V$ via their matrices of eigenvectors $A_{U,V}$:
\be
\label{eq:diagonalize U and V}
U = A_U D_U A_U^{-1}, \qquad V=A_V D_V A_V^{-1}.
\ee 
Redefining the group-valued field one last time as
\be
h(\tau) = A_V^{-1} \tilde{g}(\tau) A_U,
\ee we have the simpler, diagonal periodicity condition
\be
\label{eq:diagonal periodicity condition}
h(\beta) = D_V^{-1} h(0) D_U .
\ee The action~\eqref{eq:particle on group simple action} remains of the same form since it is invariant under constant left or right multiplications. In terms of the generators $t_0^i$ of the Cartan subalgebra $\mathfrak{g}_0$, we can write
\be
\label{eq:diagonalize holonomies}
D_U = e^{2\pi i \varphi \cdot t_0}, \qquad D_V = e^{2\pi i \alpha \cdot t_0}.
\ee The vectors $\varphi^i$ and $\alpha^i$ are related to the eigenvalues of the holonomy $a$ and chemical potential $\mu$, respectively. The classical solutions with the diagonalized boundary conditions~(\ref{eq:diagonal periodicity condition}) (for generic holonomies $\varphi,\alpha$) are
\be
h_\g(\tau) = \exp\left( 2\pi i \frac{\tau}{\beta} (\varphi-\alpha+\g)\cdot t_0\right).
\ee The vector of integers $\gamma$ takes values in the lattice $\Gamma\cong\mZ^r$ dual to the generators $t_0$, labeling closed geodesics on the Cartan torus $T^r$ where $r={\rm rank~} G$. 

We have thus found the classical solutions with a given diagonalization~\eqref{eq:diagonalize U and V} of the holonomies $U,V$. However, note that such a diagonalization is unique only up to the action of the Weyl group $W$ of $G$, which acts by reordering the eigenvalues. So, the periodicity conditions~(\ref{eq:diagonal periodicity condition}) are not general enough, and we can generate the general set by acting on the diagonal decomposition of $U$ and $V$ by the Weyl group $W$. For $w_i\in W$, this gives
\be
U&= A_U w_2^{-1} w_2 D_U w_2^{-1} w_2 A_U^{-1},
\nn \\ V&= A_V w_1^{-1} w_1 D_V w_1^{-1} w_1 A_V^{-1}.
\ee
The corresponding periodicity conditions are
\be
h(\beta) = w_1 D_V^{-1} w_1^{-1} h(0) w_2 D_U w_2^{-1} .
\ee The classical solutions for these periodicity conditions are
\be
h_{w_1,w_2}(\tau) = w_1 e^{-2\pi i \frac{\tau}{\beta} \alpha \cdot t_0 } w_1^{-1} e^{2\pi i \frac{\tau}{\beta} \gamma\cdot t_0} w_2 e^{2\pi i \frac{\tau}{\beta} \varphi \cdot t_0 } w_2^{-1}.
\ee We can observe that since for any $w\in W$, \be
\{ w (\gamma\cdot t_0) w^{-1} | \gamma\in \Gamma \} = \{ \gamma\cdot t_0 | \gamma\in \Gamma \},
\ee
one Weyl group worth of transformations do not generate new solutions. Defining an action of the Weyl group on the vectors $v\in \mathfrak{g}^*$ via
\be
w(v)\cdot t_0 \equiv v\cdot (w t_0 w^{-1}) ,
\ee we can finally list all classical solutions: 
\be
h_{\g,w}(\tau) = \exp\left(2\pi i \frac{\tau}{\beta} (\varphi-w(\alpha) +\gamma)\cdot t_0\right),
\ee where $\g\in \Z^r$ and $w\in W$.

For the case of $SU(2)$, we can pick the Cartan subalgebra to lie along $\sigma_3$, and we have the classical solutions 
\be
\tilde{g}_{n,\pm} (\tau) = A_V^{\pm} \exp\left( 2\pi i \frac{\tau}{\beta} (\varphi \mp \alpha+n) \sigma_3 \right) A_U^{-1}, \qquad n\in \mZ,
\ee where $A_V^+ = A_V$ and $A_V^{-} = A_V (i\sigma_2)^{-1}$. Concretely, we have diagonalized the holonomy and chemical potential as
\be
U&= e^{-\beta a} = A_U e^{2\pi i \varphi \sigma_3} A_U^{-1}, \nn
\\ V&= e^{\beta \mu} = A_V e^{2\pi i \alpha \sigma_3} A_V^{-1}.
\ee
The actions of these solutions are 
\be
I_{n,\pm} = \frac{8\pi^2 \Phi_r }{\beta} (\varphi \mp \alpha +n)^2.
\ee

When the parameters $\a,\varphi$ take special values, once again the moduli space of classical solutions is enhanced to a degenerate case with positive-dimensional components. The corresponding degenerate moduli space for the trumpet can be solved for along the lines of the disk case as above. The exact form of the degenerate moduli space is not strictly necessary for our purposes, hence we will omit discussing it. The main important aspect is that the moduli space is compact and finite dimensional, implying that the integral over the bosonic zero-modes in the path integral is finite, as in the case of the disk.

Having determined the classical solutions relevant for the trumpet, we now turn to computing the partition function from one-loop localization. We will specialize once again to the case of $G= SU(2)$. We can parametrize the fluctuations of the field as
\be
h(\tau) = e^{2\pi i \frac{\tau}{\beta} \alpha \sigma_3 + i \e_3(\tau) \s_3} e^{i \e_2(\tau) \s_2}e^{i \e_1(\tau) \s_1} e^{2\pi i \frac{\tau}{\beta} \varphi \s_3}.
\ee 
The action of the quadratic order fluctuations around the saddle $n$ and $w=+1$ is given by
\be
I_\text{quad} = 2\Phi_r \int_0^\beta d\tau \left[ (\epsilon'_1)^2+(\epsilon'_2)^2+(\epsilon'_3)^2 - \frac{16\pi^2(\alpha+n)\varphi}{\beta^2}(\epsilon_1^2+\epsilon_2^2)+\frac{8\pi}{\beta}(n+\alpha-\varphi)\epsilon_2\epsilon_1'\right] \, .
\ee
There is a single zero mode corresponding to constant $\e_3$, which we mod out by. For generic holonomies, $\e_{1,2}$ are massive, and we integrate over their constant modes. Expanding the fluctuations $\e_i$ in modes and integrating, we obtain the contribution of the bosonic fields to the one-loop term, up to a regularization-dependant constant  
\be
Z_{\text{1-loop}}^{\text{bosonic}} \propto \frac{\Phi_r^{1/2}}{\beta^{1/2}} \frac{1}{\sin(2\pi \alpha)\sin(2\pi \varphi)}\, .
\ee 
In addition, there is the contribution from the symplectic form 
\be
\omega_{\rm quad} = -4c \int_0^\beta d\tau \left[ \psi_1\psi_1'+\psi_2\psi_2'+\psi_3\psi_3' - \frac{8\pi \varphi}{\beta} \psi_1\psi_2 \right] \, ,
\ee 
where $c$ is some normalization constant and $\psi_i$ are the linearization around the identity of the variation $\delta g^{-1} g$, which yields a fermionic determinant of $\sin(2\pi\varphi)$, as well as an additional factor of $1/\sin(2\pi \varphi )$ coming from the ultralocal measure as explained in \cite{Kapec:2019ecr}. All together, the one-loop determinants give 
\be
Z_{\text{1-loop}} \propto \frac{\Phi_r^{1/2}}{\beta^{1/2}} \frac{1}{\sin(2\pi \alpha)\sin(2\pi \varphi)}\, .
\ee The one-loop factor for the saddle with $w=-1$ is related to the $w=+1$ saddle by sending $\alpha\to w(\alpha) = -\alpha$. 
Summing over the saddles $n$ and $w=\pm 1$, we obtain the localization result
\be
Z_\text{trumpet} \propto \sum_{n\in \Z} \frac{\Phi_r^{1/2}}{\beta^{1/2}} \frac{1}{\sin(2\pi \a) \sin(2\pi \varphi)} \left( e^{-\frac{8\pi^2 \Phi_r }{\beta}(\varphi-\a+n)^2} - e^{-\frac{8\pi^2 \Phi_r }{\beta}(\varphi+\a+n)^2}\right) \, .
\ee
We can once again compare to the canonical quantization answer from BF theory, which gives
\be
\label{eq:trumpet BF theory}
Z_\text{trumpet} &= \sum_{J=0,\frac{1}{2},\cdots} \chi_{J}(\a) \chi_J(\varphi) e^{-
\frac{\beta}{2\Phi_r} \left( J(J+1)+\frac{1}{4}\right)} \nn
\\ &= \sum_{n\in \mathbb{Z}} \frac{\Phi_r^{1/2}}{\beta^{1/2}} \frac{\sqrt{\pi}}{\sqrt{2}}\frac{1}{\sin(2\pi \alpha) \sin(2\pi \varphi)} \left( e^{-\frac{8\pi^2 \Phi_r }{\beta}(n-\alpha +\varphi)^2}-e^{-\frac{8\pi^2 \Phi_r }{\beta}(n+\alpha+\varphi)^2}\right).
\ee We can readily check that in the $\varphi\to 0$ limit we recover the disk partition function. Finally, let's analyze the partition function at $\a = 0,\frac{1}{2}$. Taking limits, we obtain 
\be
\lim_{\a\to 0} Z_\text{trumpet} &= 4 (2 \pi) ^{3/2} \sum_{n\in \Z} \frac{ \Phi_r ^{3/2}}{\beta ^{3/2}} \frac{ (n+\varphi )}{\sin (2 \pi  \varphi )} e^{-\frac{8 \pi ^2 \Phi_r  (n+\varphi )^2}{\beta }}\, , 
\\ \lim_{\a\to \frac{1}{2}} Z_\text{trumpet} &= - 4 (2 \pi) ^{3/2} \sum_{n\in \Z} \frac{\Phi_r ^{3/2}}{\beta^{3/2}} \frac{(n+\tfrac{1}{2}+\varphi )}{\sin (2 \pi  \varphi )} e^{-\frac{8 \pi ^2 \Phi_r  (n+\frac{1}{2}+\varphi )^2}{\beta }}\, ,
\ee which are perfectly regular for generic $\varphi$. Unlike the case of the disk, even at $\a=0,\tfrac{1}{2}$ the saddle-point solutions are still nondegenerate as long as $\varphi$ is generic and therefore breaking the $G_R\cong SU(2)$ symmetry to the Cartan torus $T_R\cong U(1)$. The extra power of $(\sqrt{\Phi_r/\beta})^{2}$ compared to arbitrary values of $\alpha$ comes from the two extra zero modes due to the enhancement of $T_L\cong U(1)$ to $G_L\cong SU(2)$.

\section{$\mathcal{N}=4$ JT supergravity and localization}

In this appendix, we compute the disk, trumpet, and defect partition functions of $\cN=4$ JT supergravity via localization.

\subsection{Review: Localization on the disk}
 We begin with briefly reviewing the localization calculation of the partition function for $\mathcal{N}=4$ JT supergravity on the disk. This was derived in \cite{Heydeman:2020hhw}, but it will be useful before jumping into the analogous calculation for the trumpet and the defect. 

As explained in section \ref{sec:diskN4}, $\mathcal{N}=4$ JT supergravity reduces to the $\mathcal{N}=4$ super-Schwarzian theory living on the one dimensional boundary. We have the reparametrization mode satisfying $f(\tau+\beta)=f(\tau)$ in our conventions, the $SU(2)$ mode $g(\tau+\beta) = e^{2 \pi i \alpha \sigma^3} g(\tau)$ where $g \in SU(2)$, and the $SU(2)$ fermion doublet $\eta(\tau+\beta) = - e^{2 \pi i \alpha \sigma^3}\eta(\tau)$. The action then consists of the Schwarzian action, the particle-moving-on-$SU(2)$ action studied in the previous appendix, and fermionic terms fixed by supersymmetry.

We will first quickly go over the bosonic saddle points. For the Schwarzian mode there is a unique saddle given by $f(\tau) = \tan(\pi \tau/\b)$. The saddles for the $SU(2)$ mode consistent with the boundary conditions are $g(\tau) = \exp\left(2\pi i \sigma_3 (n+\a) \frac{\tau}\beta\right)$ labeled by an integer $n$, as in appendix~\ref{sec:SU(2)-BF-appendix}. The action and one-loop determinant can be found in \cite{Heydeman:2020hhw}. 

The fermions vanish on the saddle points. Therefore, we only need to consider their quadratic fluctuations, the action for which is given by \cite{Heydeman:2020hhw}
\be
 I_{\text{ferm., quad.}} =  \Phi_r\int_0^{\beta} d\tau\, &\bigg(\eta^p \left[\frac{2\pi^2}{\beta^2}\left(1+2(n+\a)^2 \right) \partial_\tau - \partial_\tau^3\right]\bar \eta_p + \nn \\&+ \partial_\tau \eta^p \left[\frac{12\pi^2(n+\a)^2}{\b^2} + \frac{8i  \pi (n+\a)\partial_\tau}{\beta} - 3\partial_\tau^2 \right] \bar \eta_p\bigg) \, .
 \ee
 In this expression, the dependence on the temperature and the chemical potential appears from the underlying bosonic saddle we perturb around. After expanding in Fourier modes consistent with boundary conditions
 \bea \eta^1(\tau) &=&e^{i\frac{2\pi (n+\a)\tau}{\b} }\sum_{m_1 \in \dots, -\frac{1}2, \, \frac{1}2, \dots} \sqrt{\frac{\b}{2\pi}} \,\eta^1_{m_1} e^{-i\frac{2\pi m_1\tau}{\b} }\\
\eta^2(\tau) &=&e^{-i\frac{2\pi (n+\a)\tau}{\b} }\sum_{m_2 \in \dots, -\frac{1}2, \, \frac{1}2, \dots} \sqrt{\frac{\b}{2\pi}}\, \eta^2_{m_2} e^{i\frac{2\pi m_2\tau}{\b}}\,,
\ea
the quadratic action becomes 
\be
 I_{\text{ferm., quad.}} = \frac{2\pi^2 i \Phi_r}{\beta} \left[\sum_{p=1, 2}\,\sum_{m_p  \in \dots, -\frac{1}2, \, \frac{1}2, \dots} (m_p - n - 	\a) (4m_p^2-1) \eta^p_{m_p}  \bar \eta^p_{-m_p}\right]\,.
\ee
The fermion gauge zero modes determined from the symplectic form are given by the modes with $m_p=\pm1/2$ and form the eight fermionic generators of $PSU(1,1|2)$. In the expression above we can see explicitly that these modes also have zero action, as expected. The one-loop determinant is then given by 
\be
\label{eq:ferm-one-loop}
\det_{\text{ferm., one-loop}} \propto \frac{\beta^4}{\Phi_r^4} \prod_{p=1, 2}\,\,\prod_{m_p \in \dots, -\frac{5}2, -\frac{3}2, \frac{3}2, \frac{5}2, \dots} \frac{m-n -\a}{m} = \frac{\beta^4}{\Phi_r^4} \frac{\cos(\pi\a)^2}{(1 - 4(n+\a)^2)^2 }\, ,
\ee
up to a nonzero proportionality constant. We regularized this expression by dividing by the answer for $n=0$ and $\alpha=0$. We also remove by hand the fermion zero modes which are gauge modes with $m=\pm 1/2$. Finally, the temperature dependence $\beta^4$ is due to the total number---eight---of those zero modes.

\subsection{Localization on the trumpet}\label{app:N4trumpetlocalization}
The calculation in the trumpet is very similar and we will just point out the differences. Firstly, the bosonic saddles are different. The Schwarzian saddle depends on the geodesic length $b$ of the trumpet $f(\tau) \to \tanh \frac{b \tau}{2\beta}$. On the other hand, the saddle-point configurations of the $SU(2)$ field now depends on the chemical potential $\alpha$ and on the $SU(2)$ holonomy around the trumpet $\varphi$ which is now allowed to be non-trivial. The saddles for this configuration were written above in appendix~\ref{sec:appBFtrumpet}. The action on these saddles is given by 
\be 
\label{eq:on-shell-action-N=4}
I_{\cN=4,\text{bosonic}}^{\,\text{on-shell}} = \frac{2\pi^2 \Phi_r} {\b} \left(\left(\frac{b}{2\pi}\right)^2 + 4  (n\pm \a+ \varphi) ^2 \right)\,,
\ee
where the $\pm$ indicate the two families of saddles related by the action of the Weyl group, as explained in appendix \ref{sec:appBFtrumpet}. The bosonic contribution to the one-loop determinant is also straightforward.

Similar to the calculation for the disk, we can compute the quadratic action of the fermions around the bosonic saddles. Expanding in Fourier modes in a similar fashion, we obtain
\be
 I_{\text{ferm., quad.}} = \frac{2\pi^2 i \Phi_r}{\beta} \left[\sum_{p=1, 2}\,\sum_{m_p  \in \dots, -\frac{1}2, \, \frac{1}2, \dots} (m_p - n \mp 	\a) \left(4(m_p+\varphi)^2+
\left(\frac{b}{2\pi}\right)^2\right) \eta^p_{m_p}  \bar \eta^p_{-m_p}\right]\,.
\ee
We are interested in computing the dependence of the one-loop determinant on $n$, $\alpha$, $\varphi$, $\beta$, and $\Phi_r$. As in \cite{Stanford:2017thb}, to compute the $\alpha$ dependence of the one-loop determinant, we will regularize by dividing by the one-loop determinant with $n=0$ and $\alpha = 0$. We thus find that the regularized one-loop determinant is given, again up to a proportionality constant, by 
\be
\label{eq:ferm-one-loop2}
\det_{\text{ferm., one-loop}} &\propto \prod_{p=1, 2}\,\,\left(\prod_{m_p \in \dots, -\frac{1}2, \frac{1}2, \dots} \frac{m_p-n \mp \a}{m_p} \right) \sim \cos^2(\pi \alpha)\,. 
\ee
When dividing by the one-loop determinant with $n=\alpha=0$, we simply canceled the contribution from the factor $4(m+\varphi)^2+(b/2\pi)^2$. This is fixed by the localization procedure of Duistermaat-Heckman theorem that guarantees the answer is independent of the parameter $b$ and $\varphi$, see footnote 11 of \cite{Witten:2020wvy}. An independent way of verifying this would be to explicitly compute the determinant arising from the symplectic form. More importantly, for this prescription to be well defined we use the fact that for any half-integer $m$ this factor does not vanish; i.e. $4(m+\varphi)^2+(b/2\pi)^2\neq 0 $. 

Putting all these things together, we obtain the exact partition function of $\mathcal{N}=4$ JT gravity on a trumpet with geodesic length $b$ and $SU(2)$ holonomy $\varphi$ as 
\beq
\label{eq:trumpetfull}
Z_\text{Trumpet}(\beta,\alpha) \propto \sum_{n\in \mathbb{Z}} \frac{\Phi_r}{\beta} \frac{ \cot(\pi \alpha)}{\pi \sin(2\pi \varphi)}  \left( e^{-\frac{2\pi^2\Phi_r}{\beta}((\frac{b}{2\pi})^2+4(n-\alpha+\varphi)^2)}-e^{-\frac{2\pi^2\Phi_r}{\beta}((\frac{b}{2\pi})^2+4(n+\alpha+\varphi)^2)}\right) \,,
\eeq
up a prefactor independent of $\beta$, $\alpha$, $\varphi$, and $b$. This answer can be written in a more transparent way after an inverse Laplace transform; 
\beq\label{eq:N4trumpetlocalizationrho}
Z_\text{Trumpet}(\beta,\alpha) \propto \sum_{J\geq \frac{1}{2}} \left( \chi_J(\alpha) + 2 \chi_{J-\frac{1}{2}}(\alpha) + \chi_{J-1}(\alpha) \right) \chi_{J-\frac{1}{2}}(\varphi)   \hspace{0.1cm}\frac{e^{-\beta\frac{ J^2}{2\Phi_r}-\frac{\Phi_r b^2}{2\beta }}}{2\pi \sqrt{\pi \beta/\Phi_r}}\, .
\eeq
Since the trumpet only involves `long' supermultiplets and their indices vanish, the contribution of the trumpet to the index at $\alpha=1/2$ also vanishes.

\subsection{Localization on defects}\label{app:N4defectlocalization}
As explained in~\cite{Mertens:2019tcm} for example, the answer for JT gravity with defects can be obtained from the trumpet under the continuation $b\to 2\pi i \theta$, where $\theta$ parametrizes the deficit angle with $\theta=0$ being a cusp and $\theta=1$ being no defect. For generic values of $\theta$ and $\varphi$ such that $\theta \neq 1-2\varphi$, one can directly analytically continue \eqref{eq:trumpetfull} to obtain the defect partition function. In particular, the saddles for the $SU(2)$ gauge field are exactly the same as in the trumpet.

When $\theta=1-2\varphi$ and the geometric defect is related in this way to the $SU(2)$ holonomy around it, something new happens. There are new gauged fermion zero modes, only four of them in this case as opposed to the full eight on the disk. We will refer to these as supersymmetric defects. One way to check this enhancement of zero modes is to compute the symplectic form. Another way is by writing down the quadratic action for the fermions, and expanding in Fourier modes, 
\be
 I_{\text{ferm., quad.}} = \frac{2\pi^2 i \Phi_r}{\beta} \left[\sum_{p=1, 2}\,\sum_{m_p  \in \dots, -\frac{1}2, \, \frac{1}2, \dots} (m_p - n \mp 	\a) \left(4(m_p+\varphi)^2-\theta^2\right) \eta^p_{m_p}  \bar \eta^p_{-m_p}\right]\,.
\ee
When $\theta=1-2\varphi$, the second factor is given by 
\be
4(m_p+\varphi)^2-(1-2\varphi)^2 \to (2m_p+1)(2m_p+4\varphi-1).
\ee
It is clear now that the only gauge zero modes are coming from the four Fourier modes with $m_p=-1/2$ (while in the case of the disk with $\varphi=0$ all $m_p=\pm 1/2$ are zero modes). The one-loop determinant is now given by 
\beq
\det_\text{ferm., 1-loop} \sim \frac{\beta^2}{\Phi_r^2} \prod_{p=1, 2}\,\,\prod_{m_p \in \dots, -\frac{3}2, \frac{1}2, \dots} \frac{m_p-n \mp \alpha}{m_p} \sim \frac{\beta^2}{\Phi_r^2} \frac{\cos^2 (\pi \alpha)}{(1\pm 2\alpha+2n)^2}. 
\eeq
Putting everything together, we can find the partition function for supersymmetric defects with deficit angle $2\pi(1-2\varphi)$ and $SU(2)$ holonomy $\varphi$ to be
\beq
\label{eq:super-defect-part-function2}
Z_{\text{Susy-defect}}(\beta,\alpha)= \sum_{n\in\mathbb{Z}} \frac{\beta}{\Phi_r} \frac{\cot(\pi \alpha)}{8\pi^3 \sin(2\pi \varphi) } \left( \frac{e^{\frac{2\pi^2\Phi_r}{\beta}((1-2\varphi)^2-4(n-\alpha+\varphi)^2)}}{(1-2\alpha+2n)^2} -\frac{e^{\frac{2\pi^2\Phi_r}{\beta}((1-2\varphi)^2-4(n+\alpha+\varphi)^2)}}{(1+2\alpha+2n)^2}\right).
\eeq
Interestingly, this expression can be continued to $\varphi\to0$ and one recovers the answer for the disk. This is not \emph{a priori} obvious since the disk has a further enhancement of zero modes to the full $PSU(1,1|2)$ symmetry, and for example the $\varphi\to 0$ limit of non-supersymmetric defects does not match the disk. 

This expression for supersymmetric defects can also be obtained as a classical limit of the non-vacuum $\mathcal{N}=4$ super-Virasoro characters of Eguchi and Taormina by a supersymmetric generalization of \cite{Mertens:2017mtv}, as explained in \cite{Heydeman:2020hhw}, but we leave out the details here.

\bibliographystyle{utphys2}
{\small \bibliography{Biblio}{}}

\end{document}